\documentclass[a4paper,onecolumn,aps,showpacs,showkeys,nofootinbib,preprintnumbers,superscriptaddress,amsmath,amssymb,amsfonts]{revtex4}
\usepackage{graphicx}
\usepackage{amsmath}
\usepackage[latin1]{inputenc}
\usepackage[english]{babel}
\usepackage[T1]{fontenc}
\usepackage{amssymb}
\usepackage{amsfonts}
\usepackage{color}
\usepackage{float}
\usepackage{pifont}
\usepackage{bclogo}
\usepackage{epsfig}
\usepackage{colordvi}
\usepackage{psfrag}
\usepackage{color}
\usepackage{booktabs}
\usepackage{dcolumn}
\usepackage{multirow}
\usepackage{hyperref}
\usepackage{epstopdf}
\usepackage{times}
\hypersetup{
	colorlinks=true,        
	linkcolor=blue,         
	citecolor=magenta,      
}

\newcommand{\aap}{Astron. Astrophys.}
\newcommand{\aj}{Astronomical Journal}
\newcommand{\apjl}{Astrophys. J. Lett.}
\newcommand{\apjs}{Astrophys. J. Suppl. Series}
\newcommand{\apss}{Astrophysics and Space Science}
\newcommand{\araa}{Annual Review of Astronomy and Astrophysics}
\newcommand{\aapr}{The Astronomy and Astrophysics Review}

\newcommand{\jcap}{J. Cosmol. Astropart. Phys.}
\newcommand{\mnras}{Mon. Not. R. Astron. Soc.}
\newcommand{\pasa}{Publications of the Astronomical Society of Australia}

\newcommand{\physrep}{Physics Report}

\begin{document}

\title{Dark matters on the scale of galaxies}

\author{Ivan de Martino\footnote{Correspondence: ivan.demartino@unito.it}}
\affiliation{Dipartimento di Fisica, Universit\`a di Torino,  Via P. Giuria 1, I-10125 Torino, Italy} 
\affiliation{Istituto Nazionale di Fisica Nucleare (INFN), Sezione di Torino, Via P. Giuria 1, I-10125 Torino, Italy
} 
\author{Sankha S. Chakrabarty}
\affiliation{Dipartimento di Fisica, Universit\`a di Torino,  Via P. Giuria 1, I-10125 Torino, Italy} 
\affiliation{Istituto Nazionale di Fisica Nucleare (INFN), Sezione di Torino, Via P. Giuria 1, I-10125 Torino, Italy
}

\author{ Valentina Cesare}
\affiliation{Dipartimento di Fisica, Universit\`a di Torino,  Via P. Giuria 1, I-10125 Torino, Italy} 
\affiliation{Istituto Nazionale di Fisica Nucleare (INFN), Sezione di Torino, Via P. Giuria 1, I-10125 Torino, Italy
} 

\author{  Arianna Gallo}
\affiliation{Dipartimento di Fisica, Universit\`a di Torino,  Via P. Giuria 1, I-10125 Torino, Italy} 
\affiliation{Istituto Nazionale di Fisica Nucleare (INFN), Sezione di Torino, Via P. Giuria 1, I-10125 Torino, Italy
}

\author{ Luisa Ostorero}
\affiliation{Dipartimento di Fisica, Universit\`a di Torino,  Via P. Giuria 1, I-10125 Torino, Italy} 
\affiliation{Istituto Nazionale di Fisica Nucleare (INFN), Sezione di Torino, Via P. Giuria 1, I-10125 Torino, Italy
}  

\author{ Antonaldo Diaferio}
\affiliation{Dipartimento di Fisica, Universit\`a di Torino,  Via P. Giuria 1, I-10125 Torino, Italy} 
\affiliation{Istituto Nazionale di Fisica Nucleare (INFN), Sezione di Torino, Via P. Giuria 1, I-10125 Torino, Italy
} 

\begin{abstract} The cold dark matter model successfully explains both the emergence and evolution of cosmic structures on large scales and, when we include a cosmological constant, the properties of the homogeneous and isotropic Universe. However, the cold dark matter model faces persistent challenges on the scales of galaxies. 
{Indeed,} N-body simulations predict some galaxy properties that are at odds with the observations. These discrepancies are primarily related to the dark matter distribution in the innermost regions of the halos of galaxies and to the dynamical properties of dwarf galaxies. They may have three different origins: (1) the baryonic physics affecting galaxy formation is still poorly understood and it is thus not properly included in the model; (2) the actual properties of dark matter differs from those of the conventional cold dark matter; (3)  the theory of gravity departs from General Relativity.
Solving these discrepancies is a rapidly evolving research field. We illustrate some { of the solutions proposed} within the cold dark matter model, and solutions when including warm dark matter, self-interacting dark matter, axion-like particles, or fuzzy dark matter.  { We also 
illustrate some modifications of the theory of gravity: Modified Newtonian Dynamics (MOND), MOdified Gravity (MOG), and $f(R)$ gravity.}
\end{abstract}


\maketitle
\tableofcontents
\setcounter{section}{0} 

\section{Introduction}\label{sec:intro}

The fundamental nature of matter and the concept of its constituent particles have always been a central topic of philosophy and of natural sciences throughout the history of human thought. The development of new tools applied to the observation of nature has often led to the discoveries of new physical phenomena. These discoveries directly or indirectly implied the existence of particles that were previously invisible \cite{2018RvMP...90d5002B}. 

In the last century, the concept of invisible or {\it dark}  matter has reached the current connotation, although with some oversights, such as believing that the dark matter was made up of stars and/or interstellar medium too faint to be detected. It was J. H. Oort who, in 1932, first estimated the total amount of matter density near the Sun from dynamical data, and pointed out a discrepancy of a factor of up to 2 with the amount of the visible stellar populations \cite{1932BAN.....6..249O}. Although this result is often considered to be the first evidence of the existence of dark matter, the discrepancy has now been  alleviated by using more accurate observations of the stellar disk population \cite{1989MNRAS.239..605K,1989MNRAS.239..651K,Holmberg_2004}. For a comprehensive review, we refer to  \cite{2014JPhG...41f3101R}.

In 1933, F. Zwicky pointed out a discrepancy between the observed velocity dispersion along the line-of-sight of 8 galaxies in the Coma cluster ($\sim 1000$ km/s) and the one expected in a system  of $N$ massive galaxies in dynamical equilibrium  ($\sim 80$ km/s) \cite{1933AcHPh...6..110Z}.  This discrepancy implied the presence of a large amount of invisible mass; this mass was still thought to be in the form of stars and/or gas which were not yet observable. This result of Zwicky traditionally marks the birth of the dark matter problem.

A new era began in the 1970s, when V. C. Rubin and W. K. Ford measured the rotation curve of the Andromeda galaxy  (M31) out to 110 arcminutes away from the galactic centre, and estimated a mass-to-light ratio of $13\pm0.7 \, {\rm M}_\odot/{\rm L}_\odot$ at $R=24$ kpc, corresponding to a total mass of $M=(1.85\pm 0.1)\times10^{11} \, {\rm M}_\odot$  \cite{1970ApJ...159..379R}. The measure of the 21-cm line emission of neutral hydrogen also suggested that the rotation curves of spirals fall off at large radii less rapidly than they should when most of the galaxy mass is concentrated in the optically luminous component \cite{1970ApJ...160..811F,1973A&A....26..483R}. This flatness of the rotation curves led to the conclusion that galaxies are embedded in massive halos extending to large radii, as was suggested by theoretical studies of the stability of disk against the development of a bar \cite{1973ApJ...186..467O}.

In the last decades, the quest to unravel the intriguing puzzle of the dark matter has travelled along different paths. It was pointed out that the nature of dark matter could be either {\em baryonic}, or {\em non-baryonic}. Alternatively, it could be {\em gravitational}, namely the phenomena associated to dark matter actually is the signature of a break down of the standard theory of gravity. Each one of these hypotheses followed different paths to be validated. 

The search for baryonic dark matter focused on sub-luminous compact objects, such as planets, brown dwarfs,  white dwarfs, neutron stars, and black holes. These massive astrophysical compact halo objects (MACHOs) are now believed to form only a small fraction of dark matter. Microlensing surveys have been used to set an upper limit of 8\% to the contribution of MACHOs in the mass fraction of the dark matter halo of the Milky Way \cite{2000A&A...355L..39L,2007A&A...469..387T}.

Nevertheless, there is the possibility that primordial black holes, which formed before the  Big  Bang nucleosynthesis and have masses below the sensitivity range of microlensing surveys, may form a substantial fraction of the total dark matter density. This idea was {originally} discussed by Carr and Hawking \cite{10.1093/mnras/168.2.399} in 1974. However, generating a relevant abundance of primordial black holes requires a substantial degree of non-Gaussianity in the power spectrum of the primordial density perturbations \cite{Motohashi_2017,Passaglia_2019}.
Recently, the detection of gravitational waves has set a tight upper limit on the abundances of these black holes. This limit suggests that the black hole contribution to the dark matter abundance is at the level of a few per cent  \cite{PhysRevLett.120.191102}, as allowed by the constraints on non-Gaussianity obtained from the Planck satellite. \cite{2015JCAP...04..034Y,2017PhRvD..95h3006C,2019arXiv190505697P}.
{ Similar results were obtained by a recent microlensing survey of M31. In fact, if primordial black holes constitute the Milky Way and M31 dark matter halos, we expect that the M31 stars observed from Earth should generate $\sim 10^3$ microlensing events. However, in a 7 hour-long survey, only a single candidate event was identified. This result implies that the fraction of dark matter in primordial black holes in the mass range $10^{-11}-10^{-6} \, {\rm M}_\odot$} is  $\Omega_{PBH}/\Omega_{DM} < 0.001$  \cite{2019NatAs...3..524N}.

{ One or more species of elementary particles beyond the Standard Model of particle physics are expected to make up the non-baryonic dark matter: Weakly Interacting Massive particles (WIMPs), QCD axions or ultra-light bosons are some of the suggested hypothetical particles. WIMPs are expected in the theory of supersymmetry, which dates back to the 1970s and supposes that, for any given fermion, there is a boson with the same quantum numbers, and vice versa \cite{1998pesu.book....1M}.} In this case, there would be many electrically neutral and weakly interacting massive particles that could be cosmologically abundant and could play the role of dark matter. Intriguingly, a supersymmetric particle with mass in the range $\sim 1$GeV-10 TeV would give rise to a relic density consistent with the observed dark matter density \cite{Jungman_1996}.

The searches for these particles, based on methods of direct or indirect detection, are ongoing. Despite the large effort, no direct detection has been claimed to date \cite{2017PhRvL.118b1303A}  and claimed indirect detections have been questioned \cite{Bernabei_2013, 2014EPJWC..7000043B,PhysRevLett.114.151301}. However, supersymmetric particles are not the only particles that can play the role of dark matter. As mentioned above, there are other strongly motivated candidates, such as QCD axions and fuzzy dark matter \cite{2000PhRvL..85.1158H,2006PhLB..642..192A}. QCD axions are suggested by the solution of the Strong CP problem in the Standard Model \cite{Weinberg:1977ma, Wilczek:1977pj}, while fuzzy dark matter arises from the compactification of extra dimensions in the String Theory landscape \cite{2010PhRvD..81l3530A}. We will discuss these models in Section \ref{sec:particlemodels}.

Alternatively, the phenomena attributed to dark matter could originate from a modification of the theory of gravity. 
Along this line, in 1983, the Modified Newtonian Dynamics (MOND), was proposed to explain the observational evidence attributed to dark matter by modifying the second law of dynamics with the introduction of an acceleration scale  \cite{Milgrom83MOND,1983ApJ...270..371M,1983ApJ...270..384M}. Later on, more theories of gravity have been proposed \cite{2006JCAP...03..004M,2012AnP...524..545C,2017ScPP....2...16V}, all of them reporting successes and failures. Unfortunately, none of them appears to be able to explain all the relevant observations in a single general framework. In Section \ref{sec:gravitmodels} we will discuss some of these models.

In the standard cosmological model, dark matter consists of massive particles that weakly interact with ordinary matter and that decoupled from the primordial plasma when they were non-relativistic. This Cold Dark Matter (CDM) scenario encounters some difficulties in describing structures at galactic scales \cite{1994Natur.370..629M,10.1046/j.1365-8711.1999.03039.x,Boylan_Kolchin_2011,Bullock_2017,2017Galax...5...17D,2019A&ARv..27....2S}. 
These difficulties include the {\em cusp/core} problem, the problem of the {\em missing satellites}, { the {\em too-big-to-fail} problem}, and the problem of the {\em planes of satellite galaxies}. 

The cusp/core problem is the discrepancy between the flat dark matter density profile observed at the centres of dwarf and ultra-faint galaxies, and the cuspy profile predicted in collisionless N-body simulations \cite{Navarro_1996a, Navarro_1997,Ferrero_2012,2018MNRAS.474.1398G}. 
In particular, N-body simulations show { cuspy density profiles for dark matter halos of galaxy size; the halo density increases with decreasing radius $r$ as $r^{-\beta}$, with $\beta$ in the range $\sim [1 - 1.5]$.} These slopes do not match the cores favored by the observed rotation curves \cite{1985ApJ...292..371D,Flores_1994,1994Natur.370..629M,Navarro_1996a,Navarro_1997}. Nevertheless, modelling the kinematics of stars in dwarf galaxies does not lead to a clear conclusion to whether these galaxies are dominated by a core or a cusp in their innermost regions \cite{2009ApJ...704.1274W}. 
The missing satellites problem is the fact that {  the dark matter halos of galaxies like the Milky Way are predicted to have} { a number of dark matter subhalos which is an order of magnitude larger than the number of satellites} { observed around the Milky Way or other comparable galaxies} \cite{1993MNRAS.264..201K,Klypin_1999}. 
{ A related, but independent, issue is the too-big-to-fail problem \cite{Read_2006,Boylan_Kolchin_2011,Boylan_Kolchin_2012,Garrison_Kimmel_2014}, which manifests in the central densities of the most massive dark matter subhalos formed in $\Lambda$CDM simulations; these densities  are systematically larger than the central densities of the brightest classical Milky Way satellites, as inferred from their stellar kinematics. In principle, associating the classical satellites to dark matter subhalos with smaller central densities and smaller mass would erase the discrepancy; however, this association would clearly imply that the most massive subhalos would not host a galaxy. In other words, these subhalos ``failed'' to form stars even though less massive subhalos succeeded in doing so.}
Finally, the problem of the { planes of satellites refers to the fact that, in the  galaxy systems of the Milky Way, of M31, and of Centaurus A, the satellite galaxies reside in a thin plane and they generally corotate. This satellite} configuration appears unlikely in the simulations of the standard CDM model \cite{Pawlowski_2014}.

In addition to these problems, the current CDM paradigm faces other challenges. The Tully-Fisher relation suggests the existence of an acceleration scale that is adopted as a fundamental constant in MOND  \cite{McGaugh_2011}. The total angular momentum of the visible component also differs by a factor of $2-3$ from the angular momentum of CDM halos \cite{van_den_Bosch_2001, Cardone_2009}.

This review will focus only on some of the aforementioned problems that the standard dark matter paradigm encounters at galactic scales. We will present the state of the art of these problems in the context of the standard CDM model, and then review some alternative solutions that may reside either in a change of the dark matter paradigm or in models of modified gravity.  

Section \ref{sec:overviewCDM} recalls the standard cosmological model with the current constraints on the cosmological parameters. Section \ref{sec:obschallenges} discusses some of the observational challenges: the  radial acceleration relation (Section \S \, \ref{sec:RAR}), the cusp/core problem (Section \S \, \ref{sec:CCP}), the missing satellites problem (Section \S \, \ref{sec:MSP}), 
{  the too-big-to-fail problem (Section \S \, \ref{sec:TBTF})}, 
and the planes of satellite galaxies problem (Section \S \, \ref{sec:planeofsatellite}). 
Section \ref{sec:particlemodels} is devoted to possible explanations  beyond the Standard Model of particle physics such as warm and self-interacting dark matter (Sections \S \, \ref{sec:WarmCDM} and \S \, \ref{sec:SIDM}), QCD axions (Section \S \, \ref{sec:QCDaxion}), and fuzzy  dark matter (Section \S \, \ref{sec:FuzzyDM}). 
Finally, Section \ref{sec:gravitmodels} explores  possible explanations provided by alternative theories of gravity, specifically Modified Newtonian Dynamics (MOND) (Section \S \, \ref{sec:MOND}), Scalar-Vector-Tensor theory of gravity, also known as MOdified Gravity (MOG) (Section \S \, \ref{sec:MOG}), and $f(R)$-gravity (Section \S \, \ref{sec:f(R)gravity}). 
In both Sections \ref{sec:particlemodels} and \ref{sec:gravitmodels}, we discuss whether each of the specific models may resolve the issues at galactic scales that we mention in Section \ref{sec:obschallenges}.  
Finally, Section \ref{sec:summary} summarizes the main topics of the review and 
provides 
some future perspectives.

\section{Overview of the Cold Dark Matter model}\label{sec:overviewCDM}

Over the last decades, the standard model for the evolution of the Universe, the $\Lambda$CDM or {\em concordance} model \cite{1995Natur.377..600O}, was established by many independent  observations,  including the
Cosmic Microwave Background (CMB) temperature fluctuations \cite{2013ApJS..208...19H,2018arXiv180706209P,2018arXiv180706211P,2019arXiv190505697P,2019arXiv190602552P,2019arXiv190712875P}, the power spectrum of the matter density perturbations 
\cite{Percival2001, Pope2004, Tegmark2004}, 
the luminosity distances to supernovae SNeIa 
\cite{Riess1998, Perlmutter1999, Riess2004,astier, Davis2007, kowalski2008, Amanullah2010,suzuki2012}, and the expansion rate of the Universe 
\cite{Jimenez2003, Simon2005, Stern2010, Moresco2012a, Moresco2012b}. The current constraints on the cosmological parameters have reached unprecedented accuracy \cite{Genova-Santos:2020tfc}. 

In this  model, the present period of accelerated expansion is driven by the cosmological constant $\Lambda$  that provides an energy density $\Omega_{\Lambda,0}=0.686\pm 0.020$ which is in units of the critical density $\rho_c=3H_0^2/8\pi G$  \cite{2018arXiv180706209P}. The cosmological constant can be assimilated to a perfect fluid with an equation of state $p=w\rho$, between pressure $p$ and energy density $\rho$, with the constant parameter $w=-1$. The second largest contribution to the total energy-density budget of the Universe is dark matter, which is needed to explain the dynamics of galaxies and the large-scale structure. Its energy density is $\Omega_{\mathrm{DM},0} = 0.314\pm 0.020$ \cite{2018arXiv180706209P}.  
The present density of ordinary, or baryonic, matter  is $\Omega_{\mathrm{b},0}h^2 = 0.02207\pm  0.00033$ \cite{2018arXiv180706209P}. Summing the contributions of the cosmological constant, dark matter, and baryonic matter yields the curvature of the Universe $\Omega_{\mathrm{k},0}h^2 = -0.037^{+0.044}_{-0.042}$  \cite{2018arXiv180706209P}, where $h$ is the Hubble constant in units of 100 km s$^{-1}$ Mpc$^{-1}$. This curvature  makes the Universe spatially flat.  

The concordance model successfully describes the homogeneous and isotropic Universe, and the dynamics of cosmic structures. However, there are some tensions with the values of its parameters $\Omega_{\Lambda,0}$, $\Omega_{k,0}$, and the Hubble constant $H_0$. 

$\Omega_{\Lambda,0}$ is extremely small compared to the expectations from quantum field theory \cite{RevModPhys.61.1}.
Quantum chromodynamics  predicts a vacuum energy density of $\rho_\Lambda\sim10^{71}$ GeV$^4$, whereas the cosmological upper bounds on the cosmological constant give $\rho_\Lambda\sim10^{-47}$ GeV$^4$. This large discrepancy leads to a fine-tuning problem \cite{RevModPhys.61.1}. To reconcile the values of the vacuum energy density at cosmological and quantum scales, different solutions have been proposed, such as the anthropic principle or a cyclic model of the Universe \cite{2000PhRvD..61f1501R,2006Sci...312.1180S, 1987PhRvL..59.2607W,2017Galax...5...98C,2017JPhCS.880a2059C,2019PhLB..790..427C}. Comprehensive reviews on the cosmological constant and  dark energy are given in \cite{Brax_2017,2018RPPh...81a6901H,2019arXiv190703150F}.

Similarly, tensions arise in the values of the Hubble constant estimated from  the CMB and from  measurements in the nearby Universe, and also, to a less degree, in the values of $\sigma_8$, the normalization of the power spectrum of the mass density perturbations on cosmic scales, inferred from the CMB and from large-scale  structure surveys \cite{2018arXiv180706209P}. 
The Hubble constant from the CMB measurements $H_0=  67.27\pm0.60$ km s$^{-1}$Mpc$^{-1}$ \cite{2018arXiv180706209P} is at a 4.4$\sigma$ tension with the local value $H_0=73.52\pm1.62$  km s$^{-1}$Mpc$^{-1}$  \cite{2016ApJ...826...56R,2019ApJ...876...85R}. The clustering parameter $\sigma_8$ for the $\Lambda$CDM model, with fixed effective number of neutrino species $N_{\mathrm {eff}}$ and fixed total mass $\Sigma m_\nu$, obtained by the Planck collaboration \cite{2018arXiv180706209P}, is $\sigma_8=0.8149\pm0.0093$, implying $S_8\equiv \sigma_8(\Omega_m/0.3)^{0.5}=0.811\pm0.011$. On the contrary,  a tomographic weak gravitational lensing analysis of data from the Kilo Degree Survey \cite{2017MNRAS.465.1454H} led to $S_8=0.745\pm0.039$, which is at a $2.3\sigma$ tension with the Planck results.
Recently, it has also been shown that the last data release of the Planck experiment  might suggest a spatially closed, rather than flat Universe \cite{Di_Valentino_2019}. 
All these tensions remain open problems to date and the subject of intense investigation.

Dark matter is expected to consist  of stable massive
particles beyond the Standard Model of elementary particle physics. Dark matter can thus clump in self-gravitating structures embedding galaxies and clusters of galaxies. Many pieces of evidence support the existence of dark matter based on its gravitational effects, once General Relativity is assumed to be the correct theory of gravity. For example, the mass distribution within galaxy clusters can be reconstructed with the analysis of the gravitational lensing effect acting on extended light sources beyond the cluster, with the estimate of the X-ray emitting intracluster medium or with the dynamics of galaxies. These investigations  show that the baryonic fraction at the virial radius is approximately $\sim 0.18$  (e.g.  \cite{McCarthy_2007,Eckert_2013}). At smaller astrophysical scales, dark matter explains why,  in the outer regions of the disk galaxies,  the rotation velocity of stars does not fall off as expected by the presence of luminous matter alone,  $V_c \propto r^{-1/2}$. On the contrary, observations indicated that  $V_c \sim {\mathrm {const}}.$ \cite{1982ApJ...261..439R,1981AJ.....86.1825B}. 

Dark matter also allows the primordial perturbations in the density field of the baryonic matter, as mirrored in the temperature anisotropy fluctuations of the CMB, to grow and form the cosmic structures we  observe today \cite[e.g., ][]{KUHLEN201250}. 
In fact, in the history of cosmic expansion, { for models where the dark matter and dark energy are not separated from the other components of the Universe,} dark matter decouples from the primordial plasma much earlier than baryons; the primordial fluctuations in the dark matter density field thus starts growing earlier and, at recombination epoch, they are larger than the baryon density perturbations that are still coupled to the perturbations of the background radiation field that generates the CMB anisotropies. 
After the recombination epoch, baryonic matter falls into the gravitational potential wells of the dark matter halos and form the cosmic structure  (see \cite[e.g., ][]{2005PhR...405..279B,2009ARNPS..59..191B, 2010ARA&A..48..495F,2013ARNPS..63...69K,2018RvMP...90d5002B} for details).

{ In these models, 
dark matter particles are thus in thermal equilibrium with the cosmic plasma before the decoupling epoch, when they get out of kinetic equilibrium, as they become non-relativistic, at temperature $T=T_{\rm d} \ll m_{\rm DM}$, where $m_{\rm DM}$ is the mass of the dark matter particle. In the standard CDM model, the dark matter} particles are so massive that $T_{\mathrm d}\gg  T_0$, where $T_0$ is the plasma temperature at recombination epoch. They thus decouple from other particles and start moving freely at non-relativistic speeds in the early Universe. 
The comoving number density of dark matter particles freezes out when the creation and annihilation of the dark matter particles are inhibited. In general, the {\it freeze-out} cosmic temperature $T_{\rm f} \gg T_{\rm d} $. The comoving number density of dark matter particles is thus set by their annihilation cross section at this epoch \cite{Jungman_1996}
\begin{equation}
\Omega_\chi h^2 = (3 \times 10^{-27} {\rm cm}^3/{\rm sec})
/ \langle \sigma v \rangle_{\mathrm {ann}}\,.
\end{equation}
Intriguingly, for WIMPs, with mass in the range $\sim 1$GeV - 10TeV, the annihilation cross section $\langle \sigma v \rangle_{\mathrm {ann}}$ gives $\Omega_\chi h^2$  comparable to the observed dark matter density $\Omega_{\mathrm{DM},0}\sim 0.3$ \cite{2018arXiv180706209P}. This coincidence is  usually called ``the WIMP miracle''. 

Moreover, WIMPs naturally arise in the supersymmetric (SUSY) extension of the Standard Model of particle physics \cite{Jungman_1996}, where each Standard Model particle has a supersymmetric partner, and the lightest one is a good candidate for the dark matter particle. Another possibility is that WIMPs arise from compactifications of extra dimensions \cite{2003NuPhB.650..391S}.  

Nowadays, direct and indirect searches for WIMPs as well as other dark matter particle candidates are ongoing. Direct searches look for a scattering process where a dark matter particle in the Milky Way halo interacts with an atomic nucleus of a detector whose recoil generates a detectable  release of energy. Experiments of this kind are, for example, LUX and XENON 1T \cite{2011PhRvL.107m1302A,2017PhRvL.118b1303A}. 

Indirect detections of dark matter particles include  searches for neutrinos arising from the annihilation of WIMPs in the centre of the Earth or the Sun, where WIMPs can concentrate at the bottom of the gravitational potential well. This approach is adopted by experiments like IceCube and Super-K \cite{PhysRevLett.114.141301,2019ICRC...36..506B}. Similarly, WIMP annihilation in the Milky Way can enrich the cosmic rays that we detect on Earth with positrons and antiprotons
\cite{2015APh....62...12A,2013PhRvL.111q1101B}. 
Similar enrichment is expected in particle colliders, like CMS at the Large Hadronic Collider (LHC), which looks for the interactions between dark matter particles and the fermions of the Standard Model.
A comprehensive review on the results of the direct and indirect searches is given in ~\cite{PhysRevD.98.030001}. 

\begin{figure}[htb!]
\begin{center}
\includegraphics[width=0.9\textwidth]{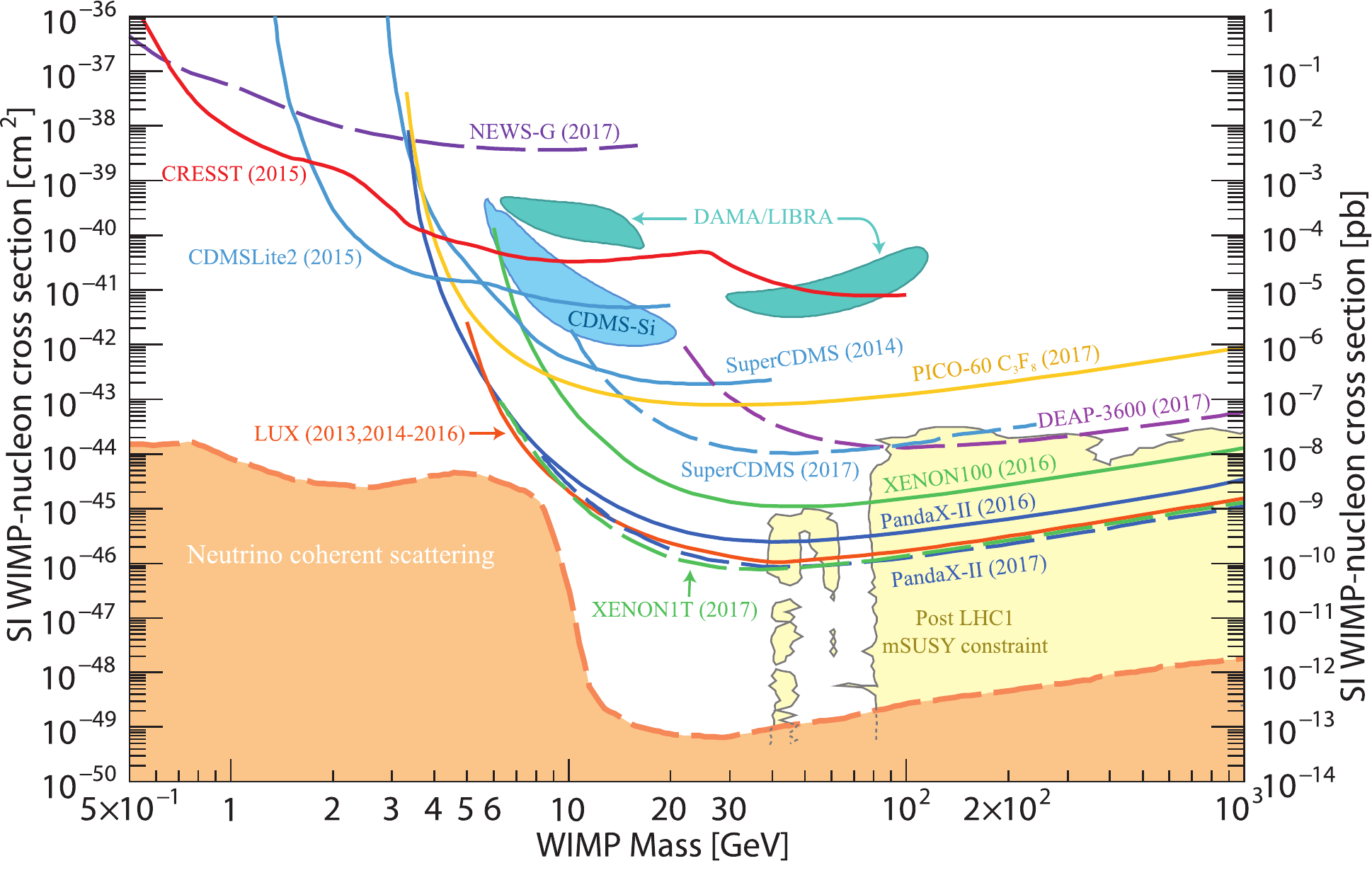}
\end{center}
\caption{
Upper limits on the spin-independent (SI) elastic dark matter-nucleon  cross-section as a function of mass of the dark matter particle. The figure is reproduced from \cite{PhysRevD.98.030001}.
\label{fig:cdm1}
}
\end{figure}

Figure~\ref{fig:cdm1} shows the allowed region, that is below each sensitivity curves, in the plane of the dark matter-nucleon cross section and the dark matter particle mass. The case of the DAMA/LIBRA experiment is representative of the complexity of this kind of investigation. The DAMA/LIBRA experiment searches for dark matter particles in the Milky Way's halo. The number of events in the 2-6 keV energy range   has an annual modulation that can mirror the relative velocity between the dark matter particle and the Earth orbiting around the Sun. This modulation has a statistical significance  of  $\sim 9\sigma$ \cite{Bernabei_2013, 2014EPJWC..7000043B}. However, the signal is only consistent with values of the interaction cross-section and of the mass of the dark matter particle that are within the regions of Figure ~\ref{fig:cdm1}  that appear to be excluded by other experiments \cite{2011PhRvL.107m1302A,2017PhRvL.118b1303A,2015APh....62...12A,2013PhRvL.111q1101B}. Alternative explanations of this signal are still intensively debated \cite{PhysRevLett.114.151301}.

Although the overall picture appears solid, the fundamental nature of dark matter still remains an intriguing mystery. In the next section, we focus on the main problems that the CDM model encounters at galactic scales and the possible solutions within the context of CDM. In Sections \ref{sec:particlemodels} and \ref{sec:gravitmodels}, we discuss some alternatives to the CDM paradigm.

\section{Observational challenges of the Cold Dark Matter model}\label{sec:obschallenges}

Despite being successful on cosmological scales, the CDM model faces persistent challenges on the scales of galaxies, which are mostly related to the dark matter distribution in the central regions of the galactic halos and to the properties of the dwarf galaxies \cite{Boylan_Kolchin_2011,2012PASA...29..395K,2015PNAS..11212249W,Bullock_2017,2017Galax...5...17D}. Here, we focus on these issues and discuss viable solutions within the CDM scenario.

\subsection{The rotation curves of disk galaxies and the baryonic scaling relations}\label{sec:RAR}

The rotation curve of disk galaxies is one of the most important pieces of evidence of the existence of dark matter: most of the luminous matter, namely stars, gas, and dust, is concentrated in the inner part of the galaxy and, if this luminous matter is the only galaxy component, we expect the rotation speed to fall off as $v^2(r)\propto r^{-1}$ in the outer regions where the luminous matter density substantially decreases. However, the outermost visible objects, namely clouds of neutral hydrogen, move as fast as the inner objects. We thus observe a fairly flat rotation curve for most disk galaxies \cite{1982ApJ...261..439R,1981AJ.....86.1825B}. 

The flatness of the rotation curve can be generated by a pressure-supported spheroidal halo embedding the entire galaxy  \cite{1991MNRAS.249..523B}. This matter is {\it dark} as it remains undetected in any electromagnetic band. 
In the CDM model, assuming spherical symmetry, the mass density profile of the dark halo is well described by the  Navarro-Frenk-White (NFW) profile \cite{NFW96}: it contains  two free parameters, the virial mass $M_{\mathrm {vir}}$\footnote{\label{foot:vir}{ The virial radius $r_{\mathrm {vir}}(z)$
at redshift $z$ is the radius of a spherical volume within which the mean mass density is $\Delta_c(z)$ times the critical density of the Universe $\rho_c(z)=3H^2(z)/8\pi G=2.775 E^2(z)h^2 10^{11}$~M$_\odot$~Mpc$^{-3}$, with $E(z)=[(\Omega_m(1+z)^3+(1-\Omega_m-\Omega_\Lambda)(1+z)^2+\Omega_\Lambda]$, and  $\Omega_m$ and $\Omega_\Lambda$ the parameters of the background Friedmann model. $\Delta_c(z)$ is the solution to the collapse of a spherical top-hat density perturbation at the time of  virialization. The virial mass is thus $M_{\mathrm {vir}} = 4\pi\Delta_c(z)\rho_c(z)r_{\mathrm {vir}}^3/3$ \cite{1998ApJ...495...80B,2001A&A...367...27W}.}} and the concentration parameter $c$ \cite{2018MNRAS.477.4187H}. 
The introduction of the dark halo in the description of galaxy dynamics comes with an additional degree of freedom: reproducing the measured rotation curve of a galaxy requires a fine-tuning of the relative contribution of the galactic disk and the dark matter halo to the gravitational pull \cite{vAlbMaxDisk85,AniyanDiskHalo18}. This fine-tuning is known as the \textit{disk-halo conspiracy}.  
{ A detailed analysis of a sample of more than a thousand disk galaxies quantifies the correlation between the luminous and dark matter and suggests the existence of a universal rotation curve  \cite{1996MNRAS.281...27P, Salucci_2007}.} 

Additional pieces of evidence quantify the discrepancy between the amount of mass estimated by its electromagnetic emission and the mass required to describe the observed kinematics of disk galaxies: the Baryonic Tully-Fisher Relation (BTFR) \cite{McGetal00BTFR,McG05BTFR}, the Mass Discrepancy Acceleration Relation (MDAR) and the Radial Acceleration Relation (RAR) \cite{McGetal16RAR,Lellietal17aRAR,Lietal18RAR}. 
We describe these relations in details below. Here, it suffices to say that they show a tight correlation between kinematic quantities and quantities associated to the baryonic component. The crucial feature of these correlation is their small intrinsic scatter, generally consistent with the observational uncertainties alone. This feature is surprising, because, in the CDM scenario, cosmic structures form hierarchically through  stochastic mergers, and we would expect a large scatter, mirroring the merging history of each galaxy. On the contrary, the lack of a relevant scatter implies that, irrespective of the merging history of the galaxy, its dark matter halo, that represents $\sim$90\% of the galaxy mass and sets its dynamical properties, adjusts its properties to those of the luminous disk, that only contains $\sim$10\% of the galaxy mass. { In other words, in spite of the dark matter halo contributing to the majority of the total mass of the galaxy, the kinematic properties of the disks are found to be strongly correlated with the luminous matter.} This correlation appears thus challenging for the CDM paradigm.

\subsubsection{Observational evidence}

Rotation curves of high surface brightness (HSB) galaxies are properly described when we adopt the NFW density profile for the dark matter halo. However, some issues emerge from the analysis of the exquisite data of the SPARC sample, that contains 171 galaxies of different luminosity and morphological type \cite{Lellietal16SPARC,2018MNRAS.477.4187H}: 
{\em{(i)}} the $c$--$M_{\mathrm {vir}}$ relation, or mass-concentration relation, is steeper than expected from cosmological N-body simulations \cite{BullocketalcMvir01,WechsleretalcMvir02,NetoetalcMvir07,Klypinetal11}; 
{\em{(ii)}} the mass-to-light ratio in the 3.6 $\mu$m band, $\mathrm{M}_*/\mathrm{L}_{[3.6]}$, 
appears unphysically negative for 51 out of 171 galaxies \cite{2018MNRAS.477.4187H}; and {\em{(iii)}} the Stellar-Population-Synthesis (SPS) models overestimate the observed correlation between $\mathrm{M}_*/\mathrm{L}_{[3.6]}$ and  the color index $(B-V)$, independently of the chosen Initial Mass Functions (IMF) \cite{2018MNRAS.477.4187H}.

Additional challenges for the CDM model arise from the baryonic-dark matter scaling relations mentioned above. 
The total baryonic mass of a galaxy, $M_{\rm d}$, and the asymptotic flat velocity of its rotation curve, $V_{\rm c}$, set by the depth of the gravitational potential well of the dark matter halo, obey the BTFR
\cite{McGetal00BTFR,McG05BTFR,McG12BTFR,Lellietal15BTFR}:
\begin{equation}
\label{eq:BTFR}
M_{\rm d}=A V_{ \rm c}^4\, ,
\end{equation}
where the normalization constant is $A = 47 \pm 6$~M$_\odot$~km$^{-4}$~s$^4$ \cite{McG12BTFR}. $A$ can also be written as $A\sim(Ga_0)^{-1}$, the product between the gravitational constant $G$ and 
an acceleration scale  $a_0=1.2\times 10^{-10}$~m~s$^{-2}$. 

\begin{figure}[htb!]
\begin{center}
\includegraphics[scale=0.7]{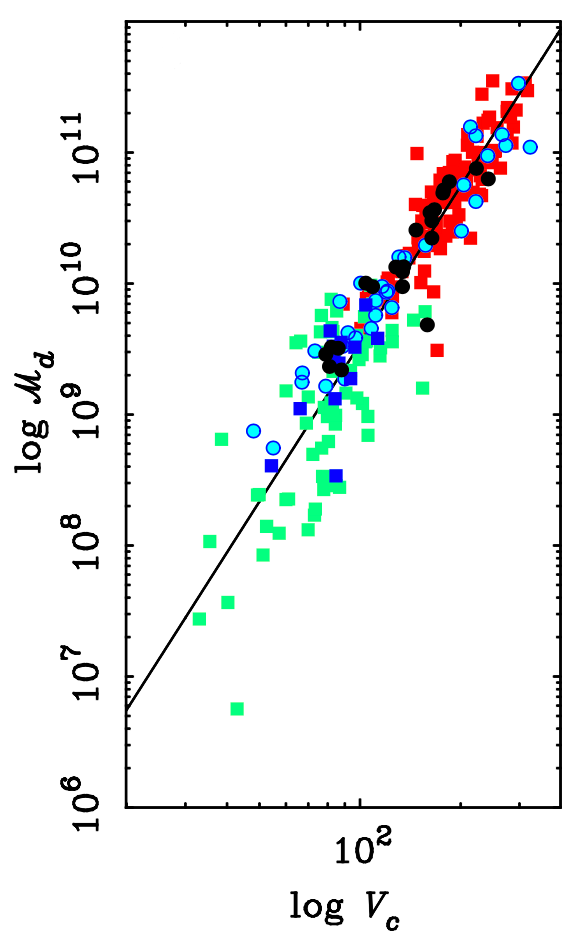}
\end{center}
\caption{
Baryonic Tully Fisher Relation (BTFR): total baryonic disk mass of galaxies against the rotation velocity $V_{\rm c}$. Circles and squares represent the data derived from~\cite{Bothunetal85}(red),~\cite{Verheijen97}(black),~\cite{Pildisetal97,EderandSchombert00} (green),~\cite{McGaughanddeBlok98}(light blue) and~\cite{Matthewsetal98} (dark blue).  The black solid line shows an unweighted linear fit. The figure is reproduced from~\cite{McGetal00BTFR}.
}\label{fig:BTFR}
\end{figure}

The observed BTFR is shown in Figure~\ref{fig:BTFR}. Circles indicate measurements of the velocities from the asymptotically flat part of the rotation curves; the squares represent measurements of $V_{\rm c}$ as half of the HI emission line width at 20\% of the peak value, whereas the stellar mass is inferred from photometric observations in the $I$-band \cite{Pildisetal97}, $H$-band \cite{Bothunetal85}, $K'$-band \cite{Verheijen97},  or $B$-band \cite{McGaughanddeBlok98,Matthewsetal98}.
The total baryonic mass of the disk is estimated as $M_{\rm d}=M_*+M_{\rm gas}$, where $M_*$ and $M_{\rm gas}$ are the total mass in stars and gas, respectively. $M_*$ is estimated as $M_*=\Upsilon_* L$, where $L$ is the total disk luminosity and $\Upsilon_*$ is a constant mass-to-light ratio. $M_{\rm gas}$ is given by $1.4M_{\rm HI}$, where $M_{\rm HI}$ is the total estimated mass of neutral hydrogen and the factor 1.4 is the standard correction to account for helium and metals.
The black solid line shows an unweighted linear fit to the points with stellar mass estimated in the $I$-, $H$-, and $K'$-bands. The $B$-band data are not included in the fit because this band is not a robust indicator for the stellar mass. 
The slope of this straight line is $b= 3.98\pm0.12$. 

For mass-to-light ratios in the 3.6 $\mu$m band, $\Upsilon_{[3.6]}$, larger than $\sim$0.2 M$_\odot$/L$_\odot$, the intrinsic scatter of the BTFR is smaller than the minimum intrinsic scatter of 0.15 dex predicted by semi-analytic models of galaxy formation in the $\Lambda$CDM model  \cite{Dutton12,DiCintioandLelli15}, based on the $c$--$M_{vir}$ relation of CDM simulations \cite{BullocketalcMvir01}. The intrinsic scatter of the BTFR reaches a minimum of $\sim$0.10 dex for $\Upsilon_{[3.6]}\gtrsim0.5$ M$_\odot$/L$_\odot$.
More importantly, the BTFR residuals show no correlations with galaxy radius and surface brightness, at odds with $\Lambda$CDM galaxy formation models  \cite{DesandWech15}. In addition, $\Lambda$CDM simulations predict a slope $\sim 3$, which implies an 8$\sigma$ tension with the observed slope $b= 3.98\pm0.12$   \cite{McGetal00BTFR}.

\begin{figure}[htb!]
\begin{center}
\includegraphics[width=0.9\columnwidth, height =5cm]{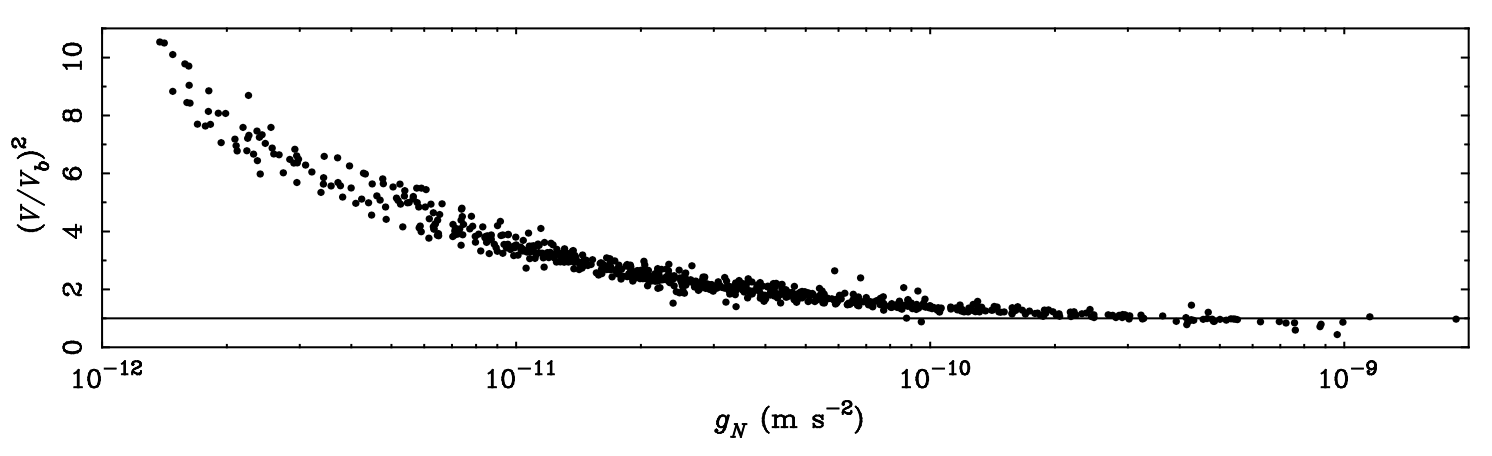}
\end{center}
\caption{
Mass Discrepancy Acceleration Relation (MDAR), where the galaxy mass discrepancy, the squared ratio between the observed velocity $V$ and the velocity due to baryons $V_{\rm b}$, is plotted against the Newtonian centripetal acceleration $g_{\rm N} = V^2_{\rm b}/R$ derived from the observed baryonic surface mass density. The black dots show several hundreds of individual resolved measured points from the rotation curves of almost one hundred spiral galaxies. The figure is reproduced from~\cite{2012LRR....15...10F}.
\label{fig:MDAR}
}
\end{figure}

{The MDAR relates, { at each radius $R$ in the galactic disk}, the Newtonian acceleration  $g_{\rm N}(R)$, generated by the observed baryonic surface mass density, to the \textit{mass discrepancy} $[V(R)/V_{\rm b}(R)]^2$, where $V$ is the observed velocity and $V_{\rm b}$ is the velocity that would be generated by the baryonic matter. The Newtonian acceleration and the mass discrepancy are anti-correlated
\cite{McG04MDAR,DiCintioandLelli15}. 
The mass discrepancy $[V(R)/V_{\rm b}(R)]^2$ remains close to one for large baryonic accelerations $g_{\rm N}(R)$, whereas it increases for accelerations smaller than the acceleration scale $a_0$ found with the normalization of the BTFR (see Figure~\ref{fig:MDAR}).

For both the BTFR and the MDAR, the scatter depends on the value of the adopted mass-to-light ratio \cite{McG04MDAR}. Specifically, the intrinsic scatter in both relations is minimized by the same mass-to-light ratio, consistent with SPS models \cite{McG04MDAR}. Intriguingly, when we use MOND (Section \ref{sec:MOND}) to describe the rotation curves \cite{SandersandMcGaugh02,McG04MDAR}, the required   mass-to-light ratio 
approximately coincides with the best mass-to-light ratio predicted by the SPS model based on a scaled Salpeter IMF \cite{BellanddeJong01}
\begin{equation}
\label{eq:ScS_IMF_BdJ01}
\Upsilon_* = \chi\Upsilon_*^{\rm Salpeter}\, ,
\end{equation}
where $\Upsilon_*^{\rm Salpeter}$ is the mass-to-light ratio provided by the Salpeter IMF and $\chi \sim 3/4$. 
}

\begin{figure}[htb!]
\begin{center}
\includegraphics[width= 0.9\columnwidth]{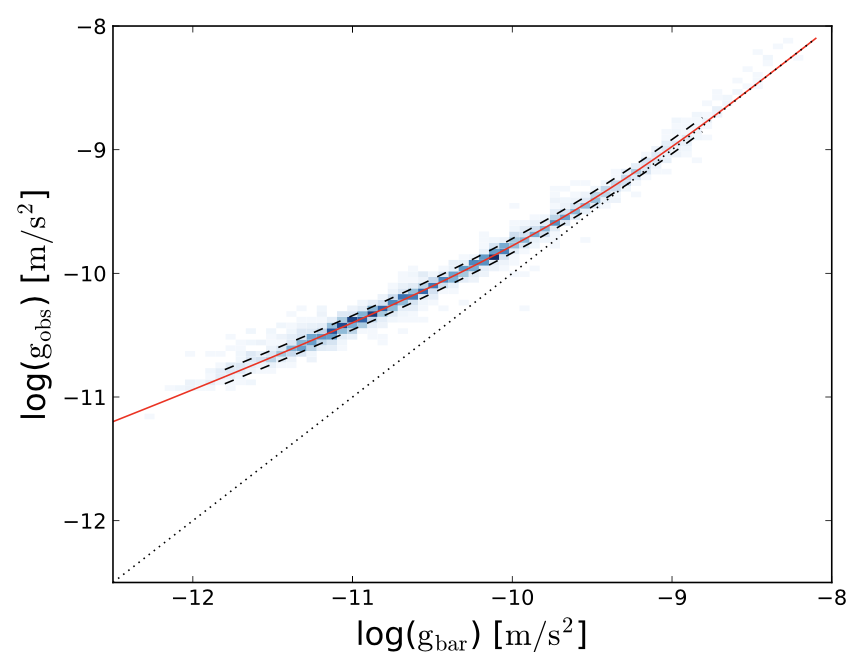}
\end{center}
\caption{
Radial Acceleration Relation (RAR), the observed centripetal acceleration versus the Newtonian acceleration due to baryonic matter alone. The blue color-scale rectangles show 2694 individual measured points from the rotation curves of 153 SPARC galaxies. The red solid line is the fitted RAR given by equation~\eqref{eq:RAR}, whereas the black dotted line represents the one-to-one relation, for comparison. The figure is reproduced from~\cite{Lietal18RAR}.
\label{fig:RAR}
}
\end{figure}

A slightly different perspective is provided by the RAR, that correlates the observed centripetal acceleration $g_{\rm obs}(R)=V^2(R)/R$, where $V$ is the measured rotation speed,  and the Newtonian acceleration $g_{\rm bar}(R)$, due to baryonic matter alone  \cite{McGetal16RAR}. Figure~\ref{fig:RAR} shows the RAR for the SPARC sample. The two quantities and their respective uncertainties are completely independent of each other~\cite{Lietal18RAR}, and follow the relation
\begin{equation}
\label{eq:RAR}
g_{\rm obs}(R) = \frac{g_{\rm bar}(R)}{1 - \exp\left(-\sqrt{\frac{g_{\rm bar}(R)}{g_\dagger}}\right)}\, .
\end{equation}

The only free parameter is $g_\dagger$. The fit with 153 galaxies from the SPARC sample yields $g_\dagger = (1.20 \pm 0.02 \pm 0.24)\times 10^{-10}$~m~s$^{-2}$ \cite{Lellietal16SPARC,McGetal16RAR}, which is within 1$\sigma$ from the acceleration scale $a_0$. 

The observed scatter is 0.13 dex, and it is comparable to the scatter 0.12 dex derived from the different mass-to-light ratios of the individual galaxies, and from the errors on the velocity measurements and the uncertainties on the inclination of the disk and the galaxy distance. 
When marginalizing over these uncertainties and over the whole sample of galaxies, the observed intrinsic scatter can be reduced to 0.057 dex.
Therefore, the scatter appears much smaller than the scatter 0.09 dex predicted by the EAGLE simulations; 
the measured scatter should actually be larger than 0.09 dex,  because the measurement errors are not included in the EAGLE prediction \cite{Lietal18RAR}.
{ The issue remains open, however, because when the observed sample includes dwarf galaxies with slowly rising rotation curves, the scatter also appears to substantially increase  at small $g_{\mathbf {bar}}$ in the observed relation \cite{SantosSantos_2020}.}

{ An additional unsettled controversy is the correlation between the residuals of the observed rotation curves from relation (\ref{eq:RAR}) and  the properties of the galaxies observed in galaxy samples \cite{2019ApJ...873..106D,2020arXiv200307377C} different from SPARC, where this correlation appears to be absent \cite{Lellietal16SPARC}.}

\subsubsection{Possible solutions within the CDM model}

The \textit{disk-halo conspiracy} can be solved within the CDM model by adopting the \textit{maximum-disk hypothesis}: the disk maximally contributes to the measured rotation curve \cite{1981AJ.....86.1825B,1987AJ.....93..816K,1987IAUS..117...67S,1990A&ARv...2....1S}. This hypothesis yields mass-to-light ratios for the disk stellar population in agreement with SPS models, and it is supported by several observational probes. 

For example, the near infrared mass-to-light ratio constrained from the Milky Way terminal velocity curve describes this curve without the need of a dark matter halo up to $\gtrsim5$ kpc
~\cite{BissantzandGerhard02}. If some dark matter is included within this distance from the Galaxy centre, the microlensing optical depth of the source stars within the  Baade Window (Galactic longitude = 1$^\circ$, Galactic latitude = -3.9$^\circ$) would probably decrease, becoming more discrepant from the observed value of this quantity derived from the MACHO measurements in~\cite{Alcocketal2000a,Popowskietal2000}.

Moreover, the mass-to-light ratios determined in the Milky Way from star counts, the radial force within the disk, and the vertical force from the disk are all consistent with each other and with the BTFR; they also properly reproduce the terminal velocity curve of the Galaxy by requiring a close to maximum disk~\cite{McGaugh_2015}. At last, only maximum disks can properly reproduce the observed ratio $\mathcal{R}\sim1.2\pm0.2$ between the corotation radius of barred galaxies and the semi-major axis of the bar; this value is the evidence that the corotation radius of barred galaxies is just beyond the end of the bar ~\cite[e.g., ][]{SellandDebMaxDisk14}.
However, the maximum-disk hypothesis only works for HSB galaxies; dwarf spheroidal and LSB galaxies appear to be dark matter-dominated systems even in their innermost regions \cite{StrigarietaldSph08,2019MNRAS.490.5451D}, { although there are indications of a correlation between the distributions of the luminous and dark matter also in these systems \cite{2017MNRAS.465.4703K}}.

{ $\Lambda$CDM hydrodynamical simulations and semi-analytical models can reproduce the normalization and slope of the BTFR  but not its small scatter  \cite{vdBandDalc00,TGetal11,Dutton12,McG12BTFR,DiCintioandLelli15,SantosSantosetal16}: the models yield a scatter of $\sim$0.15 dex, compared to the observed $\sim$0.10 dex. 
Accounting for the BTFR, including its relatively small intrinsic scatter and  the lack of correlations between their residuals and galaxy properties, requires an accurate balance between the star formation efficiency and the stellar feedback processes to regulate the relation between the properties of the baryonic matter and the properties of the dark matter halo~\cite{McG12BTFR,Lellietal15BTFR}.

The shape and the scatter of the MDAR are properly reproduced by a semi-empirical model~\cite{DiCintioandLelli15} based on the $\Lambda$CDM paradigm and on the existence of different galaxy scaling relations between (1) the concentration of the dark matter halo and its mass \cite{DuttonandMaccio14}, (2) the neutral hydrogen mass and the stellar mass \cite{Papastergisetal12}, (3) the disk  mass and its size \cite{Langeetal15}, and (4) the half-mass radius of the bulge and its mass \cite{Gadotti09}. The same model can also reproduce a BTFR with the correct normalization and slope but with an intrinsic scatter of 0.17 dex,  larger than the observed one of $\sim$0.10 dex.

The key assumption for this success is the dependence of the inner and outer slopes of the dark matter density profile on the  stellar-to-halo mass ratio  \cite{DC14a,DC14b}, which accounts for the feedback of the baryonic processes. However, for this mechanism to be effective, the stellar-to-halo mass ratio must be in the small range $[0.01\%-0.03\%]$ that might be barely appropriate only for ultra-faint or LSB galaxies but it is inappropriate for HSB galaxies. In fact,
when the stellar-to-halo mass ratio is within this range, stellar feedback leads to the formation of cores and the BTFR slope is reproduced.
However, for smaller stellar-to-halo mass ratios  stellar feedback cannot convert the cuspy NFW profile into a core and  for greater stellar-to-halo mass ratios,  galaxies are even more concentrated, up to a factor of 2.5,  than predicted by CDM-only simulations.
}

In $\Lambda$CDM, it remains difficult to explain the emergence of the acceleration scale, $a_0 = 1.2 \times 10^{-10}$~m~s$^{-2}$, that appears in all the scaling relations mentioned above. This coincidence is assumed {\it ab initio} in MOND, a modification of the law of gravity  \cite{Milgrom83MOND}. Alternatively, it could suggest the necessity of new physics affecting the dark sector \cite{Lellietal17aRAR}.

\subsection{The cusp/core problem}\label{sec:CCP}

Collisionless N-body simulations of the CDM model show that the  mass density profile $\rho(r)$ of dark matter halos is described by a steep power-law in the innermost regions: in other words, the CDM model predicts cuspy density profiles. At small radii, the NFW density profile suggested by N-body simulations in the 1990s \cite{Dubinski:1991bm,NFW96} is approximately $\rho \sim r^{\alpha}$, with $\alpha = -1$  \cite{Navarro_1997}. Later investigations suggested that the inner slope may depend on the total mass of the halo \cite{Jing_2000} and be steeper, with $\alpha = -1.5$  \cite{10.1046/j.1365-8711.1999.03039.x}.
This prediction appears to disagree with observations, because dwarf galaxies require a shallower dark matter density profile in their central regions \cite{1988ApJ...332L..33C,1989ApJ...347..760C}. 

Dwarf galaxies are known to be among the darkest galaxies observed. They show a central velocity dispersion of $\sim 10$ km s$^{-1}$, compared to the expected $\sim 1$ km s$^{-1}$ for self-gravitating systems of the same luminosity and scale radius $\sim 100$ pc at equilibrium \cite{1998ARA&A..36..435M}. Although their luminosity may vary by several orders of magnitude among different systems, from $\sim10^2 \, {\rm L}_\odot$ to  $\sim10^{10} \, {\rm L}_\odot$ \cite{McConnachie12,2016A&A...588A..89J},  they show similar velocity dispersions, suggesting that they are dominated by a similar dark matter distribution  \cite{1993AJ....105..510M}. 

In 1988, the best fit model of the measured HI rotation curve of the dwarf irregular galaxy DDO 154  \cite{1988ApJ...332L..33C}
showed that its dark matter density profile is   well described by a pseudo-isothermal model \cite[e.g., ][]{1991MNRAS.249..523B} with a core radius of 3 kpc: 
the profiles of dwarf galaxies having a core rather than a cusp is a general  result \cite{1988ApJ...332L..33C,1989ApJ...347..760C}. Furthermore,  the recently discovered ultra-faint galaxies orbiting the Milky  Way and M31 \cite[e.g., ][]{2005ApJ...626L..85W,2006ApJ...643L.103Z,2006ApJ...650L..41Z,2007ApJ...654..897B,2009MNRAS.400.1472L,2016MNRAS.459.2370T,2017ApJ...839...20C,2019MNRAS.489.2634H,2019MNRAS.488.2743T} have a half-light radius of a few kpc, which favors a core over a cuspy density profile; the core is hardly explicable in the context of the CDM model without resorting to baryonic feedback, and/or tidal effects \cite{2019MNRAS.488.2743T}.

This discrepancy between observations and simulations is known as the ``cusp/core problem'' (CCP) \cite{2017Galax...5...17D} and is still an unsolved topic of fervent debate.

\subsubsection{Observational evidence}

\begin{figure}[ht]
 \centering
 \includegraphics[width= 0.9\columnwidth]{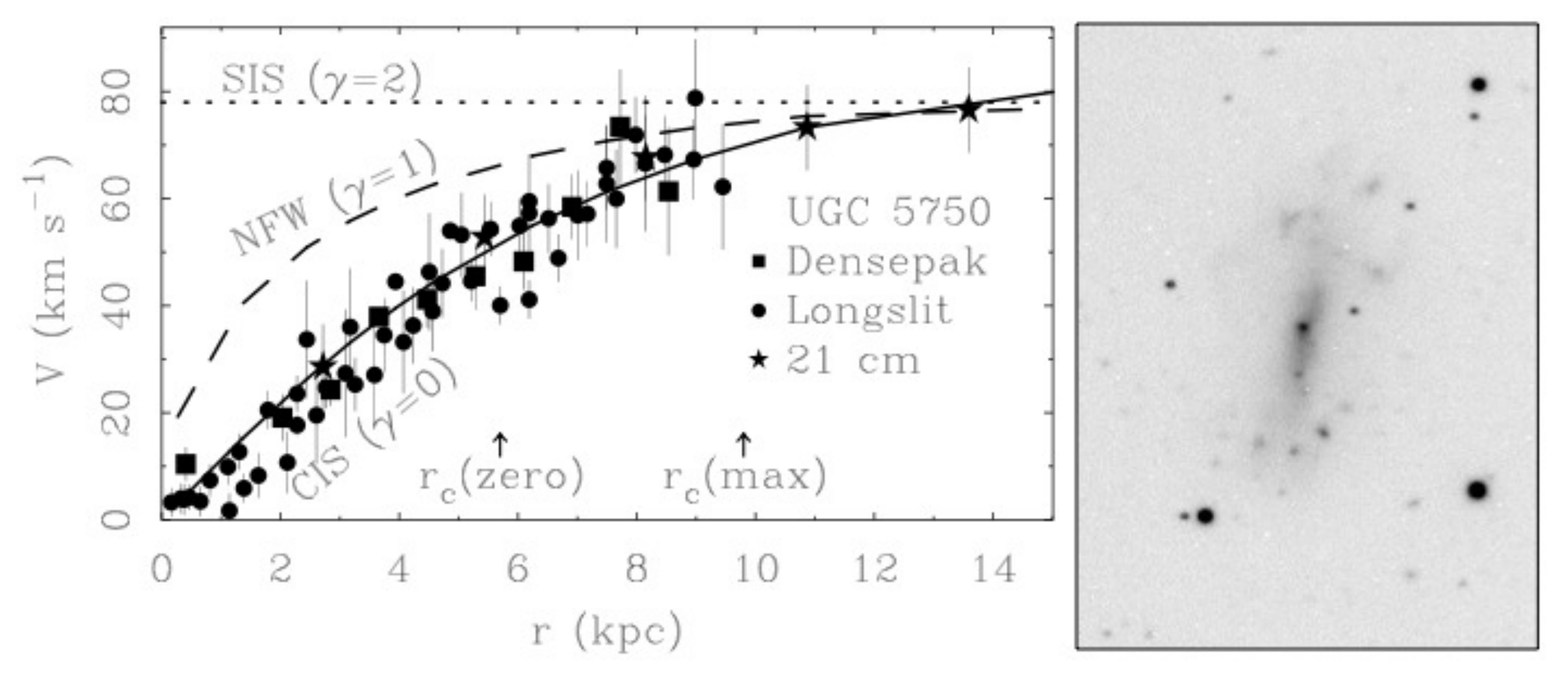}
 \caption{Rotation curve (left panel) of UGC 5750 (the LSB galaxy, showed in the right panel). The data of the rotation curve are obtained with the integrated field H$\alpha$ spectroscopy (squares) \cite{2006ApJS..165..461K}, long slit optical observations of the Balmer transition (circles) \cite{2001AJ....122.2381M,2002A&A...385..816D}, and radio observations of the 21 cm atomic hydrogen spin flip transition (stars) \cite{1993AJ....106..548V}. The isothermal sphere with a core (CIS) profile (solid line) fits the data. $r_c (\rm{zero})$ and $r_c (\rm{max})$ are the values of the core radius  obtained from the fit in the no disk and maximum disk case, respectively; the solid line is the case of no disk. Neither the NFW profile (dashed line), whose parameters are fixed and given by the $\Lambda$CDM cosmology, nor the singular isothermal sphere (SIS) profile can describe the dark matter halo of this LSB galaxy. The figure is reproduced from ~\cite{2010RAA....10.1215C}.}\label{fig:CCproblem}. 
\end{figure}

The NFW density profile is too steep  to fit the observations in the innermost regions of the dwarf and LSB galaxies \cite{McGaughanddeBlok98,C_t__2000}: the observations favor a density profile with a core \cite{10.1046/j.1365-8711.2001.04077.x,will_2018,2002A&A...385..816D,10.1046/j.1365-8711.2003.06330.x,2004MNRAS.351..903G,2000ApJ...537L...9S}. 
Figure \ref{fig:CCproblem} shows an example of how well a  density profile with a core (black solid line) fits the measured rotation curve of the LSB galaxy UGC 5750. Neither the NFW profile (long-dashed line) nor the singular isothermal sphere (SIS, dotted line) are able to fit the measured rotation curve. 

High spatial resolution rotation curves of dwarf and LSB galaxies demonstrate that these galaxies are dynamically dominated by dark matter. However, no evidence for a steep inner slope of the density profile was found on scales below $\sim 0.15$ kpc:  galaxies exhibit inner slopes $\alpha=-0.2\pm0.2$ \cite{10.1093/mnras/290.3.533}. 
Later measurements of the rotation curves of a sample of 165 low-mass galaxies yield a median inner slope  $\alpha = - 0.22 \pm 0.08$ \cite{Spekkens_2005}. Similarly,  high resolution observations of the dwarf irregular galaxy NGC 6822 \cite{2003MNRAS.340...12W}, and
HI observations of the dwarf galaxy NGC 3741 \cite{Gentile_2007} favor a core rather than the cuspy NFW  profile. 
Similar conclusions have been recently obtained with the high-resolution rotation curves of 26 dwarfs from LITTLE THINGS (Local Irregulars That Trace Luminosity Extremes, The HI Nearby Galaxy Survey) \cite{Oh_2015}. The mean slope $\alpha=- 0.32 \pm 0.24$ of this sample is consistent with the previous result found for the LSB galaxies  \cite{10.1046/j.1365-8711.2003.06330.x, Spekkens_2005}. 

Dwarf spheroidals (dSphs) also show the same cusp/core controversy. Wide-field multi-object spectrographic observations of dSphs \cite{2009ApJ...704.1274W} provide  high-quality kinematic data. A feasible analysis of those data relies on 
using the Jeans equations \cite[e.g., ][]{2007NuPhS.173...15G,2009ApJ...704.1274W,2010ApJ...710..886W}. By assuming that the dwarf galaxies contain one or more pressure-supported stellar populations in dynamical equilibrium tracing the underlying dark matter gravitational potential,  the dynamical mass distribution $M_{{\rm dyn}}(r)$, which accounts for both the stellar  and the dark matter distribution, is related to the stellar distribution through the Jeans equation (see Eq. 4-55 \& 4-56 in \cite{1987gady.book.....B}). For a spherical symmetric system, the Jeans equation reads
\begin{equation}
\frac{1}{\nu}\frac{d}{dr}(\nu \bar{v_r^2})+2\frac{\beta\bar{v_r^2}}{r}=-\frac{GM_{{\rm dyn}}}{r^2}\, ,
\label{eq:jeans}
\end{equation}
where $\nu(r)$ is the three-dimensional stellar number density, $\bar{v_r^2}(r)$ is the radial velocity dispersion, and $\beta\equiv 1-\bar{v_{\theta}^2}/\bar{v_r^2}$ is the orbital velocity anisotropy parameter of the stellar component.  For a constant $\beta$, the  solution of the previous equation is \cite{2005MNRAS.363..705M}
\begin{equation}
  \nu\bar{v^2_r}=Gr^{-2\beta}\displaystyle\int_r^{\infty}s^{2\beta-2}\nu(s)M_{\mathrm {dyn}}(s)ds\, ,
  \label{eq:jeanssolution}
\end{equation}
which must be projected along the line of sight to be compared with observations \cite{1987gady.book.....B}. 
\begin{figure}[htb!]
    \centering
    \includegraphics[width= 0.9\columnwidth]{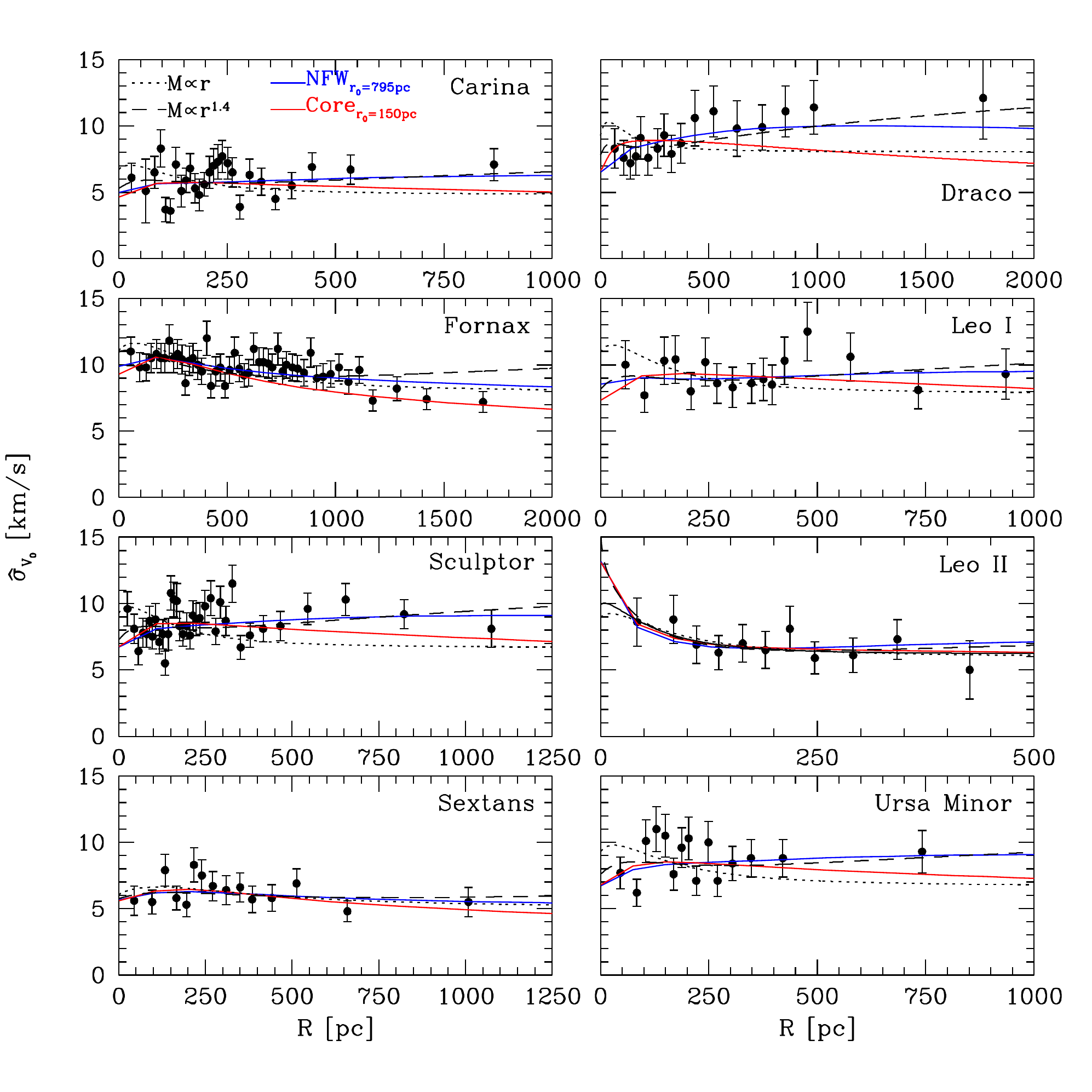}
    \caption{Projected velocity dispersion profile of the eight brightest dwarf satellites of the Milky Way. The red lines represent the best fit of the dark matter density distribution with a core, while the blue ones derive from a cuspy profile. The long-dashed and dotted lines show the isothermal and power law models, respectively. The figure is reproduced from \cite{2009ApJ...704.1274W,2010ApJ...710..886W}.}
    \label{fig:dwarf1}
\end{figure}

Under the assumption that the stellar mass is negligible compared to the halo mass, $M_{\mathrm {dyn}}$ can be identified with the dark matter mass, and Eq. \eqref{eq:jeans} is a tool to probe the dark matter model. As an example, in \cite{2009ApJ...704.1274W,2010ApJ...710..886W}, the projected velocity dispersion profiles of the Milky Way dwarf satellites were fitted using a Markov Chain Monte Carlo method.  Although the analysis did not place tight constraints on the dark matter parameters, it showed that, as illustrated in Figure \ref{fig:dwarf1}, both cuspy dark matter density distributions and distributions with a core were able to reproduce the kinematic data of these dSphs, without any statistical evidence favouring one model over the other.

However, Jeans modelling is affected by a degeneracy between the velocity anisotropy $\beta$ and the total halo mass \cite{1987gady.book.....B}, which prevents unambiguous constraints on the dark matter parameters. 
One way around would be the addition of information from multiple stellar populations \cite{Walker2011,Zhu2016a,Zhu2016b}, or higher velocity moments \cite{Lokas2003}. Using the multiple stellar populations would help to lift the degeneracy but would not improve the constraints on the dark matter parameters.  Using higher velocity moments would be more appropriate when proper motion measurements are available  \citep{Read2017,Webb2018}. For example, the Sculptor and Fornax galaxies are known to have  two distinct stellar components. Using the kinematics data of both populations,  the degeneracy between the mass and the anisotropy parameter is removed, and the dispersion velocity profiles is fitted by a dark matter halo with a core \cite{2008ApJ...681L..13B,Walker2011}. 
At odds with these results, LSB  \cite{10.1111/j.1365-2966.2004.08359.x} and late-type dwarf galaxies  \cite{van_den_Bosch_2001}  appear to have dark matter density distributions consistent with the cuspy NFW profile.

Caution should be taken in the analysis of the data. In fact, if the asphericity of the distribution of the stellar populations is not taken into account, the methodology applied in \cite{Walker2011} for dSphs may mistakenly favour a density profile with a core even if a cuspy profile is present  \cite{2018MNRAS.474.1398G}. This overlook can also introduce a bias in the estimate of the dwarf mass of the order of 10\% for a galaxy with a viewing angle, defined as the angle between the line of sight and the major axis of the distribution of the metal-poor sub-population,  of $90^\circ$ \cite{2018MNRAS.474.1398G}. 
Similarly, a more careful analysis, based on modelling the stellar orbits and leading to a non-parametric estimation of the dark matter density distribution, yields a cuspy profile with $\alpha = -1.0 \pm  0.2$ for the Draco dSph \cite{Jardel_2013}.

Despite the increasing number of studies, and the increasing accuracy of data, whether dSphs exhibit a cuspy dark matter density profile or a profile with a core is still unclear. A recent analysis of Draco found an inner dark matter density of $\rho_{DM}(150 {\mathrm{pc}}) = 2.4_{-0.6}^{+0.5} \times 10^8 \, {\rm M}_\odot$~kpc$^{-3}$, which is consistent with a CDM cuspy  density profile \cite{2018MNRAS.481..860R}. Nevertheless, 
the discovery of { ultra-diffuse} galaxies such as Crater II \cite{2017ApJ...839...20C}, and Antlia II \cite{2019MNRAS.488.2743T}, and the discovery of galaxies that appear to be devoid of dark matter, such as NGC 1052-DF2 and NGC 1052-DF4 \cite{2018ApJ...856L..30V,2018ApJ...864L..18V,2018Natur.555..629V,2018RNAAS...2...54V,2019ApJ...874L...5V,2019ApJ...874L..12D}, can challenge the standard CDM paradigm.
Specifically, galaxies such as DF2 and DF4 appear to be at $2.6\sigma$  and $4.1\sigma$ tension with the standard model, respectively: according to $\Lambda$CDM simulations,
the probability of finding DF2-like galaxies at a distance 11.5 Mpc from the observer is at most $10^{-4}$; this probability drops to $4.8 \times 10^{-7}$ at a distance 20.0 Mpc \cite{2019MNRAS.489.2634H}. { However, more accurate dynamical models can substantially alleviate this tension \cite{2018ApJ...863L..17N,2019MNRAS.484..510N,2019arXiv190708035N}.  In addition, a recent analysis \cite{10.1093/mnras/stz771} suggests that the distance to DF2 is 13 Mpc rather 20 Mpc, as previously estimated; the closer distance increases to 75\% the dark matter content of the galaxy mass.  Similarly, properly taking into account the uncertainty on the velocity measurements of the small sample of globular clusters of DF2 suggests that its mass-to-light ratio is at the low end of the distribution of the mass-to-light ratios of dwarf galaxies \cite{Martin_2018}, but its velocity dispersion and mass are still consistent with the universal mass profile of the Local Group dwarf galaxies \cite{2009ApJ...704.1274W}.}

\subsubsection{Possible solutions within the CDM model}

Possible solutions to the CCP, in the context of the CDM scenario, can reside either in neglected physical processes, mostly affecting the baryonic matter, or in systematic effects and/or observational limits.

There are processes that, in principle, can convert a cuspy density profile into a profile with a core. 
For example, an inner Lindblad-like resonance, which couples the rotating bar to the orbits of the star through the cusp, was suggested to cause an angular momentum transfer from the bar-pattern to the dark matter halo \cite{Weinberg_2002}; on turn, this transfer flattens the density profile in the central region of the halo. 
However, later investigations show opposite results: the generation of a bar in the disk would actually make steeper, albeit slightly, the inner dark matter density profile  \cite{Dubinski_2009}. 

The most popular solutions rely on  
supernova feedback and dynamical friction. Winds driven by supernovae can be an effective mechanism to transform the cusp of the dark matter density profiles into a core  \cite{Navarro_1996a,2006Natur.442..539M}. 
A similar effect can be generated by stellar winds \cite{2002MNRAS.333..299G,Mashchenko_2008}.
Supernova and stellar winds produce energy   feedback that can drastically modify the shape of the dwarf galaxies by forcing  the gas and the dark matter particles to move outwards,  change the gravitational potential well and flatten the density profile \cite{Mashchenko_2008}. 

In less massive galaxies, starbursts heat the gas that expands and inhibits the star formation. Radiative cooling makes the gas collapse again and a starburst is reignited \cite{2014ApJ...793...46O}. This cyclic starburst periodically develops density waves whose resonance with the dark matter particles generates a cusp-core transition in the halo dark matter density \cite{2014ApJ...793...46O}.

Decreasing the steepness of the density profile might not always alleviate the tension between the data and the CDM model. The profiles with a core derived from the high-resolution HI observations of LSBs can not explain the gravitational lensing signal \cite{2002ApJ...566..652L,2005ApJ...629...23C}: to match this signal, the dark matter density profile should be steeper \cite{2010RAA....10.1215C}. 

Recent high-resolution simulations of isolated dwarf galaxies show that cores of size comparable to the stellar half-mass radius may form if the star formation lasts for  $\sim$4 Gyr for a dwarf mass $M_{200}=10^8 \, {\rm M}_\odot$ and $\sim$14 Gyr for a mass $M_{200} = 10^9 \, {\rm M}_\odot$ \cite{10.1093/mnras/stw713}. However, the Auriga and EAGLE simulations suggest that  gas outflows are unable to modify the dark matter profile of dwarf galaxies, and that repeated outbursts cannot explain the transition from a cusp to a core \cite{2019MNRAS.486.4790B}.

An additional process advocated for solving the CCP is dynamical friction between gas clumps  with individual mass $10^5-10^6 \, {\rm M}_\odot$ \cite{El_Zant_2001}.
This friction between the clumps would  transfer angular momentum from the gas  to the dark matter particles that, on turn, would move away from the central region of the halo and flatten its density profile. This effect should be efficient in the early phase of the galaxy formation when the halo size is  smaller  \cite{El_Zant_2001,Romano_D_az_2008}. 
The main difficulty to make this solution efficient is to have gas clumps sufficiently massive \cite{2009ApJ...691.1300J}.  
The dynamical friction  is found to be three times weaker \cite{2009ApJ...691.1300J} than previously assumed in \cite{El_Zant_2001}, and, to increase the efficiency, the mass of the clumps should reach $\sim7\%$ of the total halo mass. Gas naturally forms clumps through the Jeans instability and the clumps can in principle be as massive as $\sim10^6 \, {\rm M}_\odot$ \cite{2006MNRAS.370.1612K}.  However, in dwarf galaxies, star formation and cloud fragmentation may be inefficient  \cite{10.1093/mnras/stu2217}, and gas clumps may form via thermal instability due to galaxy mergers. This process leads, for a total mass of the halo of the order $\sim10^7 \, {\rm M}_\odot$, to clump masses of the order $\sim10^4 \, {\rm M}_\odot$ \cite{Arata_2018}, which are two orders of magnitude less massive than required to solve the CCP \cite{2009ApJ...691.1300J}.

The globular clusters of a dwarf can also play the role of the gas clumps. This process was applied to the five globular clusters of the Fornax dSph. High-resolution simulations \cite{2020MNRAS.492.3169B} have been used to demonstrate that if  globular clusters are embedded in a dark matter mini-halo, which make them more massive and fall more rapidly towards the galaxy center,  and  were accreted within the last 3 Gyr, early accretion favours the disruption of globular clusters due to repeated interactions with the galaxy; thus, their crossing in the central region of Fornax can generate the transition from a cusp to a core, due to momentum  exchange  between the globular clusters and the galaxy dark matter halos, with the frequency of the crossing which sets the size of the core. 

The CCP may not originate from neglecting some physical processes, but rather from systematic effects \cite{de_Blok_2010} due to observational limits, including non-circular motions and pointing effects hiding the signature of a cusp in the innermost region of the galaxies. 

Systematic bias towards shallower slopes of the density profile can originate from non-circular motions of gas and stars  of at least $\sim 20$~km s$^{-1}$, offsets of $\sim 3-4$~arcsec from the correct center because of inaccurate telescope pointings, or a separation between the dynamical and photometric centers of  at least $\sim 0.5-1$~kpc \cite{10.1046/j.1365-8711.2003.06330.x}. 

However, the existence of this pointing effects was questioned by two independent observation campaigns of the same sample of galaxies made by independent groups with two different telescopes. One set of observations was made with the 193-cm telescope at the Observatoire de Haute Provence \cite{2002A&A...385..816D}, while the other set of observations was made at  the 3.6-m Telescopio Nazionale Galileo at Canary Islands \cite{Marchesini_2002}. Pointing effects  in these two independent data  sets are of the order of $0.3$~arcsec, an order of magnitude below the offset required  to  bias a cuspy   density profile towards a core 
\cite{de_Blok_2010}. 

An additional  source of error might come from the too simplistic modelling. The gas is usually supposed to move on circular orbits. If this assumption is invalid and the gas motion is disturbed, the steepness of the slope of the density profile can be underestimated \cite{2003ApJ...583..732S,10.1046/j.1365-8711.2003.06330.x}. However, using a sample of galaxies from THINGS, non-circular motions have been estimated to be of the order of few kilometres per seconds for dwarf galaxies \cite{Trachternach_2008}. These velocity perturbations are too small to overlook the presence of a cusp \cite{Eymeren2009,de_Blok_2010}. { Nevertheless, high-resolution simulations suggest that, in cuspy and triaxial CDM halos, non-circular motions can substantially affect the observed rotation curves \cite{SantosSantos_2020} and that correcting for these motions in observations is far from trivial. Therefore, the possible relevant role of non-circular motions in shaping the observed rotation curves cannot yet be ruled out. }

\subsection{The missing satellites problem}\label{sec:MSP}

{The stellar mass function derived for field galaxies and satellite galaxies in the Local Group is significantly less steep at low masses than the mass function expected for dark matter halos in the CDM model: $dn/dM_* \propto M_* ^{\alpha}$, with $\alpha \simeq -1.5$ for faint-end galaxies and  $\alpha \simeq -1.9$ for dark matter halos.
Indeed, high-resolution CDM cosmological simulations of Milky Way-like dark matter halos} 
predict thousands of subhalos with { mass sufficiently large ($M \gtrsim 10^7 \, {\rm M}_\odot$) to host galaxies. However,} this number is $\sim 20$ times larger than the number of satellites observed around the Milky Way and M31 \cite[e.g.,][]{1993MNRAS.264..201K,Klypin_1999,Moore_1999_apj,Bullock_2017}.
Recently, $\sim$50 ultra-faint galaxies with stellar mass as small as 300 ${\rm M}_\odot$ have been detected within the virial volume of the Milky Way \cite{DWetal15}, and many more may be  discovered using the Gaia satellite \cite[e.g.,][]{Antojaetal15,Ciucaetal18}. However, it is very unlikely that their number  will reach the thousands satellites predicted by the CDM simulations \cite{Bullock_2017}. This discrepancy between the predicted and observed numbers of satellites is known as the \textit{missing satellites} problem (MSP). 

\begin{figure}[htb!]
    \centering
    \includegraphics[width= 0.9\columnwidth]{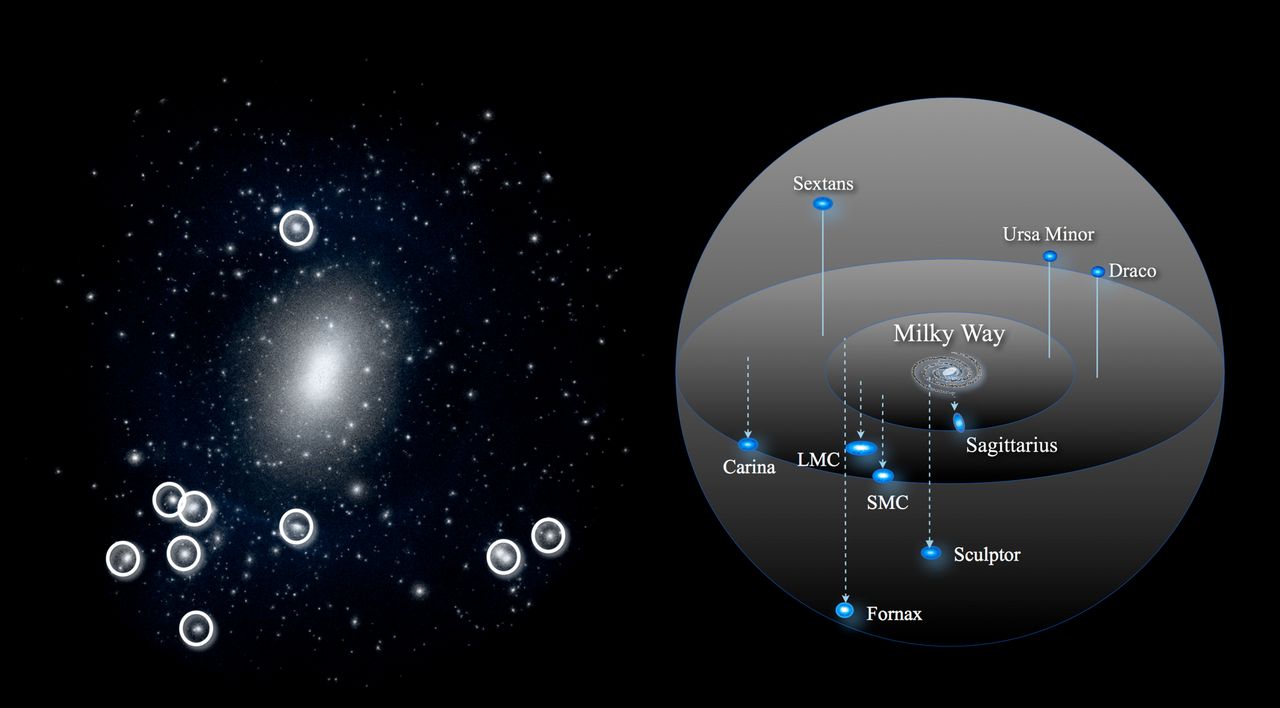}
    \caption{{\em Left panel}: Projected dark matter distribution 
    (600 kpc on a side) of a $10^{12}\, {\rm M}_\odot$ dark matter halo from the ELVIS $\Lambda$CDM simulations \cite{Garrison_Kimmel_2014}. The number of small subhalos strongly exceeds the number of known Milky Way satellites (missing satellites problem; Section \ref{sec:MSP}). The circles highlight the nine most massive subhalos. {\em Right panel}: Spatial distribution of the closest nine of the eleven most luminous (classical) satellites of the Milky Way (the diameter of the outer sphere is 300 kpc). For these satellites, the central mass inferred from stellar kinematics is a factor of $\sim$5 lower than the mass predicted for the central regions of the subhalos highlighted  in the left panel, preventing the association of the classical satellites to the most massive subhalos of the dark matter halo of Milky Way-like galaxies (too-big-to-fail problem; Section \ref{sec:TBTF}). The figure is reproduced from \cite{2015PNAS..11212249W}.}
\label{fig:MSP}
\end{figure}

Figure~\ref{fig:MSP} gives a visual representation of {the MSP}: on the left panel, thousands of subhalos predicted by the CDM simulations are shown, whereas on the right panel we can see the classical Milky Way satellites \cite{2015PNAS..11212249W}.  

{ The MSP implies that either the CDM model produces too many subhalos with low mass or the formation of galaxies in these halos becomes less and less efficient as the halo mass declines. 
While the former view calls for a modification of the CDM paradigm, the latter suggests the need for an appropriate treatment of baryonic physics.} 

\subsubsection{Possible solutions within the CDM model}

The MSP can be tackled within the CDM model by using the \textit{abundance matching} (AM) technique. This technique matches the cumulative distribution of an observed property of galaxies with the predicted cumulative distribution of the mass of their dark matter halos \cite{Bullock_2017}: for example, by adopting the mean star formation rate as the observed property, the MSP in the Milky Way appears to be solved for satellite masses larger than $10^9$~M$_\odot$  \cite{ReadandErkal19}.

For smaller masses, the solution rests upon the suppression of star formation by  UV reionization, when the mass of the dark matter halo is smaller than $M_{\rm vir} \thickapprox 10^9 \, {\rm M}_\odot$ \cite[e.g.,][]{Efstathiou92,Sawalaetal16,Bullock_2017},
or by atomic cooling in the early Universe for mass smaller than $M_{\rm vir} \thickapprox 10^8 \, {\rm M}_\odot$~\cite[e.g.,][]{ReesandOstriker77,Bullock_2017}. Both these effects become dominant in ultra-faint galaxies with stellar mass $M_*\lesssim 10^5 \, {\rm M}_\odot$ \cite{Bullock_2017}.

\begin{figure}[htb!]
    \centering
    \includegraphics[width= 0.9\columnwidth]{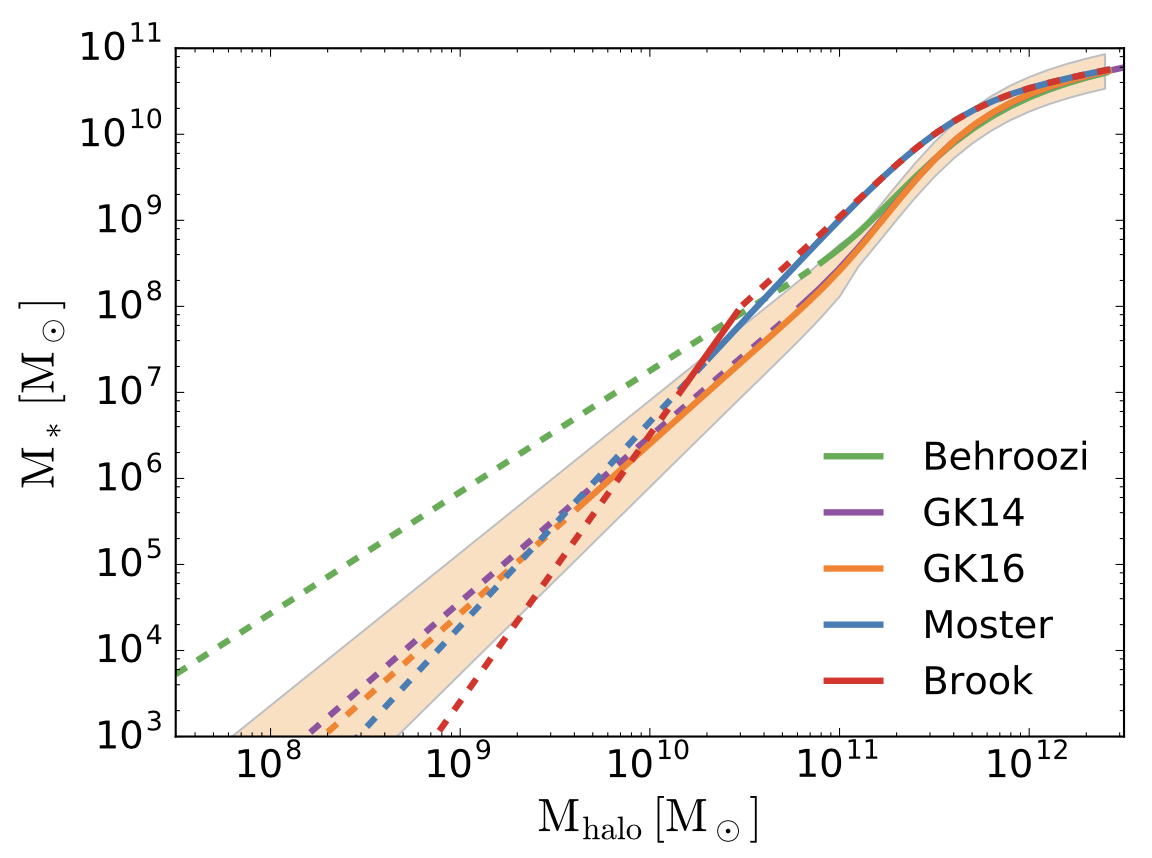}
    \caption{{ Abundance matching (AM) relations from the models by Behroozi~\cite{Behroozi_2013}, Garrison-Kimmel GK14~\cite{Garrison-Kimmel_2014}, Garrison-Kimmel GK16~\cite{Garrison-Kimmel_2017}, Moster~\cite{Moster_2013}, and Brook~\cite{Brook_2014}. The solid lines show the models in a region of the parameter space  where observational data are available; the dashed lines show the extrapolation of the models where the observational information is missing. The orange shaded area show the 1$\sigma$ log-normal scatter around the model GK16. The figure is  from~\cite{Dooley_2017}.}}
    \label{fig:AM_Mstar}
\end{figure}

AM can be used to quantify this expectation. { Figure~\ref{fig:AM_Mstar} shows the correlation between the stellar mass,  $M_*$, and the total mass of the galaxy, $M_{\mathrm {halo}}$. In observations, the stellar mass $M_*$ can be obtained by fitting broad-band photometric data with a Spectral Energy Distribution model \cite[e.g.,][]{2009ApJ...696..348W,2011Ap&SS.331....1W,2013MNRAS.435...87M}, whereas the total galaxy mass $M_{\mathrm {halo}}$ can be inferred from either gravitational lensing or HI
rotation curves \cite[e.g.,][]{2006MNRAS.368..715M,2010ApJ...710..903M,2017MNRAS.466.1648K}.}
{ Figure~\ref{fig:AM_Mstar} also shows various models that adopt  different galaxy data sets and the dark matter halo mass function from different $N$-body simulations to derive
the expected $M_*-M_{\mathrm {halo}}$ relation. In addition, they adopt different assumptions on the star formation rate and the star formation history of the galaxies to infer $M_*$ from the galaxy luminosities. Therefore, the models can be different, especially at low stellar and halo masses. 
In Figure~\ref{fig:AM_Mstar}, the solid lines show the predictions derived for stellar masses $M_* \gtrsim 10^8 \, {\rm M}_\odot$, where observational data are available, whereas the dashed lines show the extrapolation of the models at smaller masses, where the observational information is missing. The orange shaded area shows the 1$\sigma$ log-normal scatter around one of the models shown in \cite{Garrison-Kimmel_2017}. For $M_* \gtrsim 10^8 \, {\rm M}_\odot$, all the models are consistent with each other. However, at smaller stellar masses, the incompleteness of the observational surveys and the increased stochasticity of the star formation mechanism in the models { make the estimation of the stellar mass uncertain and, consequently, the AM relation less constrained:} 
for example, a stellar mass $M_* \sim 10^6 \, {\rm M}_\odot$ would correspond to $M_{\mathrm{halo}} \sim 10^9 \, {\rm M}_\odot$ for the Behroozi model, but to $M_{\mathrm{halo}} \sim 5\times10^9 \, {\rm M}_\odot$ for the Brook model. These differences translate into an additional uncertainty on the number of low mass satellites in the Local Group.  
This uncertainty is further increased by the fact that this halo mass, $M_{\rm halo} \thickapprox 10^9 {\mathrm M}_\odot$, is close to the mass scale, $\approx10^8 \, {\rm M}_\odot$, below which star formation is expected to be suppressed. }

\begin{figure}[htb!]
    \centering
    \includegraphics[width= 0.9\columnwidth]{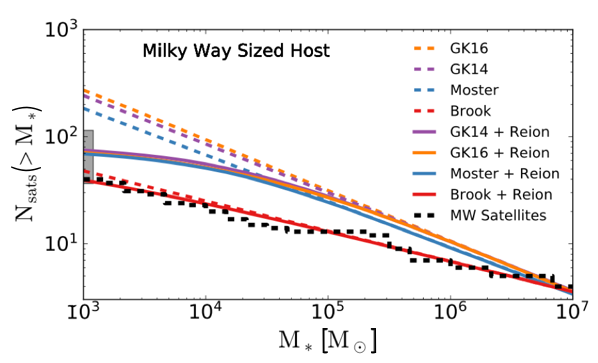}
 \caption{{ Mean number of satellite galaxies with mass larger than a given stellar mass threshold, $N_{\rm sats}(>M_*)$, as a function of the threshold mass, $M_*$, around a Milky Way-sized galaxy hosted by a dark matter halo with virial mass  $M_{\rm halo}=1.4\times 10^{12} \, {\rm M}_\odot$, as predicted by various AM models applied to the dark-matter only {\it Caterpillar} simulations.
 The colored dashed lines indicate the predictions of the AM models shown in Fig.~\ref{fig:AM_Mstar}; the solid lines show the predictions of the same models after the inclusion of reionization, that suppresses the formation of low-mass galaxies.
 The grey box represents the prediction of 37-114 satellites with stellar luminosity $L_* > 10^3 L_{\odot}$
 by ~\cite{Hargis_2014}, derived by combining a number of toy models applied to dark matter-only simulations with the sample of dwarfs corrected for completeness, observed by the Sloan Digital Sky Survey.
 For comparison, the complementary cumulative stellar mass function (derived from $V$-band luminosities, assuming a mass-to-light ratio equal to 1 ${\rm M}_\odot/{\rm L}_\odot$) of the 40 known satellite galaxies of the Milky Way with $M_* > 10^3 \, {\rm M}_\odot$ is shown with a black dashed line. 
 The figure is reproduced from~\cite{Dooley_2017}. Here, we corrected the misprint that is present in the label of the ordinate axis of the original figure.}}
\label{fig:Cum_count_Mstar}
\end{figure}

{ Dooley et al.\ \cite{Dooley_2017} applied the  above-mentioned AM models, both with and without the inclusion of reionization effects, to a set of 
Milky Way-like dark matter halos from the dark-matter only {\it Caterpillar} simulations.  For each AM model, Dooley et al.\ estimate, for any dark matter halo with a given mass, the expected number of satellite galaxies with a given stellar mass. The results of this exercise, for a dark matter halo with mass $M_{\rm halo}=1.4\times 10^{12} \, {\rm M}_\odot$, is shown in
Figure~\ref{fig:Cum_count_Mstar}:} the { Brook AM model introduced in \cite{Brook_2014} (red line in Figure \ref{fig:AM_Mstar})  agrees with the { complementary} cumulative\footnote{{ In the relevant literature,  adopting the expression {\it cumulative distribution function} for $N(>X_*)$, namely the number of objects with a physical properties $X$ larger than a threshold $X_*$, dates back to the pioneering paper by Moore et al.\ \cite{Moore_1999_apj}. However, in statistics, the cumulative distribution function actually is $N(<X_*)$. $N(>X_*)$ is the {\it complementary cumulative distribution} function. Here, we prefer to adopt this more rigorous terminology.} } { stellar mass function}, $N_{\rm sats} (>M_*)$, of the { 40 detected} Milky Way's dwarf satellites { with $M_* > 10^3 \, {\rm M}_\odot$ (black dashed line), irrespective of the inclusion of reionization
({ solid and dashed} red lines)}. { On the other hand,} the other AM models shown in Figure~\ref{fig:AM_Mstar} overestimate the observed { complementary} cumulative stellar mass function of the Milky Way satellites, { even when reionization suppresses the formation of low-mass galaxies}.} 
{ A better agreement between these models and the { complementary} cumulative stellar mass function of the Milky Way satellites is obtained by assuming { a lower halo mass for the Milky Way.} In this case, the Brook model would clearly underestimate the number of satellites. 
The numbers of Milky Way satellites with $M_* > 10^3 \, {\rm M}_\odot$ predicted by the other models in Figure \ref{fig:AM_Mstar} are consistent with other analyses~\cite{Hargis_2014,DWetal15} and would solve the MSP if more than 80 satellites with $M_* > 10^3 \, {\rm M}_\odot$ were detected.}
Figure ~\ref{fig:Cum_count_Mstar} also shows that { all the models agree with each other and with the Milky Way { complementary} cumulative stellar mass function for $M_* \gtrsim 4.5\times10^5 \, {\rm M}_\odot$. This agreement suggests that there might be no MSP for halos with stellar mass $M_*>4.5\times10^5 \, {\rm M}_\odot$ or, equivalently, for halos with total mass $M_{\rm halo} \gtrsim 10^{10} \, {\rm M}_\odot$ \cite{Dooley_2017}.}

Although the stellar mass $M_*$ has been successfully used for the AM method, it suffers from several biases. For example,   the suppression of star formation due to either the infall of the galaxy toward a larger galaxy \cite[e.g.,][]{2012ApJ...757...85G,2013MNRAS.433.2749G} or tidal stripping \cite[e.g.,][]{2006MNRAS.366..429R,Read_2006,2016ApJ...827L..15T} introduces an intrinsic scatter in  the $M_*-M_{\mathrm {halo}}$ relation \cite[e.g.,][]{2015MNRAS.452.1861C,2015NatCo...6.7599U,2017MNRAS.467.2019R} which must be taken into account to correctly predict the number of satellites \cite[e.g.,][]{2005ApJ...624..505Z,2006RPPh...69.3101B}. One possible solution relies on replacing the stellar mass with the mean star formation rate, $\langle {\rm SFR} \rangle$, averaged over the star-forming time of the galaxy, as anticipated at the beginning of this section  \cite{ReadandErkal19}: for isolated galaxies, the stellar mass rises monotonically with the halo mass \cite{2010ApJ...710..903M,2017MNRAS.466.1648K,2017MNRAS.467.2019R}, and $\langle {\rm SFR} \rangle$ is reasonably expected to behave similarly. By adopting $\langle {\rm SFR} \rangle$ rather than $M_*$, the scatter in the AM relation at a given $M_{\mathrm {halo}}$ decreases, and  the  accuracy  of  the  AM mass estimator increases \cite{ReadandErkal19}. 

The  $\langle {\rm SFR} \rangle - M_{200}$\footnote{{ The radius $r_{200}(z)$ at redshift $z$
 is the radius of a spherical volume within which the mean mass density is $200$ times the critical density of the Universe $\rho_c(z)$. The mass enclosed within $r_{200}$ is thus  $M_{200} (z)= 4\pi 200 \rho_c(z)r_{200}^3/3$.
$M_{200}$ is approximately equal to the viral mass $M_{\mathrm{vir}}$, defined in the footnote \ref{foot:vir}, in a Universe with $\Omega_m=1$ \cite{2001A&A...367...27W}.}} relation derived in  \cite{ReadandErkal19} is based on  surveys of faint galaxies \cite{Blanton_2005}, whose stellar mass function is complete down to $M_*\sim2\times10^7\, {\rm M}_\odot$ \cite{Baueretal13,Hill_2017}. 
The $\langle {\rm SFR} \rangle - M_{200}$ relation is derived by combining estimates of the star formation rate from the data and the halo mass function from the model. The mass $M_{200}$ of the subhalos are identified with the mass of the dark matter halos just before falling into the main host halo.  

\begin{figure}[htb]
    \centering
    \includegraphics[width=0.9\columnwidth]{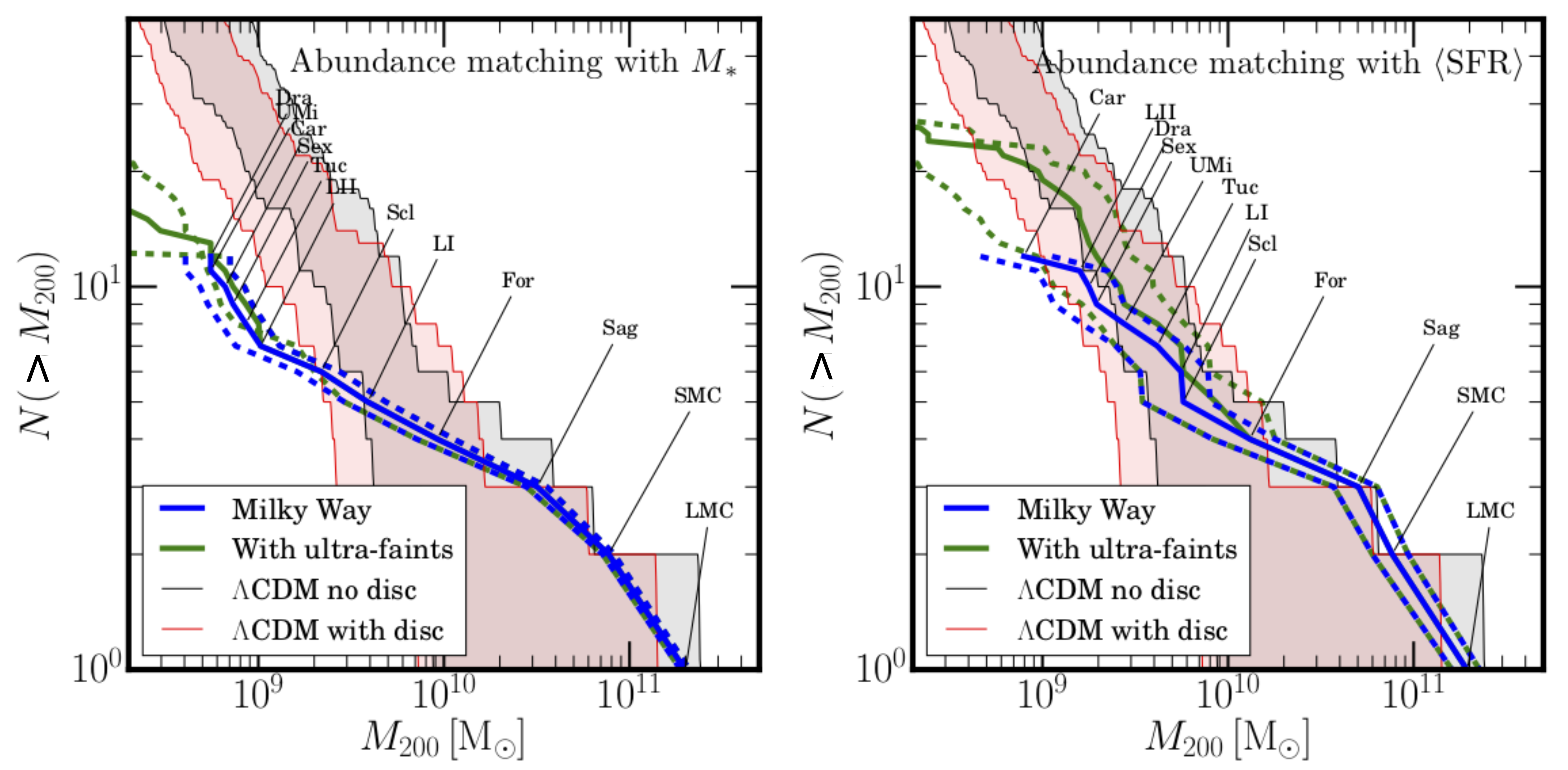} 
    \caption{{ Complementary} cumulative subhalo mass function of the Milky Way satellites in a sphere of radius 280 kpc centred on the Galaxy. The mass function is derived from either the $M_*-M_{200}$ AM relation (left panel) or the $\langle {\rm SFR} \rangle - M_{200}$ AM relation (right panel). The peak masses before the infall are taken as the subhalos masses $M_{200}$. The names of the individual galaxies used to build the complementary cumulative  subhalo mass functions are indicated in the plots. The median relations are shown as blue solid lines, whereas the $\pm 68$\% confidence intervals are delimited by the blue dashed lines. The green lines have the same meaning as the blue lines but they also include the sample of ultra-faint dwarf galaxies of~\cite{McConnachie12}. The grey shaded areas represent the variation of the complementary cumulative subhalo mass functions resulting from ten dark-matter-only ``zoom-in'' simulations of the Milky Way in the $\Lambda$CDM model. The red shaded areas have the same meaning as the grey shaded areas but they include a model for the stellar disk of the Milky Way. The figure is from~\cite{ReadandErkal19}. Here, we corrected the misprint that is present in the label of the ordinate axis of the original figure.}
    \label{fig:Cum_count_Mstar_MeanSFR}
\end{figure}

Figure~\ref{fig:Cum_count_Mstar_MeanSFR} shows the { { complementary} cumulative subhalo mass function of a complete volume-limited sample of the bright Milky Way satellites within 280 kpc from the Galaxy center. In the left panel, the subhalo mass function is derived from the $M_*-M_{200}$ AM relation, whereas the right panel shows the subhalo mass function based on the $\langle {\rm SFR} \rangle - M_{200}$ AM relation \cite{ReadandErkal19}. The { complementary} cumulative  subhalo mass function based on the $M_*-M_{200}$ relation (left panel) shows the existence of the MSP for masses $M_{200} \lesssim 2 \times 10^9 \, {\rm M}_\odot$: there are not enough quenched satellites, such as Sculptor and Leo I, in the Milky Way to be consistent with the predictions of the $\Lambda$CDM model. On the contrary, the complementary cumulative subhalo mass function based on the $\langle {\rm SFR} \rangle - M_{200}$ AM relation (right panel) suggests that no MSP exists for masses above $M_{200}\sim 10^9 \, {\rm M}_\odot$.}

{ The subhalos with pre-infall mass $M_{200} \sim 5 \times 10^8 - 5 \times 10^9 \, {\rm M}_\odot$ correspond to  ultra-faint dwarf galaxies \cite[e.g.,][]{ReadandErkal19,Kimetal17,Jethwaetal18,Contentaetal18}; the number of Milky Way satellites  with mass $M_{200} \lesssim 5 \times 10^8 \, {\rm M}_\odot$ is  smaller than the number predicted by $\Lambda$CDM, and it is still unclear whether a MSP really exists below this mass scale. These satellites might be fainter than the completeness limit of the current galaxy surveys \cite[e.g.,][]{Koposovetal08,Koposov_2015,Bechtoletal15}, or they might be completely dark \cite[e.g.,][]{Readetal2006c}.}

\subsection{ The too-big-to-fail problem} \label{sec:TBTF}

\begin{figure}[!ht]
 \centering
 \includegraphics[width=0.8\columnwidth]{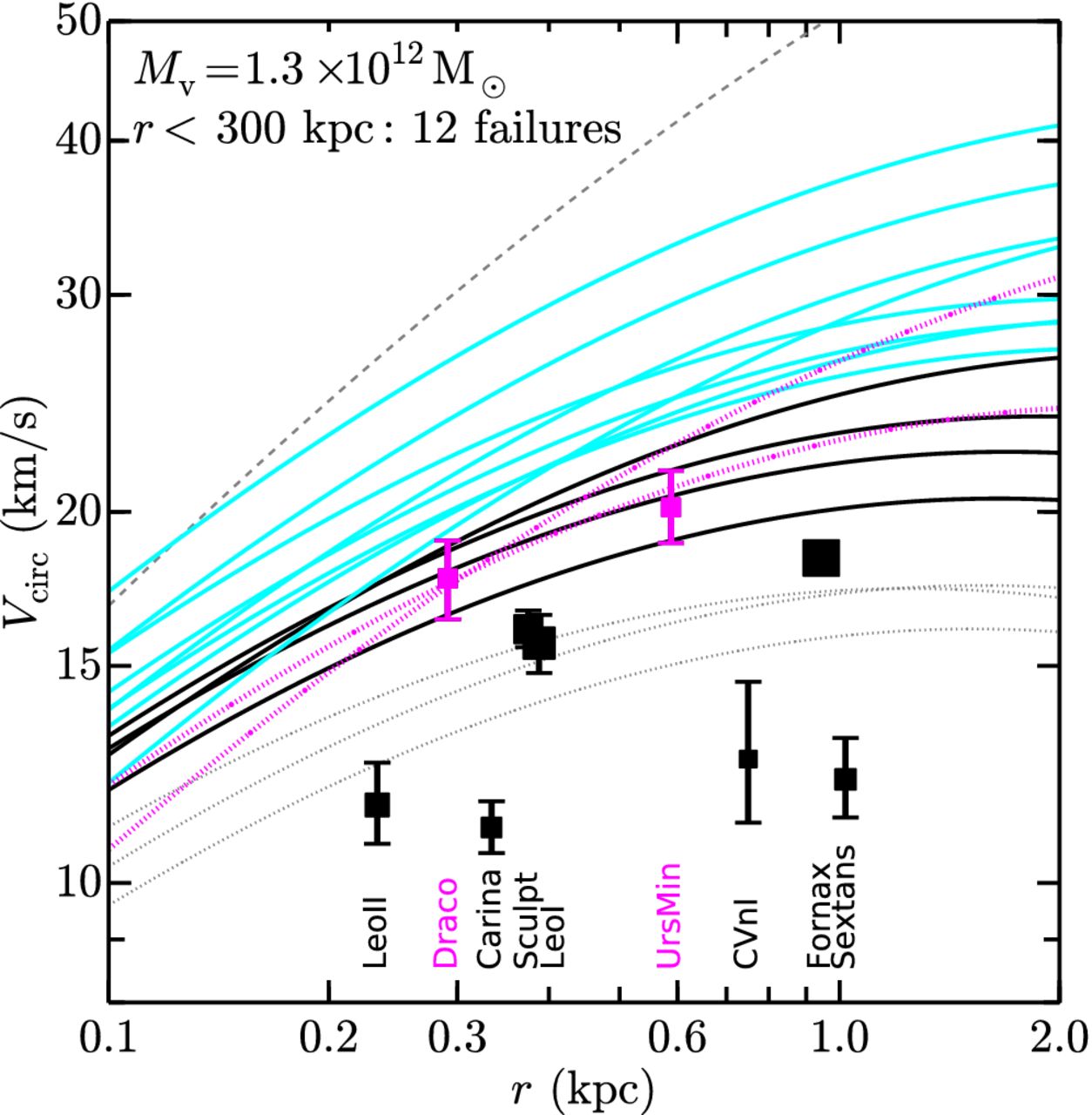}
 \caption{  
 Circular velocity profiles of simulated dark matter subhalos compared to measured velocity dispersions of Milky Way dSph satellites to illustrate the too-big-to-fail problem in the Milky Way system.
 The circular velocity profiles, shown with lines of different colors and styles, are derived by Garrison-Kimmel et al.\ \cite{Garrison_Kimmel_2014} for the resolved, massive dark matter subhalos characterized by maximum circular velocity $V_{\rm peak}>30$ km s$^{-1}$ at the time when the halo reaches its maximal mass;
 these subhalos are identified in the ELVIS simulations within 300 kpc of the centre of a dark matter halo with virial mass $M_{\mathrm vir} = 1.3 \times 10^{12}\, {\rm M}_\odot$. 
 The circular velocities, $V_{\rm circ}(r_{1/2})$ measured at the half-light radius of the Milky Way dSph satellites brighter than $2 \times 10^5\,  {\rm L}_\odot$, are compiled by Wolf et al.\ \cite{Wolf_2010} and are plotted as filled squares whose size is proportional to the logarithm of the dSph stellar mass.
 The cyan solid lines represent the circular velocity profiles of the so-called ``strong massive failures'', subhalos that are too dense to host {\it any} of the observed Milky Way dSphs (i.e. whose  $V_{\rm circ}(r_{1/2})$ is above the $1\sigma$ constraints for all the dwarfs in the sample).
 The black solid lines indicate the ``massive failures'', i.e. additional subhaloes that are not accounted for by the dense galaxies in the observational sample. The subhalos selected to host the high-density galaxies, Draco and Ursa Minor, are indicated by dotted magenta lines, with their associated galaxies plotted as magenta squares.
 The dotted lines plot the subhalos that are consistent with at least one of the remaining seven dwarfs in the sample.
 The grey dashed line indicates the sole subhalo expected to host a Magellanic Cloud ($V_{\rm max} > 60$  km $s^{-1}$). 
 Not plotted are 40 resolved ($V_{\rm max} > 15$ km s$^{-1}$) subhalos with $V_{\rm peak} < 30$ km s$^{-1}$. 
 Overall, 
 there are 12 unaccounted-for massive failures, including eight strong massive failures that are too dense to host any bright Milky Way dSph. The figure is reproduced from \cite{Garrison_Kimmel_2014}.}
\label{tbtf_rc}
\end{figure}

{  Dissipationless $\Lambda$CDM simulations of Milky Way-like dark matter halos predict that the most massive subhalos of the Milky Way are too dense to host any of the brightest ($L_{\rm V} > 10^5 \, {\rm L}_\odot$) Milky Way dSph satellites \cite{Read_2006,Boylan_Kolchin_2011,Boylan_Kolchin_2012}.
The density discrepancy can be erased by assuming that the most massive, dense subhalos are dark and the brightest dwarfs reside in subhalos that are a factor of $\sim 5$ less massive. 
Clearly, this scenario generates a new, serious conflict, because the largest dark matter subhalos, characterized by the deepest potential wells, are expected to be able to retain gas and form stars:  
in other words, they are ``too big to fail'' to form stars and should thus host observed dwarfs.
Solving this latter conflict requires going back to the simplest scenario, where the most massive subhalos do host the brightest dSphs, and where the density conflict appears.
This issue was dubbed as the {\it too-big-to-fail} (TBTF) problem by Boylan-Kolchin et al.\ \cite{Boylan_Kolchin_2011}.

An illustration of the TBTF issue  (together with the MSP; see Section~\ref{sec:MSP}) is presented in Figure~\ref{fig:MSP} for the Milky Way: in the left panel, the nine most massive subhalos of a dark matter distribution in a simulated Milky Way-like dark matter halo are highlighted with circles; the right panel shows the actual distribution of the nine, most closeby ``classical'' satellite galaxies of the Milky Way. 
The overdensity of the most massive subhalos with respect to the ``classical satellites'' prevents their reciprocal association, generating the TBTF conflict.

It is important to note that the TBTF problem is an issue distinct from the MSP discussed in  Section~\ref{sec:MSP}. It can be seen as a problem of ``missing dense satellites'' 
related to the internal mass distribution of subhalos, rather than to their abundance \cite{Garrison_Kimmel_2014}.
Therefore, the TBTF problem is largely independent of the exact relationship between halo mass and stellar mass (see, e.g., the AM technique in Section \ref{sec:MSP}).
The relevance of the mismatch between estimated and predicted central densities 
depends upon the specific realization of the dark matter halo substructure. However, the comparison of the observational data with a series of simulations 
shows that, for the Milky Way, the discrepancy appears too large to be a statistical fluke \cite{Boylan_Kolchin_2011,Boylan_Kolchin_2012}.

The density discrepancy manifests as a kinematic discrepancy between estimated and predicted circular velocities.
Specifically, for dispersion-supported systems as the Milky Way dSph galaxies, the circular velocity  at any given radius, $V_{\rm circ}(r)$, is indicative of the dynamical mass enclosed within that radius and thus of the matter density within that radius: $V^2_{\rm circ}(r) =G\, M(<r)/r$.
As shown by Wolf et al.~\cite{Wolf_2010}, an estimate of the luminosity-weighted square of the line-of-sight velocity dispersion enables to accurately determine the dynamical mass of a spherical, dispersion-supported system within a characteristic radius, approximately equal to the deprojected half-light radius, $r_{1/2}$, through the relation $M_{1/2} \equiv M(<r_{1/2})=3\, G^{-1} \langle \sigma^2_{\rm los} \rangle r_{1/2}$; the relevant feature of this estimate is that it is nearly independent of the spatial variation of the stellar velocity dispersion anisotropy, $\beta (r)$, as long as the radial profile of the velocity dispersion is fairly flat near $r_{1/2}$, as typically observed. 
From the relation between circular velocity and dynamical mass given above, it follows that 
the circular velocity at the half-light radius can be written, in terms of the observable line-of-sight velocity dispersion, as $V_{\rm circ}(r_{1/2})=\sqrt{3\, \langle \sigma^2_{\rm los} \rangle}$.

Because dSphs are dark-matter dominated at all radii, the dynamical mass $M_{\rm dwarf}(<r_{1/2})$ is approximately equal to the dark matter mass enclosed within the half-light radius. 
A necessary, although not sufficient, condition for a dark matter subhalo to host such a dispersion-supported galaxy is that, at the {\it observed} half-light radius  of the galaxy, $r=r_{1/2}$, the mass $M_{\rm dwarf}(<r_{1/2})$ agrees with the mass of the subhalo at the same radius, $M_{\rm sub}(<r_{1/2})$ \cite{Boylan_Kolchin_2011}. This agreement, in turn, requires that the observed and simulated values of $V_{\rm circ}$ must agree with each other at $r=r_{1/2}$.

For the brightest Milky Way dSphs, this requirement is fulfilled only by subhalos that are not among the most massive subhalos produced in $\Lambda$CDM simulations.
On the contrary, the most massive subhalos are grossly inconsistent with the kinematic properties of the dSphs: they are characterized by $M_{\rm sub}(<r_{1/2})$ systematically larger than $M_{\rm dwarf}(<r_{1/2})$,
and consequently by $V_{\rm circ}(r_{1/2})$ systematically higher than observed.
For instance, in the Aquarius simulations 
\cite{Springel_2008}, at least 10 massive subhalos are characterized by circular-velocity profiles that are not consistent with {\it any} of the observed dSph circular velocities, $V_{\rm circ}(r_{1/2})$:
these subhalos have their maximum circular velocity (at $z=0$) systematically higher ($V_{\rm max} > 25$ km s$^{-1}$) than those of the halos that best fit the observational data of the Milky Way dSph galaxies ($V_{\rm max} \sim 12-25$ km s$^{-1}$) \cite{Boylan_Kolchin_2012}. Similar results are obtained by Garrison-Kimmel et al.\ \cite{Garrison_Kimmel_2014} with the ELVIS simulations: 
here, the number of dense halos unaccounted for by observations ranges from 2 to 25 within 300 kpc from the Milky Way. Figure~\ref{tbtf_rc} illustrates this result.

Even though it was originally identified for the Milky Way system, the TBTF issue also appears in the M31 system, where the dSph satellites have circular velocities $V_{\rm circ}(r_{1/2})$
systematically lower than expected for the most massive subhalos of $\Lambda$CDM simulations of Milky Way-like dark matter halos \cite{Tollerud_2014}. 
Furthermore, the conflict emerges in isolated field galaxies of the Local Group, indicating that this is not a satellite-specific problem: none of these galaxies is denser than the densest satellite of the Milky Way or M31 \cite{Kirby_2014}, and a comparison of the observational data with the ELVIS simulations shows that the number of halos  
that are too dense to be consistent with the observations 
 increases from the above-mentioned 2--25 to a number of 12 to 40, when moving to the outskirts of the Local Group \cite{Garrison_Kimmel_2014}. 
 
 The TBTF issue also appears beyond the Local Group. By applying the AM technique to the number density of galaxies in the ALFALFA sample as a function of their rotational velocity as inferred from the width of their HI emission line, Papastergis et al. \cite{Papastergis_2015} show that either the dwarf galaxies are hosted by halos that are significantly more massive than indicated by their rotation velocity or the number density of galaxies at the scale of dwarfs is much larger than observed.
 Clearly, neither solution is consistent with the observations.
 
 Similar tensions are suggested by photometric observations alone. The systems of galaxies surrounding both M94 and M101 appear to contain galaxies that are substantially less massive than expected in the CDM framework. M94 is a group with a central Milky Way-mass galaxy that displays only two satellites, each with mass $M_*\lesssim 10^6 \, {\rm M}_\odot$, instead of the $\sim 10$ satellites expected for similar systems \cite{Smercina_2018}. The luminosity function of the galaxies of the M101 group is similar to the luminosity function of the Milky Way system, suggesting a similar lack of intermediate-mass galaxies \cite{Danieli_2017}. In the ultra-faint dwarf galaxy regime, the M101 group also appears to exacerbate the MSP (see Section~\ref{sec:MSP}), by containing only half of the satellites of the Milky Way system \cite{Bennet_2020}. 
}

\subsubsection{  Possible solutions within the CDM model}
{ 

A systematic overestimate by a factor of $\sim 2$  of the mass of the dark matter halo of the Milky Way can in principle remove the TBTF problem for the Galaxy \cite{Boylan_Kolchin_2012}. 
If the Milky Way mass were $M \sim 5\times 10^{11}$~M$_\odot$, the number of extra massive subhalos that do not correspond to observed dwarfs would be $\sim 3$ and might be consistent with statistical fluctuations. On the contrary, mass estimates
of  $M \sim 1\times 10^{12}$~M$_\odot$ \cite{Xue2008, Brown2010, Gnedin2010, Cautun2020} or  $M \sim 2\times 10^{12}$~M$_\odot$ \cite{Sakamoto2003, Sohn2018} imply  larger numbers of extra subhalos, $\sim 6$
or $\sim 12$, respectively, and generate the TBTF problem.
However, there are currently no indications of the existence of systematic errors that can reduce the estimate of the Milky Way mass by the required factor.
In addition, the TBTF problem appears in the M31 and in other galaxies; therefore, the overestimate of the Milky Way mass would not fully remove the tension.

The most promising way out is to resort to the effects of  baryonic physics. N-body/hydrodynamical simulations suggest that energetic feedback from supernovae can substantially reduce the amount of dark matter mass in the central region of  the subhalos, thus decreasing the observable circular velocity \cite{Brooks2014,Dutton2016,Buck2019}. 
The mass depletion in the central regions appears to be less relevant  in the APOSTLE simulations \cite{Fattahi2016, Fattahi2018, Sawalaetal16} when the EAGLE models for galaxy formation \cite{Crain2015, Schaye2015} are implemented; in these simulations, the TBTF problem is instead mostly solved by subhalo disruption and mass loss due to tidal stripping. 
Similar results are obtained in the  FIRE simulations of Local Group-like and isolated Milky Way-like volumes \cite{Garrison-Kimmel2019}.
However, in these simulations,   Local Group-like volumes still contain $\sim 10$ more subhalos than observed  with stellar mass $M_*>10^5$~M$_\odot$, showing how  reproducing the properties of dwarfs both in high- and low-density regions remains challenging.

The most relevant limitation of the N-body/hydrodynamical simulations is the mass resolution of the baryonic particles. This limitation makes the final results dependent on the details of the recipes implemented for the baryonic physical processes, mostly gas cooling, star formation, stellar and supernova feedback, and chemical enrichment of the interstellar medium. For example, the APOSTLE simulations typically have particle mass $\sim 10^5$~M$_\odot$, whereas in the simulations of Brooks and Zolotov \cite{Brooks2014} the particle mass is $\sim 2\times 10^4$~M$_\odot$ and in the FIRE simulations it is $\sim 3.5-7\times 10^3$~M$_\odot$ \cite{Garrison-Kimmel2019}. Therefore, the physical processes are  self-consistently resolved only on scales comparable or larger than the total stellar mass of the least massive classical dwarfs. Self-consistently resolving the baryonic physics on smaller scales is still beyond the current capability of the simulations and the full implication of the processes on these scales remain to be assessed.  

The TBTF problem might be partially solved without resorting to these baryonic subgrid processes: tidal effects of the baryonic disk of the host galaxy can be sufficient to reduce the
circular velocities at 300 pc by 20-30\% for most subhalos, as demonstrated by dark-matter only simulations where a growing disk
potential is embedded at the centre of the dark matter halo \cite{Robles_2019}. However, this solution generates an anticorrelation between the central density of the subhalos and their pericentre values that lacks in the observations; in addition, this solution cannot be applied to field dwarf galaxies.

Ostriker et al.\ \cite{Ostriker_2019} also suggest that resorting to baryonic physics is actually unnecessary. They identify the TBTF problem with the lack, in galaxy groups, of moderately luminous galaxies, 1--2 magnitudes fainter than the brightest member. 
According to their analysis, N-body simulations, including the EAGLE and IllustrisTNG simulations, provide magnitude gaps between the most and the second most luminous galaxies comparable to or even larger than observed. It follows that the TBTF problem would vanish. Because these large gaps are also present in dark-matter only simulations, where a mass gap replaces the magnitude gap, Ostriker and collaborators conclude that these gaps are driven by gravitational dynamics rather than baryonic physics. However, it remains to be shown that this photometric interpretation of the TBTF problem also removes the original discrepancy between the circular velocities of the real dwarfs and the circular velocity of the most massive dark matter subhalos in galaxy groups. 
}

\subsection{Planes of satellite galaxies} \label{sec:planeofsatellite}

The orbits of the satellite { galaxies of the Milky Way, of M31, and of} Centaurus A (CenA), whose properties are well-measured, tend to be aligned in 
{ significantly flattened configurations}  
\cite{Pawlowski_2012, Ibata_2013, Conn_2013, Tully:2015zfa, Crnojevi__2014, Crnojevi__2016, Muller:2016ddt, Muller:2018hks}. Moreover, the satellites have significant kinematic correlations. These spatial alignments and kinematic coherence are extremely rare in CDM simulations \cite{Pawlowski:2014una}. This 
{ puzzling fact} is referred to as the \textit{planes of satellite galaxies} problem (PSP). Unlike some other small scale problems of the CDM model, this issue may not be tackled by incorporating the baryonic feedback alone, because the satellites are at larger distances compared to the size of the baryonic disk of the host galaxy. For an extensive review of this topic, the reader may refer to ~\cite{Pawlowski:2018sys}. Here, we briefly mention the observational evidence of such alignments, how rarely they occur in the simulations performed so far, and some suggested solutions within the CDM framework.

\subsubsection{Evidence of the orbital alignment of the satellites}

\begin{figure}[htb!]
 \centering
 \includegraphics[width=0.9\columnwidth]{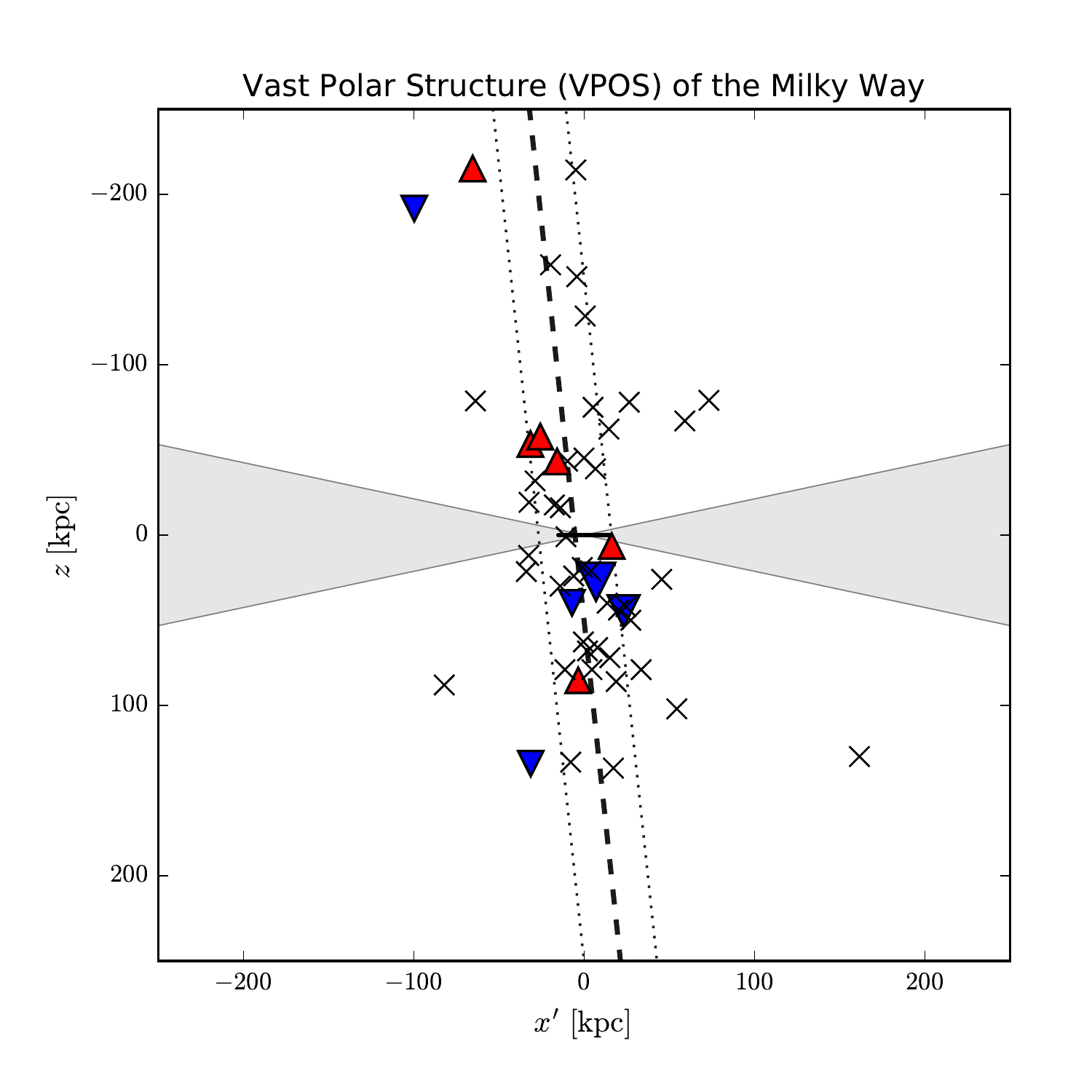}
 \caption{Edge-on view of the \textit{vast polar structure} (VPOS) around the Milky Way and the disk (solid black line at the center). The orientation and the width of the best fit satellite plane are indicated by the dashed and dotted lines, respectively. The colored triangles, red upward (blue downward) for receding from (approaching towards) an observer at rest with respect to the host galaxy, indicate coherent kinematics of the co-orbiting satellites. The satellites with no proper motion measurements are plotted as crosses. The grey area corresponds to the region $\pm 12^\circ$ from the Milky Way disk which is obscured by galactic foreground. The figure is reproduced from \cite{Pawlowski:2018sys}.}
\label{MW}
\end{figure}

 For a long time, it has been known  that the well measured satellite galaxies of the Milky Way lie almost on the polar great circle \cite{10.1093/mnras/174.3.695, 1976RGOB..182..241K}. A more recent study confirms the existence, around the Milky Way, of a \textit{vast polar structure} (VPOS) that includes distant globular clusters and stellar streams \cite{Pawlowski_2012}. Depending on the selection of the samples, the plane of VPOS has an r.m.s. thickness of $20$ to $25$ kpc, an axis ratio of $0.18$ to $0.30$ and an inclination of $73^\circ$ to $87^\circ$ with respect to the galactic disk \cite{Pawlowski:2018sys} (Figure~\ref{MW}). New samples of fainter satellites from the Sloan Digital Sky Survey \cite{York:2000gk} and the Dark Energy Survey \cite{Abbott:2005bi} also support this spatial correlation \cite{Pawlowski:2015vef, Pawlowski:2018sys}. In addition, kinematic measurements imply that at least $8$ out of $11$ well-measured satellites co-rotate in the plane of VPOS \cite{Pawlowski:2018sys, Pawlowski_2013, Pawlowski_2017}. The most recent data from Gaia also confirm both these spatial and kinematic correlations of the Milky Way satellites \cite{GaiaDR2:2018_mw, Pawlowski:2019bar}. 

\begin{figure}[htb!]
 \centering
 \includegraphics[width=0.9 \columnwidth]{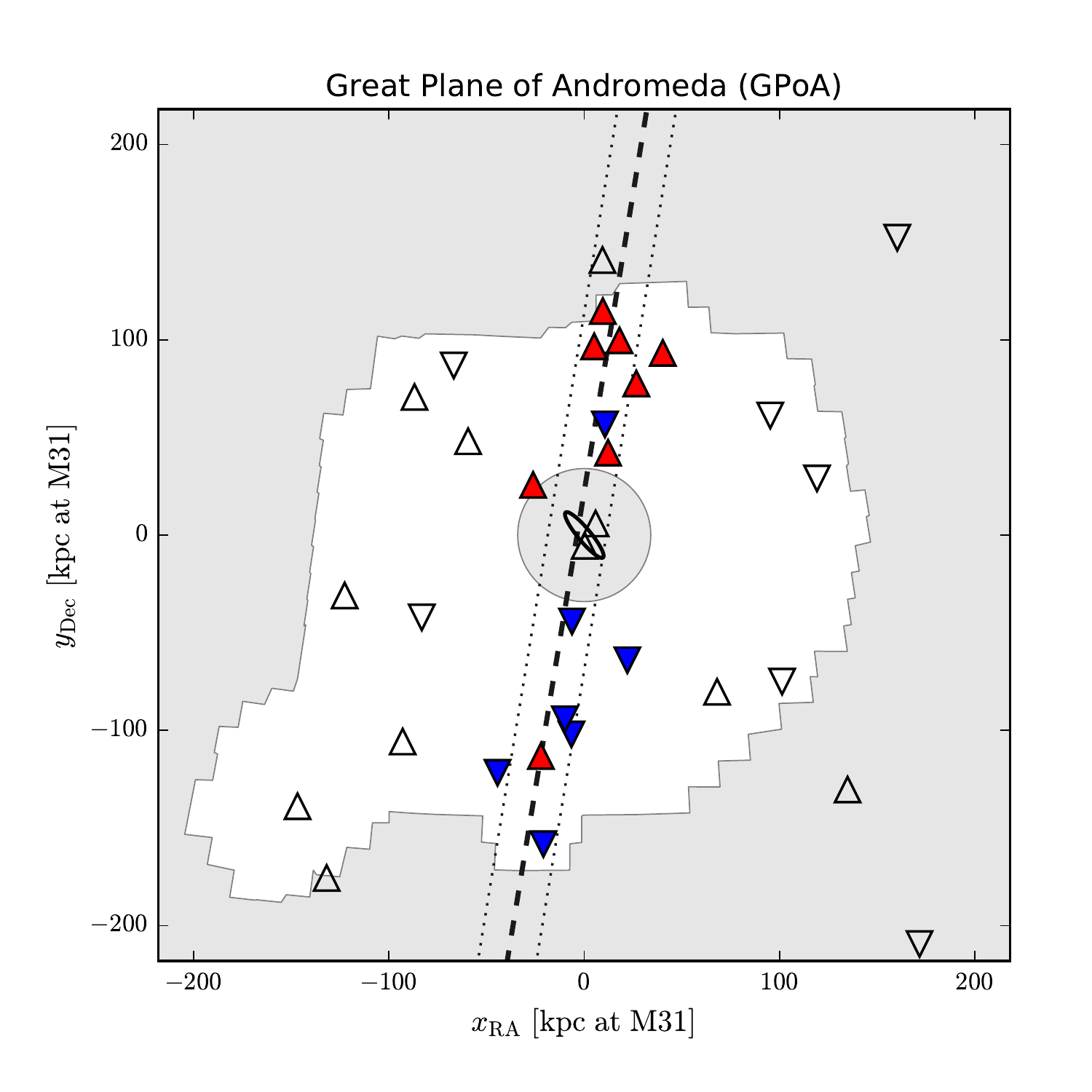}
 \caption{The \textit{giant plane of Andromeda} (GPoA) and the Andromeda (M31) disk (solid black ellipse at the center) as viewed from the Sun. The orientation and the width of the best fit satellite plane are indicated by the dashed and dotted lines, respectively. The colored triangles, red upward (blue downward) for receding from (approaching towards) an observer at rest with respect to the host galaxy, indicate coherent kinematics of the co-orbiting satellites. The satellites that are not part of the GPoA are plotted as open triangles whose orientations, upward for receding, indicate their line-of-sight velocities. The grey area corresponds to the region outside the Pan-Andromeda Archaeological Survey. The figure is reproduced from \cite{Pawlowski:2018sys}.}
 \label{M31}
\end{figure}

 Due to the lack of measures, the plane of satellites of M31 was not apparent until the Pan-Andromeda Archaeological Survey \cite{McConnachie:2009up} discovered new satellites, and provided distance measures. These data show that $15$ out of $27$ satellites lie on a plane, the \textit{giant plane of Andromeda} (GPoA), which is aligned with the Giant Stellar Stream in the M31 halo \cite{Ibata_2013, Conn_2013}. The GPoA has an r.m.s. thickness of $12.6$ kpc, an axis ratio of $0.1$ and an inclination of about $50^\circ$ with respect to the M31 disk \cite{Pawlowski:2018sys, Pawlowski:2013kpa}. The GPoA is viewed almost edge-on from our vantage point, and line-of-sight velocities of the satellites suggest a strong kinematic correlation (see Figure~\ref{M31}). Thirteen out of the fifteen in-plane satellites follow the coherent kinematic trend and suggest a co-rotating plane of satellites. Interestingly, most of the in-plane satellites are on the Milky Way side of the GPoA whereas off-plane satellites are randomly distributed \cite{Conn_2013}. The angle between the two satellite planes of the Milky Way and M31 is between $40^\circ$ to $50^\circ$ and they have similar spin directions. 

In addition to the satellite planes of the two major galaxies mentioned above, 14 out of the sample of 15 dwarf galaxies located within the Local Group but beyond the virial volumes of the Milky Way and M31, are part of two planes called the \textit{Local Group Plane 1 and 2} \cite{Pawlowski:2013kpa, Pawlowski:2014yxa}. The first and second planes contain 9 and 5 galaxies, respectively. Both planes are about $300$ kpc from the Milky Way and M31, and are strongly flattened, with r.m.s. thicknesses of about $60$ kpc and axis ratios of about $0.1$ . The probability of having these configurations purely by coincidence is less than 0.003 in the CDM model \cite{Pawlowski:2014yxa}.  

\begin{figure}[htb!]
 \centering
 \includegraphics[width=0.9 \columnwidth]{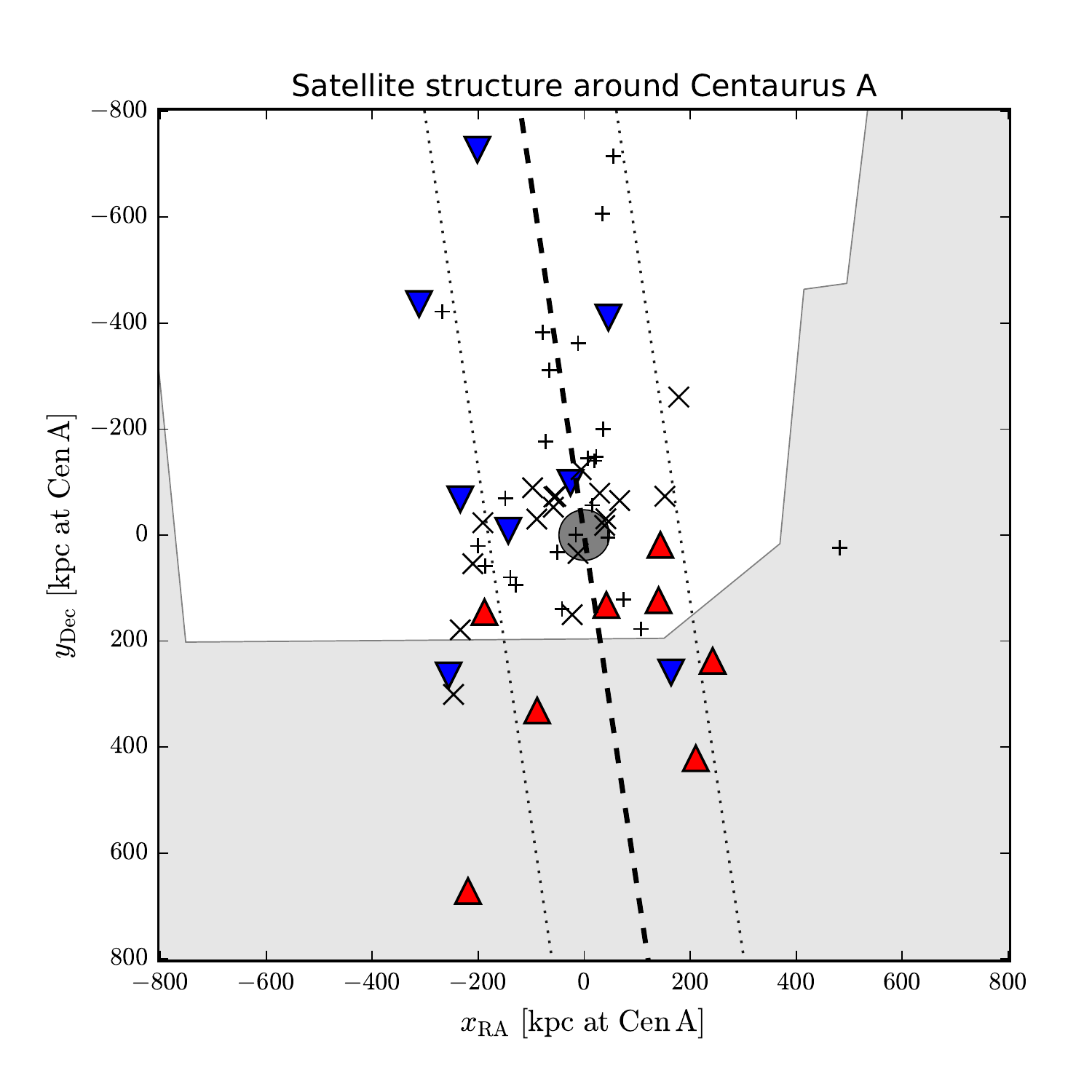}
 \caption{\label{CenA}The \textit{Centaurus A satellite plane} (CASP) and the Centaurus A disk (dark grey circle at the center) as viewed from the Sun. The orientation and the width of the best fit satellite plane are indicated by the dashed and dotted lines, respectively. The colored triangles, red upward (blue downward) for receding from (approaching towards) an observer at rest with respect to the host galaxy, indicate coherent kinematics of the co-orbiting satellites. The grey area show the volume outside the survey. The figure is reproduced from \cite{Pawlowski:2018sys}.}
\end{figure}

 Observing satellite galaxies outside the Local Group is  challenging. Nevertheless, several independent studies of the satellites around CenA strongly suggest preferential alignments of their orbits \cite{Tully:2015zfa, Crnojevi__2014, Crnojevi__2016, Muller:2016ddt, Muller:2018hks}. Earlier studies with smaller samples suggested the existence of two parallel planes of satellites with r.m.s. thicknesses of about $60$ kpc, axis ratios of about $0.2$ and a separation of $300$ kpc between the planes. A recent study \cite{Muller:2016ddt} with larger samples claims the existence of only one satellite plane, the \textit{Centaurus A satellite plane} (CASP), with similar orientation but larger thickness. The CASP is found to be almost perpendicular to the plane of the galactic disk, similar to the satellite plane of the Milky Way. From our position, the CASP is seen almost edge-on and $14$ out of $16$ satellites that have line-of-sight velocity measures, are found to be co-rotating in the plane \cite{Muller:2018hks} (Figure~\ref{CenA}). 

\subsubsection{Comparison with the CDM predictions}

The planar distribution of the orbits of the $11$ best-measured Milky Way satellites was initially thought to be inconsistent with the isotropic distribution predicted by the CDM model \cite{Kroupa:2004pt}. However, even in the CDM paradigm, the preferential accretion along the cosmic filaments  \cite{Libeskind:2005hs} and the triaxiality of the halo of the host galaxy \cite{Zentner:2005wh} might be responsible for the anisotropy of the spatial distribution of the satellites. In  high-resolution simulations \cite{Wang:2012bt}, only $5$-$10\%$ of the simulated samples resemble the spatial flatness of the observed satellite distribution. 

However, the combination of the spatial and kinematic coherence has probability $\sim 10^{-3}$ of appearing in CDM simulations \cite{Pawlowski:2014una}. In addition, finding two host galaxies (such as the Milky Way and M31) in the simulations, each with a co-orbiting satellite plane, is extremely rare with a probability $\sim 10^{-6}$ \cite{Pawlowski:2014una}. Increasing the resolution of the simulations does not seem to alleviate the issue \cite{Pawlowski:2018sys}. In the literature, there are disagreements about the severity of the tension between the plane of satellites and the CDM predictions which mostly stem from the different selection criteria of the satellite halos adopted in the simulations \cite{Pawlowski:2018sys}. 

\subsubsection{Possible solutions within the CDM model}

Although several solutions have been suggested for the PSP within the CDM model, none of them has been unanimously accepted, because each solution either fails to reproduce all the observational aspects or it is based on poorly-investigated assumptions. 

 The accretion along the filaments of the cosmic web results into anisotropic spatial distributions of the satellites \cite{Zentner:2005wh, Libeskind:2005hs, Lovell:2010ap, Libeskind:2010pd}, because the massive subhalos tend to be accreted along the spine of the filament. There is suggestive evidence that the satellite planes of M31 and CenA have the correct alignments as expected from this mechanism \cite{Libeskind:2015lwa}, but this same mechanism appears unable to generate the satellite plane of the Milky Way. Preferential alignments due to the cosmic filaments increase the degree of anisotropy in the subhalo distribution, but are not strong enough to reproduce the observed planar structure \cite{Libeskind:2014qqa, Pawlowski:2018sys}. Moreover, when both the spatial and kinematic correlations are required, only $0.6\%$ of the simulated samples are similar to the observed ones \cite{Pawlowski_2013, Pawlowski:2012ti}.  

 Certain amount of spatial and kinematic correlations among the satellites are expected if the satellite plane is the result of the infall of a group of satellites \cite{LyndenBell:1995zz, Metz:2009ys}. Tight alignments of the Magellanic stream with the VPOS of the Milky Way and the alignments of the Giant Stellar stream with the GPoA of M31 provide further support to this mechanism \cite{Lake:2008zt, Nichols:2011df}. However, there are several unsolved issues. Firstly, the observed number of luminous satellites may require more than one falling group \cite{Li:2007mf, Wetzel:2015xia, Shao:2017iye, Wang:2012bt}. Secondly, the falling group or groups must also be sufficiently compact and narrow to survive the tidal disruption from the host galaxy and result into the observed planar structure of the satellites \cite{Metz:2009ys, 2011MNRAS.416.1401A, 2016MNRAS.462.3221A}. Thirdly, support from the Magellanic or Giant Stellar stream is not quite robust because some satellites in the planes are located in regions which, according to the simulations, are found to be inaccessible by such streams even if the streams are allowed to complete more than one orbit \cite{Nichols:2011df}. 

 One of the most promising solutions is that the satellite planes consist of tidal dwarf galaxies (TDGs) that originated from the debris of interacting or colliding major galaxies \cite{2000A&A...358..819W, Bournaud:2006qz, 2007MNRAS.375..805W}. They are common in hydrodynamical simulations \cite{Ploeckinger_2017}, and can account for the strong phase-space correlation required to match the observations \cite{Kroupa:2004pt, Metz:2007cg, Pawlowski_2012}. However, the low metallicity of the observed satellites disfavors their formation from a recent merger since, in that case, they would be more metal-rich as their parent galaxies \cite{Recchi_2015}. This tension might disappear if the satellites formed  $\sim 10$ Gyrs ago from pre-enriched large-mass galaxies \cite{Pawlowski_2012}. However, this solution might not be viable: according to hydrodynamical CDM simulations, TDGs are expected to be baryon dominated whereas the observed satellites appear to be dark matter dominated if they are in virial equilibrium \cite{McConnachie12}. Clearly, if the satellites are out of equilibrium, their observed velocity dispersion profiles might not necessarily imply the presence of dark matter  \cite{Metz:2007cg, KROUPA1997139, Yang:2014yia}, but this hypothesis would contradict the assumption of their early formation.  

One of the solutions that go beyond the CDM framework is to assume new properties of the dark matter, e.g. self-interacting dissipative dark matter (see Section \ref{sec:SIDM}), which allows the existence of a dark disk in the host galaxy, so that dark matter-dominated TDGs may form during the merger events \cite{Foot:2013nea, Randall:2014kta}. However, the presence of a dark matter disk in the Milky Way is disfavored by Gaia data \cite{Schutz:2017tfp, Buch:2018qdr}. Alternatively, if we adopt MOND as the theory of gravity (see Section \ref{sec:MOND}), the formation of TDGs is enhanced due to the additional self-gravity  \cite{Tiret:2007fy, Renaud:2016blx}. However, it is not clear yet whether TDGs produced in MOND would show sufficient non-Newtonian behavior to account for the observed dark matter-like signatures \cite{Pawlowski:2018sys}.

\section{Possible solutions beyond the standard Cold Dark Matter }\label{sec:particlemodels}

Several astrophysical aspects of dark matter are related to its particle nature, like the minimum mass of a dark matter halo created by the hierarchical formation of structure 
\cite{Buckley2018}. The two parameters that regulate the structure formation are the characteristic free-streaming wavenumber $k_{\mathrm {fs}}$ and the characteristic interaction or decay rate $\Gamma$ of the dark matter particles \cite{Buckley2018}. The wavenumber $k_{\mathrm {fs}}$ is set by the mean velocity of the dark matter particles  at the time of decoupling, and depends upon the decoupling temperature $T_{\mathrm {kd}}$ and the mass of the dark matter particle; $\Gamma$ is set by the dark matter-dark matter scattering rate or the lifetime of the dark matter particles.  

\begin{figure}[htb!]
    \centering
    \includegraphics[width= 0.9\columnwidth]{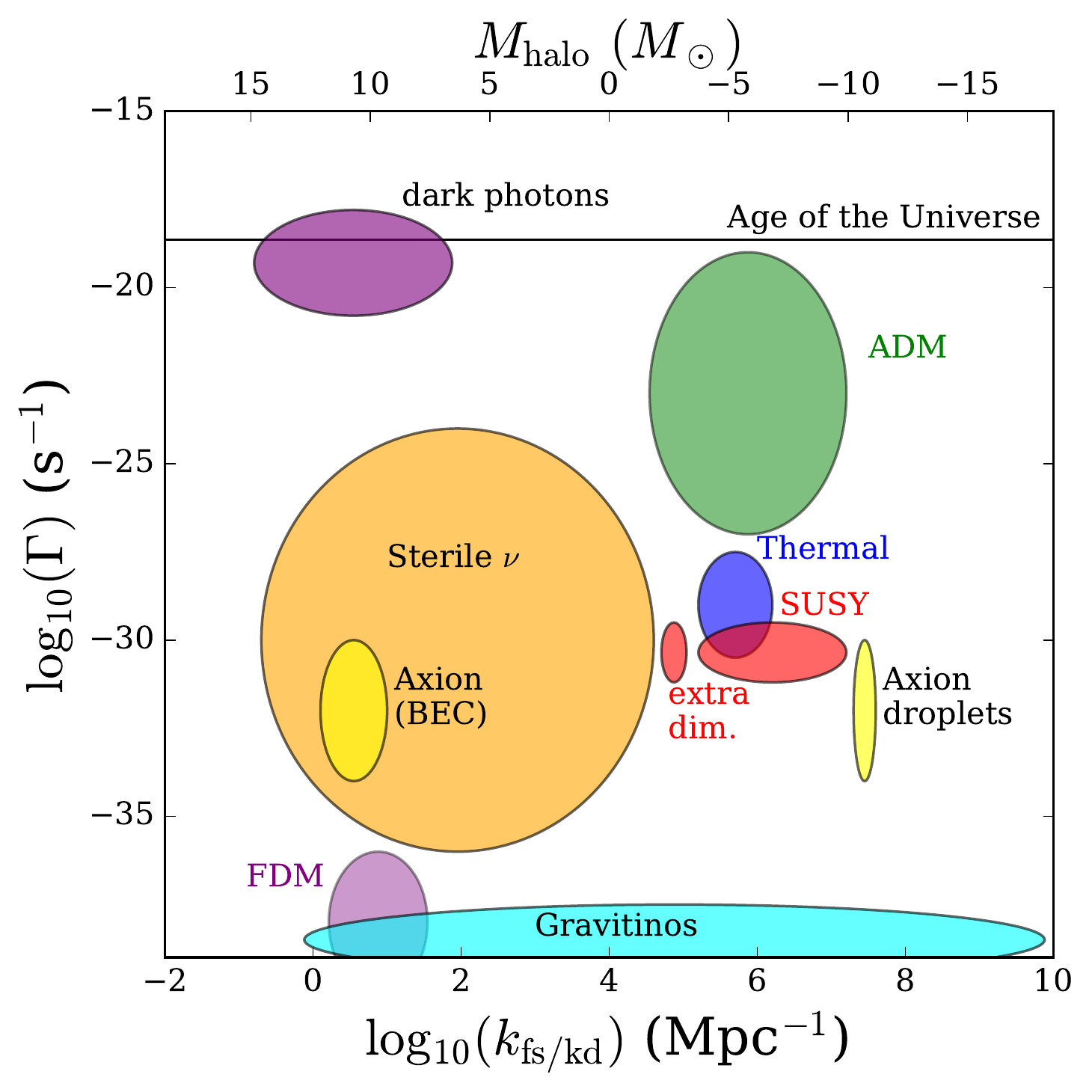}
    \caption{Allowed parameter space of the currently most popular dark matter candidates. The bottom axis shows the characteristic free-streaming wavenumber of the model, which is set in the primordial Universe. The top axis shows the corresponding halo mass, and the y-axis shows the characteristic interaction or decay rate which quantifies the evolutionary effects of dark matter. ADM, BEC, and FDM stand for  Asymmetric  Dark  Matter, Bose-Einstein Condensate, and Fuzzy Dark Matter, respectively. The figure is  from \cite{Buckley2018}.}
    \label{fig:dm_general}
\end{figure}

Figure \ref{fig:dm_general} shows the allowed regions in the $\Gamma-k_{\mathrm {fs}}$ parameter space for different dark matter candidates. { Deviations from the CDM paradigm arise when the gravitational collapse of dark matter is inhibited or modified above a characteristic comoving wavenumber. This wavenumber translates into a characteristic halo mass  below  which  the  number of halos  is  reduced. Alternatively, the  deviations can  be  driven  by  the  interactions  of the  dark  matter particles  with the Standard Model particles throughout the evolution of the Universe. Both effects can erase  existing  structures  or  change  the velocity  distributions of the dark  matter particles and the densities of their structures. For WIMPs, whose decoupling temperature varies from 15 MeV to 1500 MeV and $k_{fs}\sim 1 {\mathrm {pc}}^{-1}$ for $m_\chi$=100 GeV, one gets a minimum halo mass of the order of $\sim [10^{-8}-10^{-2}]$~M$_\odot$. Sterile neutrinos, which are a {warm dark matter} (WDM) candidate with mass ranging from $0.4$ to $10^5$ keV, have  $k_{\mathrm {fs}}\sim 0.5(m_\chi/{\mathrm {keV}})$~Mpc$^{-1}$ which translates into a minimum halo mass in the range $M_{\mathrm {halo}}\sim [10^{-6}-10^{11}]$~M$_\odot$. Similarly, gravitinos lead to  $M_{\mathrm {halo}}\sim [10^{-17}-10^{14}]$~M$_\odot$ while their mass is set by the supersymmetry breaking scale which is in the range 100 eV to 100 TeV.  Fuzzy  Dark  Matter  models  also erase structures below the de-Broglie wavelength of the particle, of the order of the kiloparsec, which leads to a minimum halo mass $M_{halo}\sim 10^{10}$~$_\odot$ for a particle mass $\sim10^{-22}$~eV.}

In the following sections, we will review the state of the art of some dark matter models, and discuss both successes and failures of each model in solving the challenges faced by the standard CDM model discussed in Section \ref{sec:obschallenges}.

\subsection{Warm Dark Matter model}\label{sec:WarmCDM} 
In contrast to CDM, WDM particles decouple when they are still relativistic; they thus erase primordial fluctuations on sub-galactic scales, and produce a cut-off in the primordial power spectrum \cite{2011PhRvD..84f3507S}. WDM particles can play the role of CDM in the cosmological evolution of the Universe \cite{Schneider_2014} and may also alleviate some of the problems of the CDM model at galactic scales as we explain below. 

One of the most powerful tools to investigate the suppression of the primordial power spectrum on small scales is the Lyman-$\alpha$ forest. A lower limit to the mass of the WDM particles were initially set to $m_\chi > $ 750 eV \cite{Narayanan_2000} by fitting   the Lyman-$\alpha$ forest in quasar spectra. More recent analyses of the Lyman-$\alpha$ forest and of the Milky Way satellites have increased the bound to  several keV  \cite{PhysRevLett.97.191303, PhysRevD.73.063513, PhysRevLett.100.041304,2014MNRAS.442.2487K}. 

\begin{figure}[htb!]
    \centering
    \includegraphics[width= 0.9\columnwidth]{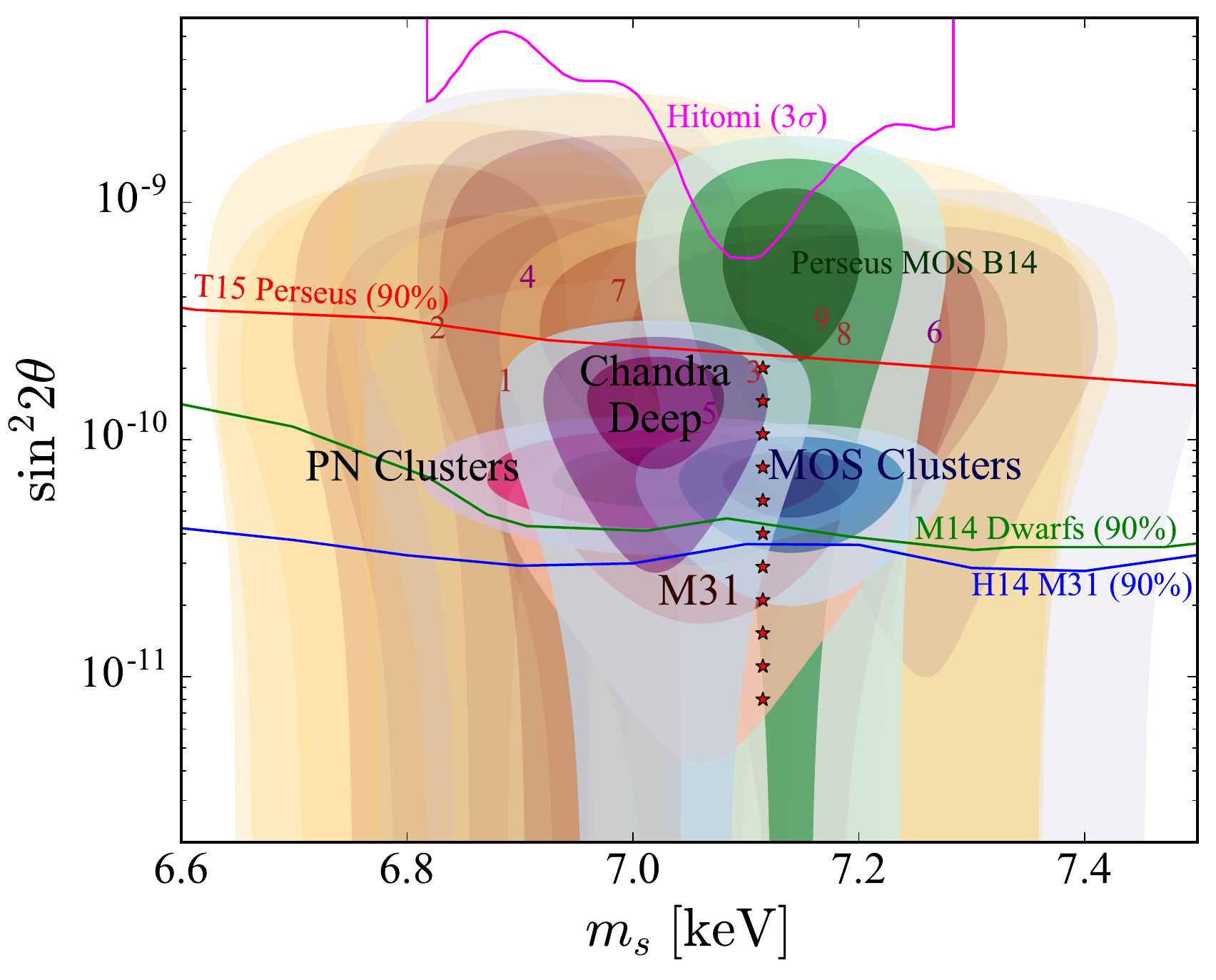}
    \caption{
    The allowed parameter space for sterile neutrinos obtained from the X-ray emission. The colored regions are 1, 2 and 3$\sigma$ confidence levels. Numbers from 1 to 9 mark the centre of the 2$\sigma$ region detections obtained using nine galaxy clusters \cite{2015arXiv150805186I}.  The lines show constraints at the 90\% confidence level from Chandra  \cite{2014PhRvD..89b5017H}, Suzaku \cite{Tamura_2015} and dwarf galaxies dataset \cite{Malyshev_2014}. The stars represent several models of sterile neutrino. The figure is reproduced from \cite{ABAZAJIAN20171}.}
    \label{fig:wdm3}
\end{figure}

One of the most stringent bounds on the mass of WDM particles comes from the high-resolution HIRES/MIKE spectrographs: $m_\chi>4.65$ keV  \cite{Y_che_2017}. The most promising candidate of WDM is sterile neutrino, with mass $m_s$, which is mixed with an ordinary neutrino \cite{1994PhRvL..72...17D,2002APh....16..339D}. For small mixing angles such as $\sin^2 2\theta \sim 10^{-7}$, the total amount of sterile neutrinos is only a small fraction of the ordinary neutrinos.  In Figure \ref{fig:wdm3}, we give a visual representation of the allowed parameter space for sterile neutrinos in the plane $\sin^2 2\theta - m_s$. 
{ The claimed detections are based on the so-called 3.55 keV line emission, which is attributed to the decay of dark matter particles, and are obtained by using Chandra X-ray observations of  galaxies in the Local Group \cite{2014PhRvD..89b5017H}, studies on dwarf galaxies \cite{Malyshev_2014}, and  Suzaku observations of the Perseus galaxy cluster \cite{Tamura_2015}. These investigations imply a sterile neutrino of mass $\sim 7 $ keV and $\sin^2(2\theta)= [2,20]\times10^{-11}$. In contrast to these results, the full-sky Fermi Gamma-ray Burst Monitor data \cite{Ng_2015} do not reveal any significant signal for sterile neutrino decay lines in the energy spectrum, and improve previous upper limits by an order of magnitude (for a comprehensive review on sterile neutrino we refer the reader to \cite{ABAZAJIAN20171})}.

{ Additional constraints have been obtained by using gravitationally lensed quasars: under the assumption of a thermal relic dark matter particle,  modelling the image positions and
the flux-ratios of several gravitationally lensed quasars implies a lower limit of $m_\chi > 5.58$ keV \cite{10.1093/mnras/stz3177} or $m_\chi>5.2$ keV \cite{10.1093/mnras/stz3177}, consistent with the lower limits $m_\chi > 5.3$ keV or $m_\chi > 3.5$ keV, depending on the assumed thermal history of the intergalactic medium, derived from the analysis of the Lyman-$\alpha$ forest \cite{Bolton2008,2017PhRvD..96b3522I,Garzilli2019}.}

\subsubsection{Solutions to the observational challenges}

\begin{figure}[htb!]
    \centering
    \includegraphics[width= 0.45\columnwidth]{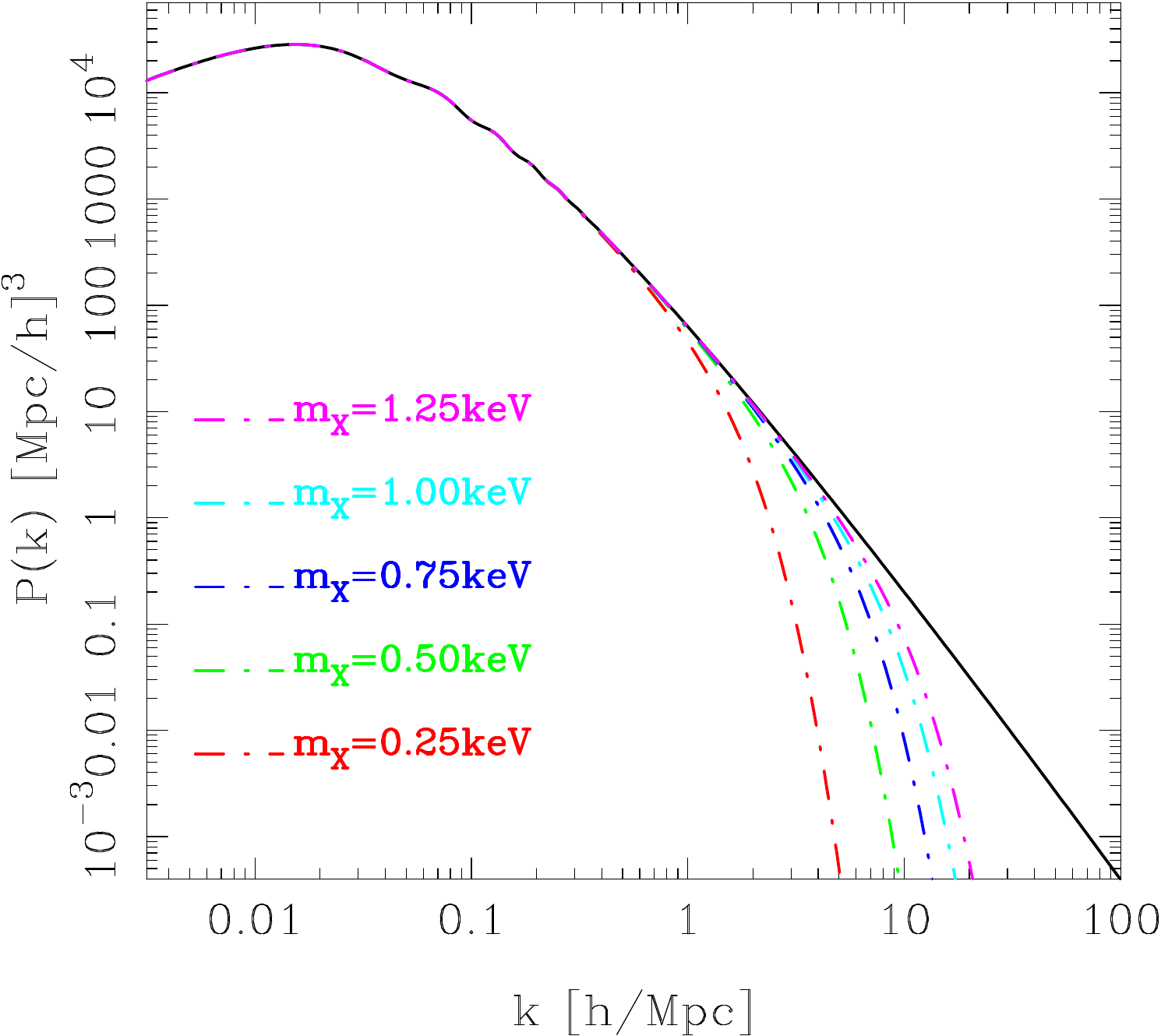}
    \includegraphics[width= 0.47\columnwidth]{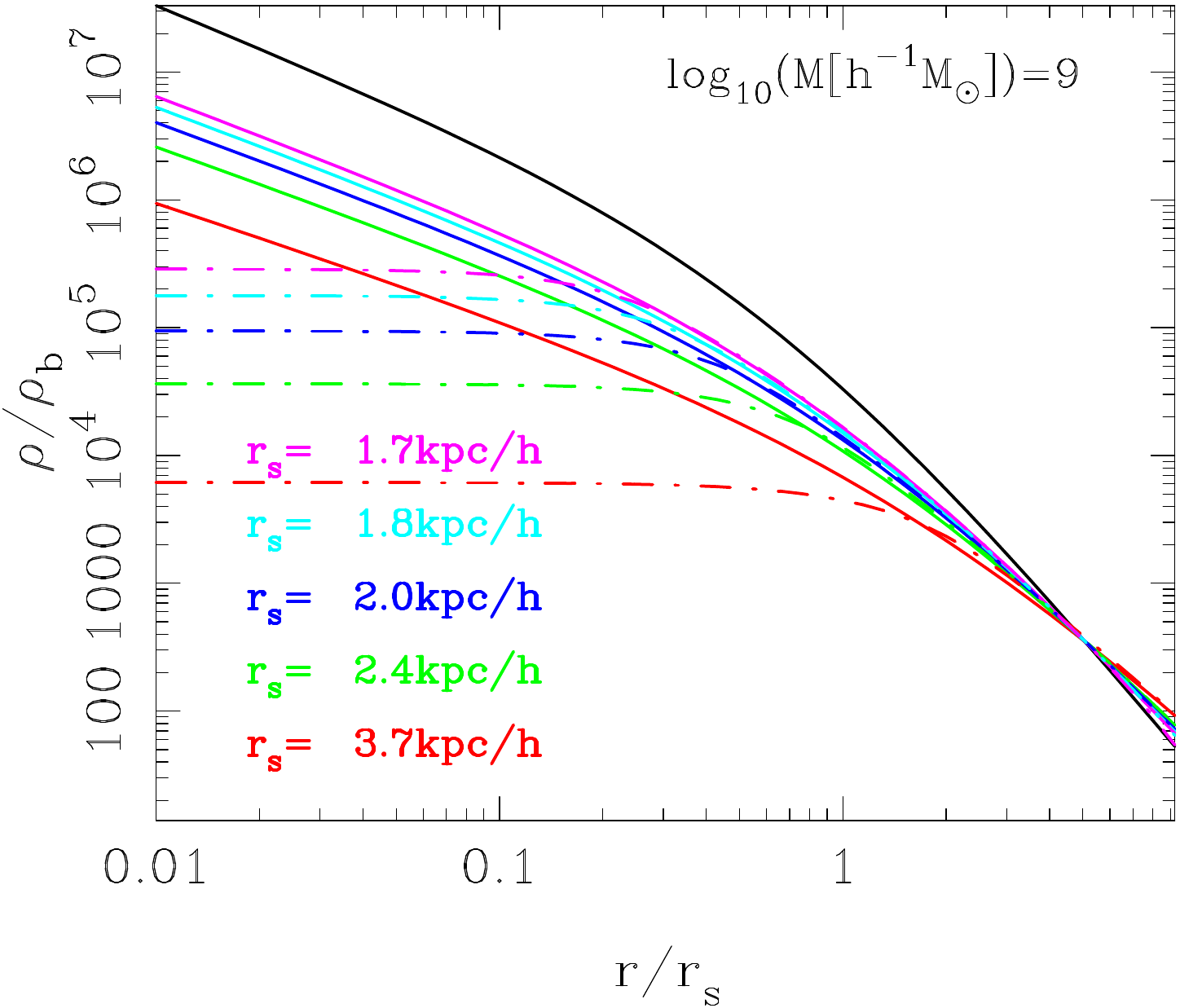}
    \caption{  {\em Left panel}: comparison of the linear matter power spectrum as a function of wavenumber
in the WDM and CDM scenarios. The solid black line shows the CDM model. The dot-dashed lines represent the WDM power spectra for different values of the particle mass.  {\em Right panel}: density profiles of CDM and WDM halos  with mass: $M = 10^9 \, {\rm M}_\odot$. The black line shows the CDM profile; the red, green, blue, cyan, and 
magenta lines show the WDM profiles for particle mass $0.25,\,0.5,\,0.75,\,1.0$, and $1.25$~keV, respectively. The
solid colored lines show the case where there are no thermal relic
velocities; the dot-dash lines denote the case where it is assumed the existence of a relic thermal velocity distribution for
the WDM particles. In the panel, it is indicated the scale radius of the halo for the various WDM particle mass considered. The figure is reproduced from \cite{2011PhRvD..84f3507S}.}
    \label{fig:wdm2}
\end{figure}

 WDM particles are moving freely when they are relativistic; they can thus travel distances of the order of the horizon size.
 It follows that density fluctuations are suppressed on scales below the inverse of a
characteristic comoving wavenumber \cite{2001ApJ...556...93B}:
\begin{equation}
k_{\rm WDM}\sim 15.6 \frac{h}{\mathrm {Mpc}} \left(\frac{m_X}{ \mathrm{keV}}\right)^{4/3}\,\left(\frac{0.12}{\Omega_{\mathrm DM} h^2}\right)^{1/3}\, ,
\end{equation}
with obvious meaning of the symbols.
This suppression of the density perturbations may lead to a solution of  some of the problems that the CDM model encounters at galactic scales (see Section \ref{sec:obschallenges}).
As shown in the left panel of Figure \ref{fig:wdm2}, the matter power spectrum drops below a certain length scale $k^{-1}$ depending upon the mass of the WDM particles. For example, the power spectrum is suppressed below  $k^{-1}\sim 100$ kpc for a particle mass of $\sim$1~keV \cite{Viel_2005}. Therefore, the subhalo mass function can be brought into agreement with satellite counts and the MSP would be solved \cite{2011PhRvD..84f3507S,Zavala_2009,2011ApJ...739...38P}.

In addition, the gravitational collapse leads to a cuspy halo profile with a lower central concentration compared to  CDM halos \cite{2011PhRvD..84f3507S}. This feature is shown in  the right panel of Figure \ref{fig:wdm2}, where the NFW profile (black solid line) is compared with the cuspy WDM density distribution (solid colored lines) for a halo having a mass of $M = 10^9 \, {\rm M}_\odot$.
Moreover, the existence of a relic thermal velocity distribution for the WDM particles may convert the cusp in the density profile into a core (dot-dashed colored lines), providing a solution to the CCP \citep{PhysRevLett.42.407}. Nevertheless, the cores appear to be smaller than required to explain the data on LSB galaxies \cite{2011JCAP...03..024V,Maccio2012}.
These results have been widely validated by many N-body simulations of structure formation within the WDM scenario (see, for example, \cite{2001ApJ...556...93B,PhysRevD.83.043506,Lovell2012,Maccio2012,2012JCAP...10..047A,Schneider2012,Angulo2013,2013ApJ...767...22K}). 
{  Finally, a WDM particle with mass in the range 1.5--2 keV can potentially solve the TBTF problem for satellite dwarf galaxies, as Milky Way-sized dark matter halos have fewer and less dense massive subhalos in WDM than in CDM \cite{Lovell2012,Schneider_2014}. Furthermore, the  TBTF problem  for field dwarf galaxies can also be solved by a WDM particle with mass of $\sim 1$ keV \citep{Papastergis_2015}. However, WDM particles with such small masses are in conflict with a number of observational constraints (see Section \ref{sec:WarmCDM}). 
Masses of the WDM particles that are consistent with the largest estimate of $\sim 7$ keV enable WDM simulations to alleviate the TBTF problem only when baryonic feedback and tidal process are accounted for, similarly to CDM simulations; the results of these simulations show a higher degree of consistency with the observational data than the results produced in CDM simulations
\citep{Lovell_2017,Kang_2020}.
}

Unfortunately, some difficulties are still far from being solved. Due to the high efficiency in removing power from the matter power spectrum, the mass of the WDM particles must be tuned to the scale of dwarf galaxies \cite{PhysRevD.83.043506}.  Moreover, studies based on N-body simulations have shown that WDM is almost indistinguishable from CDM when the bounds from the Lyman-$\alpha$ forest are taken into account and, therefore, it might not be capable of alleviating the CDM problems \cite{Schneider_2014}. 

\subsection{Self-interacting dark matter}\label{sec:SIDM}

A model of warm and self-interacting dark matter was firstly introduced in 1992 \cite{Carlson:1992fn}, and subsequently constrained few years later \cite{1994ApJ...431...41M,deLaix:1995vi}. 
In 1999, Spergel and Steinhardt \cite{Spergel:1999mh} proposed the idea of cold and self-interacting dark matter in the concordance cosmology to solve two small scale issues of CDM: the core-cusp and the missing satellites problems (see Sections \ref{sec:CCP} and \ref{sec:MSP}). The newly proposed self-interacting dark matter (SIDM) particles behave like collisionless CDM at larger length scales, where the rate of collisions becomes negligible due to the smaller density. The typical collision rate of SIDM particles in the central region of a dwarf galaxy is \footnote{{ In  \cite{Tulin:2017ara}, the numerical value was mistakenly reported as 0.1 Gyr$^{-1}$.}} \cite{Tulin:2017ara}
\begin{equation}
R_{\text{coll}} = \rho_{\text{DM}} v_{\text{rel}} \left( \frac{\sigma}{m} \right) \approx 1.0 \; \text{Gyr}^{-1} \left( \frac{\sigma / m}{1 \; \text{cm}^2 / \text{g}} \right) \left( \frac{\rho_{\text{DM}}}{0.1 \; M_\odot /\text{pc}^3} \right) \left( \frac{v_{\text{rel}}}{50 \; \text{km/s}} \right)\, , \label{SIDM_coll_rate}
\end{equation}
where $m$ and $\rho_{\text{DM}}$ are the mass and mass density of the dark matter particles, $v_{\text{rel}}$ and $\sigma$ are the relative velocity and the scattering cross section, respectively . SIDM models are commonly parametrized by the cross-section per unit mass ${\sigma}/{m}$, which, in general, is a function  either of the relative velocity $v_{\text{rel}}$ of the dark matter particles, or of the total mass of the virialized halo $M_{\text{halo}}$. In addition, $v_{\text{rel}}$ and $M_{\text{halo}}$ are related by the fact that the velocity dispersion of the dark matter particles is larger in more massive halos. Here, we will briefly discuss the small-scale challenges addressed by SIDM and where SIDM stands in particle physics. For a detailed review, the reader may consult Tulin \& Yu  \cite{Tulin:2017ara}. 

\subsubsection{Solving small-scale issues with SIDM}

\begin{figure}[htb!]
 \centering
 \includegraphics[width= 0.9\columnwidth]{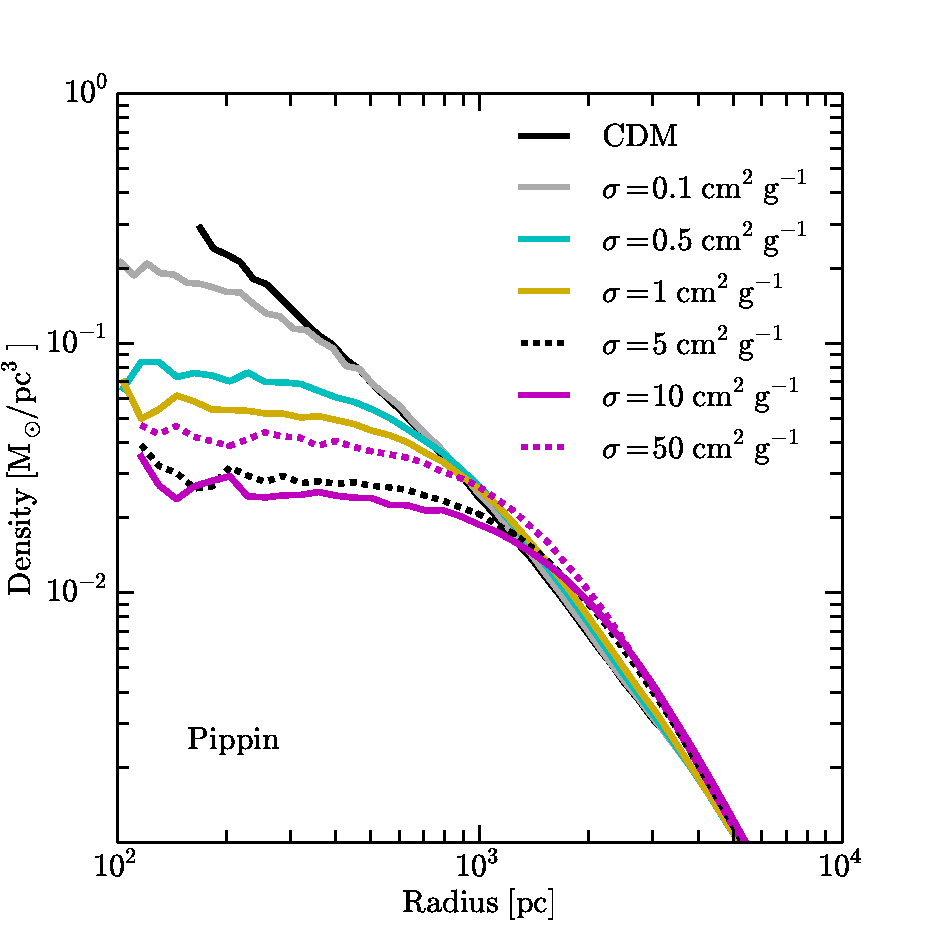}
 \caption{\label{halo_den} Inner density profile of a dark matter halo with $M_{\text{halo}} = 0.9\times 10^{10} \, {\rm M}_\odot$ {  in collisionless CDM and in SIDM} for different values of $\sigma / m$. The figure is reproduced from \cite{Elbert:2014bma}.}
\end{figure}

Cosmological simulations of SIDM halos without the baryonic feedback predict constant density cores \cite{Dave:2000ar, Yoshida:2000uw, Rocha:2012jg, Elbert:2014bma} (see Figure~\ref{halo_den}) contrary to the cuspy profiles from the simulations of collisionless CDM, as discussed in Section \ref{sec:CCP}. The SIDM simulations solve the cusp-core problem if ${\sigma}/{m} \gtrsim 0.5 \; \text{cm}^2/\text{g}$ for galactic halos with $M_{\text{halo}} \sim 10^{11} \, {\rm M}_\odot$ \cite{Firmani:2000ce, Elbert:2014bma}. However,  galaxy clusters with $M_{\text{halo}} \sim 10^{14} \, {\rm M}_\odot$ require ${\sigma}/{m} \sim 0.1 \; \text{cm}^2/\text{g}$ \cite{Meneghetti:2000gm, Firmani:2000ce, Firmani:2000qe, Tulin:2017ara}. The difference in ${\sigma}/{m}$ at galaxy and cluster scales implies velocity-dependent cross-sections which is a crucial aspect of the SIDM models.  

Although SIDM with large scattering rates tends to reduce the number of subhalos, the MSP (see Section \ref{sec:MSP}) is \textit{not} easily resolved by SIDM \cite{Rocha:2012jg, Colin:2002nk, Vogelsberger:2012ku, Zavala_2013}, unless non-minimal SIDM interactions are assumed \cite{Tulin:2017ara}. Halos found in SIDM-only simulations are more spherical in the inner regions compared to their CDM counterparts; moreover, observed ellipticities of dark matter halos inferred from gravitational lensing have put several stringent constraints on the SIDM parameters at both galaxy and cluster scales \cite{Peter:2012jh}. However, there are disagreements among various constraints \cite{Peter:2012jh, Feng:2009hw} which may be resolved by properly accounting for the effects from baryons \cite{Tulin:2017ara}. From SIDM-only simulations, the values of ${\sigma}/{m}$ required to solve the MSP are most likely ruled out by constraints from the measured ellipticities of dark matter halos (see \cite{Tulin:2017ara} and references therein). However, as we discuss below, the baryonic contributions may significantly change this scenario on galaxy scales. On the scale of galaxy clusters, values of ${\sigma}/{m}$ in the range $0.3$ to $10^4 \; \text{cm}^2/\text{g}$ are excluded based on the fact that, unlike the number of satellite galaxies, the number of galaxies in a cluster is comparable to the number found in the simulations of the formation of the large-scale structure  \cite{Gnedin:2000ea, Rocha:2012jg}. 

Simulations of SIDM and baryons \cite{Vogelsberger:2014pda, Fry:2015rta}, with sufficient resolution at small scales, are computationally expensive. To overcome this drawback, a semi-analytical approach, known as the Jeans method, has been pursued to study the baryonic contributions in the SIDM paradigm \cite{Rocha:2012jg, Kaplinghat:2013xca, Kaplinghat:2015aga}. The Jeans method determines the inner density profile of the halo by assuming the inner halo to be isothermal and in hydrostatic equilibrium with the baryons. However, the outer part of the halo matches the NFW profile since SIDM particles are effectively collisionless there. For a galaxy like the Milky Way, the results using the Jeans method suggest that the inner halo shape is oblate rather than spherical as found in SIDM-only simulations, and the radius of the central core is smaller by an order of magnitude compared to the core size in SIDM-only simulations \cite{Kaplinghat:2013xca}. These modifications upon including the effects of baryons demand corrections of the constraints on SIDM models based on the observed halo shapes. 

\begin{figure}
 \centering
 \includegraphics[width= 0.9\columnwidth]{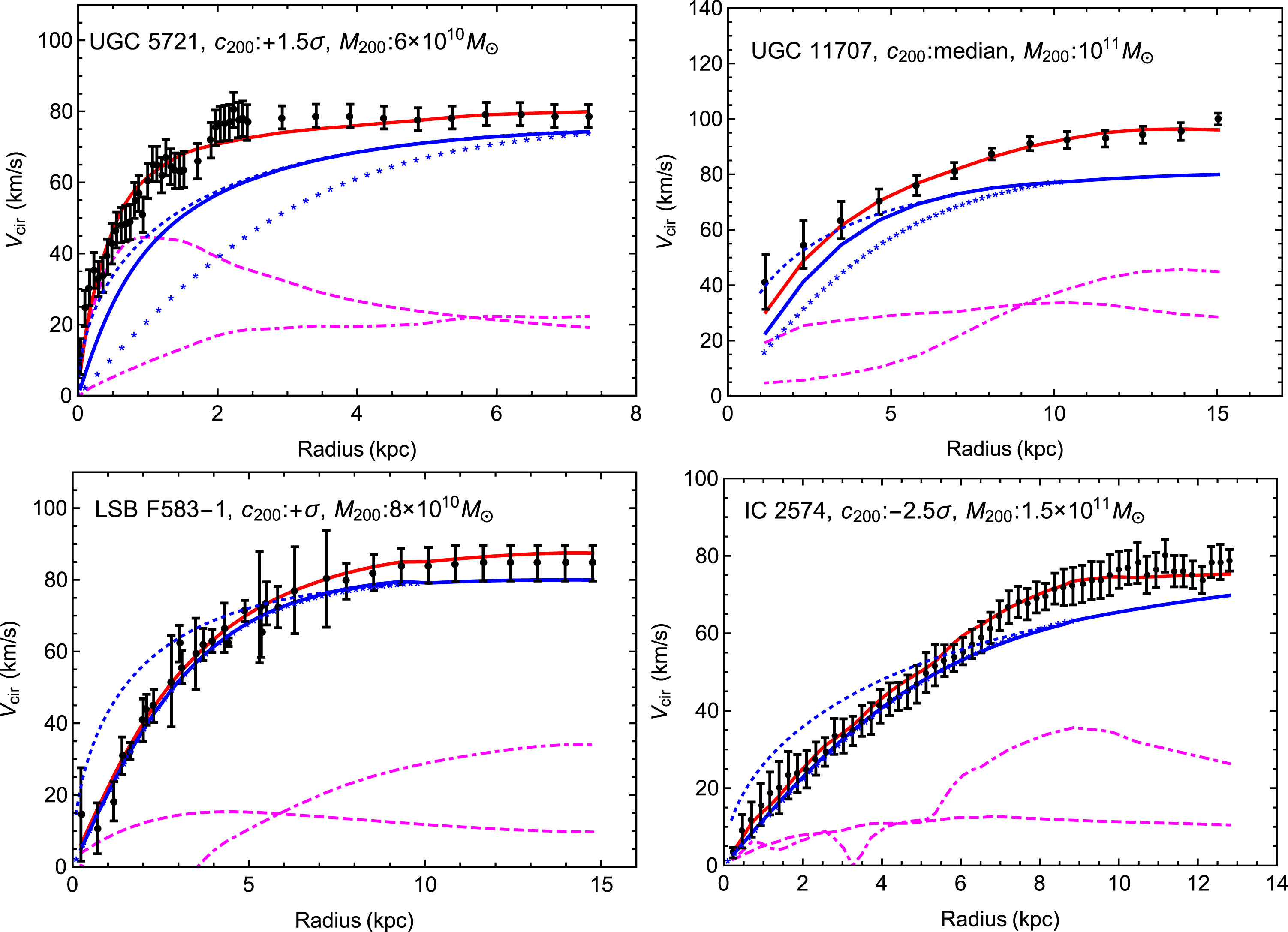}
 \caption{\label{diversity} Observed rotation curves of four disk galaxies. Solid red lines indicate the total rotation curves for SIDM with ${\sigma}/{m} = 3 \; \text{cm}^2/\text{g}$ which include contributions from the dark matter halo (solid blue), stars (magenta dashed), and gas (magenta dot-dashed). The corresponding CDM halos (dashed blue) and SIDM halos (blue stars) neglecting the baryons are also shown. The concentration parameters $c_{200}$ are indicated in terms of the standard deviation from the median cosmological concentration of halos of mass $M_{200}$. The figure is reproduced from \cite{Tulin:2017ara, Kamada:2016euw}.}
\end{figure}

In addition, if baryonic effects are taken into account by the Jeans method, SIDM  may solve the issues related to the disk-halo conspiracy \cite{Oman:2015xda} (see Section~\ref{sec:RAR}), i.e. it can reproduce diverse inner rotation curves from a single value of ${\sigma}/{m}$ for galaxies with similar maximum rotation speed $V_{\text{max}}$ \cite{Kamada:2016euw, Ren:2018jpt} (see Figure~\ref{diversity}). Collisionless CDM, even with baryonic feedback, does not solve this issue because of the large dark matter density in the central region. Thanks to the smaller dark matter density provided by the SIDM scenario, the baryons are more effective at setting $V_{\text{max}}$. 

{  The self-interactions between the dark matter particles also alleviates the TBTF problem (see Section \ref{sec:TBTF}) for satellite and field dwarf galaxies. Even though self-interaction has little effect on the abundance or total mass of the subhalos, it
effectively decreases the central density of the most massive subhalos by removing mass from these regions, characterized by cuspy density profiles. 
Velocity-independent SIDM models are consistent with the kinematics of the Milky Way dSphs  
for values of the cross-section per unit mass ${\sigma}/{m} \approx 1$ cm$^{2}$/g \citep{Zavala_2013, Dooley:2016ajo}; lower values of ${\sigma}/{m}$ generate a population of subhalos too dense to be consistent with the observations. On the other hand, velocity-dependent SIDM models successfully solve the TBTF \citep{Loeb_Weiner_2011, Vogelsberger:2012ku,Dooley:2016ajo}. However, the class of velocity-dependent SIDM models remains largely unconstrained.}

In principle, SIDM with dissipative or inelastic scattering may also alleviate the problem of the planes of satellite galaxies (see Section \ref{sec:planeofsatellite}) by allowing the observed satellites to be dark matter dominated tidal dwarfs \cite{Foot:2013nea, Randall:2014kta}. In CDM simulations, dwarf galaxies formed due to tidal disruption during the merging of two galaxies are baryon dominated. On the contrary, a dissipative dark matter scenario allows a thin dark matter disk in a halo, and merging galaxies with such dark disks may produce dark matter dominated dwarf galaxies with strong phase-space correlations.
However, Gaia data do not currently support  the presence of a dark matter disk in the Milky Way  \cite{Schutz:2017tfp, Buch:2018qdr}.

The collisional nature of SIDM at high relative velocities ($\sim 1000$ km/s) can be probed by merging clusters, such as the Bullet Cluster, in different ways \cite{Markevitch:2001ri}. Clusters, that nearly survive the merger, must have scattering depth $\tau = {\sigma \Sigma}/{m} < 1$, where $\Sigma$ is the surface density of dark matter \cite{Markevitch:2003at}. Additional constraints come from the amount of dark matter that is stripped off during the merger: we can infer this mass by estimating the mass-to-light ratios of the two merging clusters \cite{Markevitch:2003at}. Another probe of the self-interaction is to look for any possible offset between the centroid of the luminous part of the merging cluster and that of its dark matter content inferred from weak-lensing \cite{Markevitch:2003at}: self-interactions between the dark matter contents of two colliding clusters would generate a drag force that may exceed the gravitational binding between the dark matter and the luminous components. Combining the constraints from all the three methods for the Bullet Cluster shows that ${\sigma}/{m}$ is less than $0.7 \; \text{cm}^2/\text{g}$ \cite{Markevitch:2003at}. For a list of such constraints for other mergers, see  \cite{Tulin:2017ara} and references therein.  

\subsubsection{SIDM in particle physics}

SIDM may solve several small-scale issues of the CDM paradigm, provided ${\sigma}/{m} \sim 1 \; \text{cm}^2/\text{g}$ on the scale of galaxies and $\sim 0.1 \; \text{cm}^2/\text{g}$ on the scale of galaxy clusters. We now briefly describe the role of SIDM in the context of particle physics. 

The simplest model of SIDM could be a real scalar field $\phi$ with quartic self-interactions $({\lambda}/{4!}) \phi^4$. There are several ways to build this kind of models \cite{Bento:2000ah, McDonald:2001vt, Burgess:2000yq}. However, the contact-type self-interactions have a major drawback, namely the cross section is independent of the particle velocity and this feature generates
dark matter cores both in galaxies and in galaxy clusters. Unfortunately, real galaxy clusters do not show any presence of a core. 

\begin{figure}[htb!]
 \centering
 \includegraphics[width= 0.9\columnwidth]{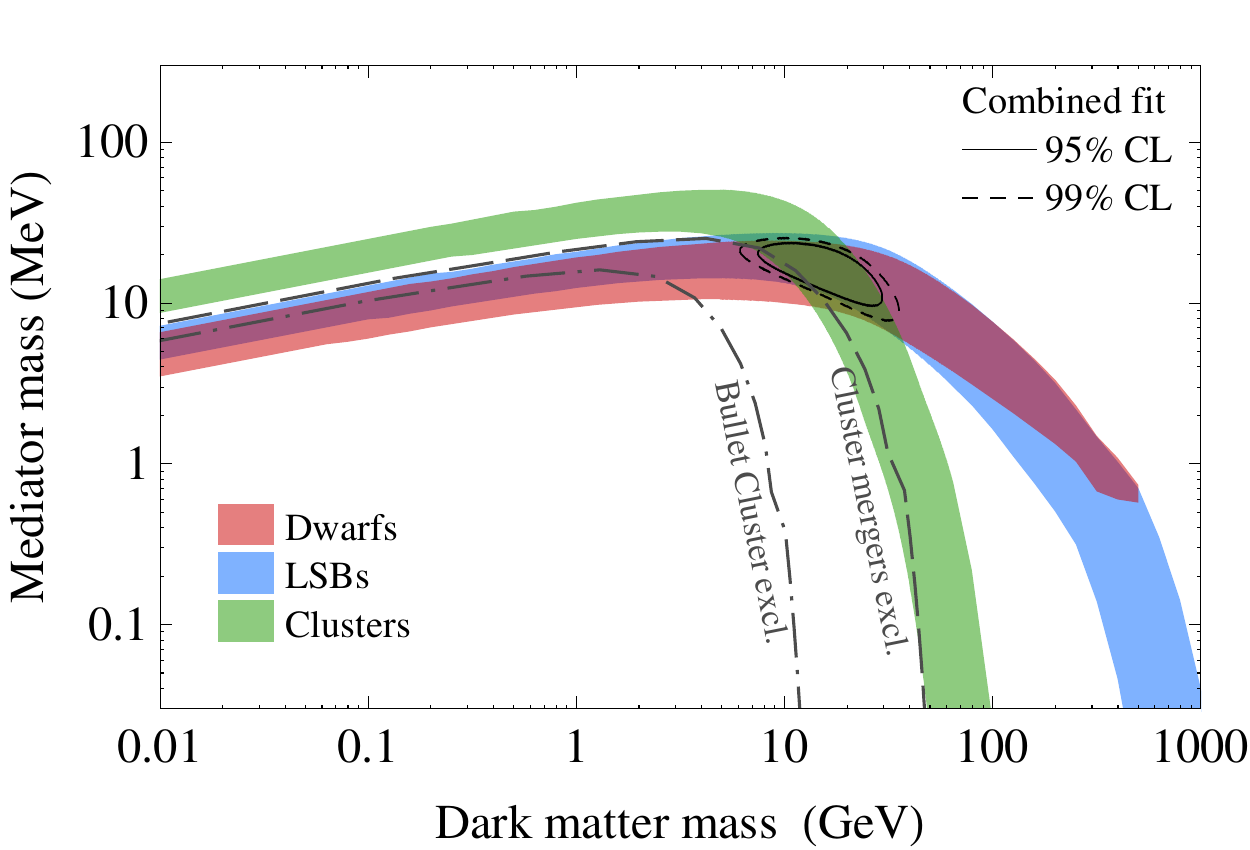}
 \caption{\label{darkphotonmodel} Parameter space spanning the  mass $m_\chi$ of the SIDM particle and the vector mediator mass $m_\phi$ with self-coupling $\alpha_\chi = \alpha = {1}/{137}$. Regions preferred by dwarfs (red), LSB spiral galaxies (blue) and galaxy clusters (green), each at $95\%$ confidence level, are indicated. Combined $95\%$ ($99\%$) region is shown by the solid (dashed) contour. The region estimated to be excluded by the Bullet Cluster (other observed merging clusters) lies below the dot-dashed (long-dashed) curve. The figure is reproduced from \cite{Kaplinghat:2015aga}.}
\end{figure}

The issue of the velocity independence can be tackled by considering a model where the dark matter particles ($\chi , \bar{\chi}$) are fermions and are self-interacting via a light mediator $\phi$, which can be a scalar or a vector field \cite{Tulin:2013teo, Buckley:2009in, Tulin:2012wi}. The self-coupling strength in the dark sector is characterized by $\alpha_{\chi}$, which is similar to the fine structure constant in electrodynamics. In the limit of $\alpha_\chi m_\chi \ll m_\phi$, the differential cross-section for elastic self-scattering is given by \cite{Tulin:2012wi, Tulin:2013teo}
\begin{equation}
\frac{d\sigma}{d\Omega} = \frac{\alpha_{\chi} ^2 m_{\chi} ^2}{\left[ \frac{1}{2} (1-\cos \theta) m_{\chi} ^2 v_{\text{rel}} ^2 + m_{\phi} ^2 \right]^2} \label{diff_mediator}\, ,
\end{equation}
where $m_{\chi}$ and $m_{\phi}$ are the mass of the dark matter particle and the dark mediator, and $\theta$ is the scattering angle. In the limit of $m_\chi v_{\text{rel}} \ll m_{\phi}$, the cross-section is constant, whereas, for $m_\chi v_{\text{rel}} \gg m_{\phi}$, the cross-section varies as $1/v_{\text{rel}} ^4$. The model is consistent with both galaxy and cluster scales, if the transition between the two regimes occur around $v_{\text{tran}} \sim 300$ km/s or equivalently, $m_\phi / m_\chi \sim v_{\text{tran}} / c \sim 10^{-3}$ \cite{Kaplinghat:2015aga}. More accurate cross-sections have been calculated to study the full parameter space of this model \cite{Feng:2009hw, Buckley:2009in, Tulin:2012wi, Tulin:2013teo, Braaten:2013tza}. The dependence of the cross section on the velocity varies with the limits of the parameters \cite{Tulin:2012wi}. Figure~\ref{darkphotonmodel} shows the preferred regions of the $(m_\chi , m_\phi)$ space for a fixed coupling $\alpha_\chi = 1/137$ \cite{Kaplinghat:2015aga}. 

Other models of SIDM include composite states of strongly interacting dark sector particles ($m \sim 1$ TeV), e.g. dark hadrons in a QCD-like dark sector, and bound states of different dark sector particles, e.g. dark atoms and dark matter with excited states (see ~\cite{Tulin:2017ara} and references therein). The SIDM paradigm is rich in phenomenology and each of the particle candidates of SIDM can be probed by both terrestrial searches for dark matter as well as astrophysical observations. 

\subsection{QCD axions} \label{sec:QCDaxion}

QCD axions have the double virtue of solving the Strong CP problem in the Standard Model (SM) of particle physics \cite{Peccei:1977hh, Peccei:1977ur, Weinberg:1977ma, Wilczek:1977pj} and of being a potential candidate for dark matter \cite{1983PhLB..120..133A, 1983PhLB..120..127P, Dine:1982ah}. Starting with a brief description of their origin in particle physics, we will discuss the cosmology of axions, the constraints on their mass and when they behave differently from WIMPs.  

\subsubsection{Emergence of QCD axions}

Quantum chromodynamics (QCD) describes the \textit{strong} interactions between quarks and gluons. The CP violating parameter $\bar{\theta}$ in QCD results into a non-zero electric dipole moment of the neutron \cite{Crewther:1979pi} which has an experimental upper bound \cite{Baker:2006ts}: 
\begin{equation}
d_n \approx 5 \times 10^{-16} \; \bar{\theta} \; \text{e-cm} \;\; < 2.9 \times 10^{-26} \; \text{e-cm} \;\; , \label{dnbound}
\end{equation} 
i.e. the parameter $\bar{\theta}$ is constrained to be unusually small, $\bar{\theta} \lesssim 10^{-10}$. Given that the CP violating term is required to solve the $U_A (1)$ \textit{problem} \cite{Weinberg:1975ui} and that the CP violation occurs in the \textit{weak} sector of the SM \cite{PhysRevD.98.030001}, $\bar{\theta}$ is expected to have a value of $\mathcal{O} (1)$ between $0$ and $2\pi$. Why the value of $\bar{\theta}$ is so small, i.e. why CP is so weakly violated in the \textit{strong} sector of the SM, is known as the \textit{Strong CP problem}. Several solutions to the Strong CP problem have been proposed (see \cite{Peccei:2006as, Kim:2008hd, Hook:2018dlk} and references therein). Among them, the most attractive and popular solution is the Peccei-Quinn (PQ) solution \cite{Peccei:1977hh, Peccei:1977ur} which implies the existence of a new pseudo-scalar boson \textit{axion} \cite{Weinberg:1977ma, Wilczek:1977pj}. In the PQ solution, the CP-violating parameter is promoted to be a dynamical variable ${a(x)}/{f_a}$ where $a(x)$ is the axion field and $f_a$ is the \textit{axion decay constant} with a dimension of energy. As the axion field $a(x)$ relaxes to its vacuum expectation value $\langle a \rangle$ to minimize the energy, the value of $\bar{\theta}$ becomes zero and that solves the Strong CP problem. The constant $f_a$ is given by the energy scale at which the PQ symmetry is broken. For QCD axion, the mass is related to the decay constant by \cite{Sikivie:2006ni}:
\begin{equation}
m_a \approx 6 \times 10^{-6} \; \text{eV} \; \left(\frac{10^{12} \text{GeV}}{f_a} \right) \; . \label{axion_mass}
\end{equation}
However, for axion-like particles (ALPs), both $m_a$ and $f_a$ are considered to be independent parameters. ALPs not obeying Eq.~(\ref{axion_mass}) do not solve the Strong CP problem. The existence of ALPs or ultra-light ALPs (ULALPs) is motivated from string theory \cite{2010PhRvD..81l3530A}. In the literature, the term \textit{axions} may refer to QCD axions or ALPs or ULALPs depending upon the context. 

After QCD axions were proposed to solve the Strong CP problem, they were quickly recognized to be a dark matter candidate \cite{1983PhLB..120..133A, 1983PhLB..120..127P, Dine:1982ah}, i.e. it was shown that they could be abundantly produced in the early Universe and their energy density behaves as that of CDM. Couplings of QCD axions with the SM particles can also be made very small since $f_a$ can be allowed to be much larger than the QCD energy scale \cite{Kim:1979if, Shifman:1979if, Dine:1981rt, Zhitnitsky:1980tq}. Here, we will briefly mention the key features of axion cosmology. The reader may consult ~\cite{Sikivie:2006ni, Marsh:2015xka} for extensive reviews. In the early Universe, after the PQ symmetry is broken at temperature $T_{PQ} \sim f_a$, axions could be produced by both thermal and non-thermal mechanisms. However, the present number density of non-thermal or \textit{cold} axions is much larger than that of the thermal axions. \textit{Cold} axions are produced by the non-thermal mechanisms such as vacuum realignments, decays of axion strings, and decays of axion domain walls \cite{Sikivie:2006ni}. {  At any given cosmic time $t$,} the velocity dispersion of the cold decoupled axions is, {  in units of the speed of light $c$ }\cite{Sikivie:2006ni},
\begin{equation}
\beta_a (t) \sim 3 \times 10^{-17} \left( \frac{10^{-5} \; \text{eV}}{m_a} \right)^{\frac{5}{6}} { \frac{a(t_0)}{a(t)} }\, , \label{cold_vel_disp}
\end{equation}
where $a(t)$ is the scale factor of the expanding Universe and $t_0$ is the age of the Universe today. This velocity dispersion corresponds to an effective temperature of $0.5 \times 10^{-34} \; \text{K} \left( 10^{-5} \; \text{eV}/m_a \right)^{2/3}$ today. 

\begin{figure}
 \centering
 \includegraphics[width= 0.9\columnwidth]{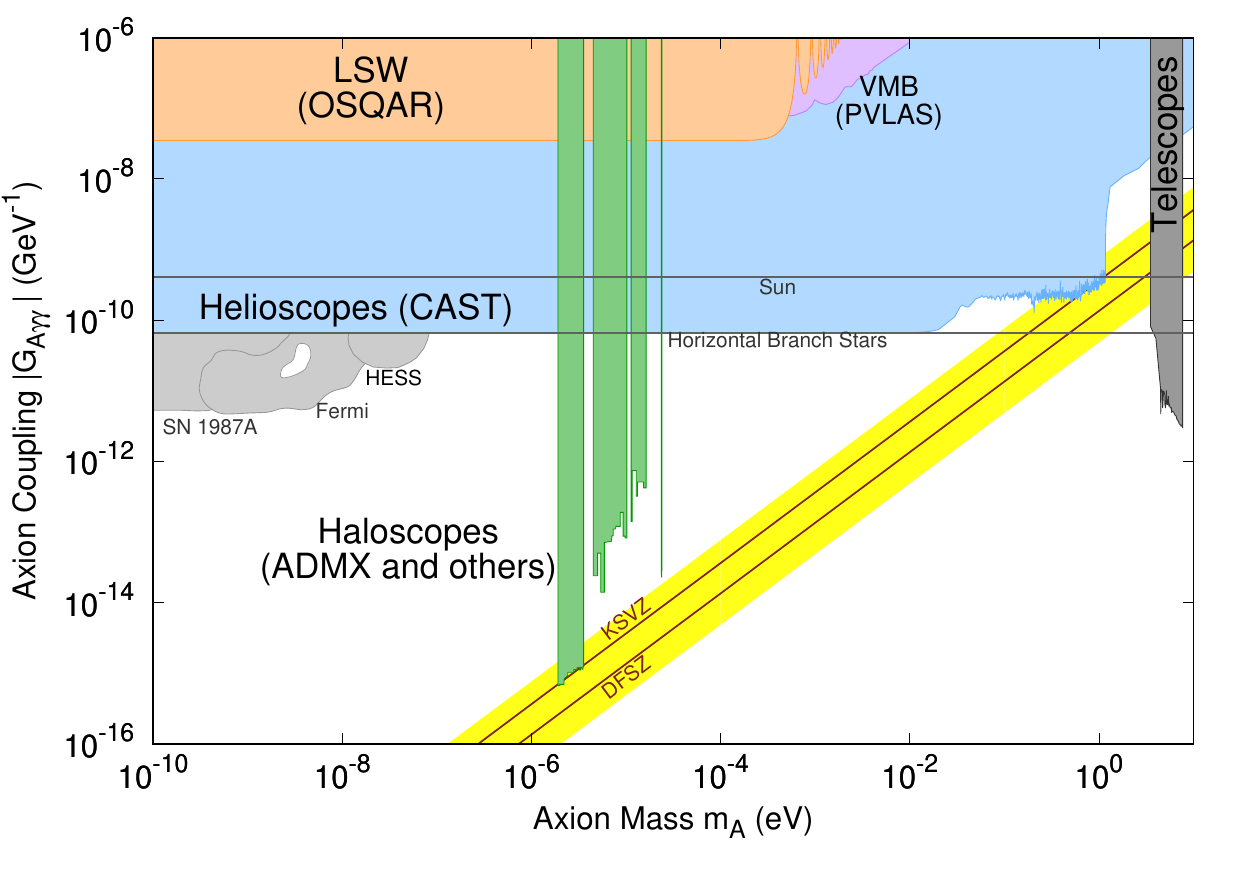}
 \caption{\label{axion_search}The current constraints on QCD axions and ALPs. The yellow band represents the various models of QCD axions. The constraints from the Sun and the Horizontal Branch stars are the regions above the corresponding lines.
 { The regions ruled out by various experiments are indicated by the colored areas.} The figure is reproduced from \cite{PhysRevD.98.030001}.}
\end{figure}

Axion mass is constrained by cosmology, astrophysics, as well as terrestrial experiments. If $f_a$ is larger than about $10^{12}$ GeV, there would have been a large amount of axions to overclose the Universe \cite{Sikivie:2006ni}. This implies a lower bound on the mass of QCD axions: $m_a \gtrsim 6 \times 10^{-6}$ eV. This bound is rather soft due to a number of uncertainties in the axion energy density from decays of strings and domain walls. Astrophysical phenomena, such as stellar evolution and the supernova SN1987A, also put constraints on the axion mass (see, \cite[e.g., ][]{Raffelt:2006cw, Armengaud:2019uso}). Axions were thought to be \textit{invisible} until it was shown by Sikivie \cite{Sikivie:1983ip} that, in a cavity with a strong inhomogeneous magnetic field, an axion can be resonantly converted into two photons when the tuning frequency of the cavity matches the axion mass. The cavity technique is used to design both axion haloscopes and helioscopes. Axion haloscopes, such as ADMX \cite{Du:2018uak}, are designed to search for axions in the Milky Way halo, while axion helioscopes, such as CAST \cite{Anastassopoulos:2017ftl} and IAXO \cite{Armengaud:2019uso}, are designed to probe axions emitted from the Sun via Primakoff conversions. Heavier QCD axions with $m_a \gtrsim 50$ keV have been ruled out by beam dump experiments and rare meson decays (see, \cite[e.g., ][]{Kim:1986ax}). Searches for QCD axions exploit their coupling $G_{a \gamma \gamma}$ to two photons which is related to the mass by \cite{PhysRevD.98.030001}
\begin{equation}
G_{a \gamma \gamma} = C \; 10^{-15} \; \text{GeV}^{-1} \; \left( \frac{m_a}{6 \times 10^{-6} \; \text{eV}} \right)\, , \label{photon_coupling}
\end{equation}
where $C$ is a model dependent number of order one. The current constraints on axion mass and their coupling to photons are shown in Figure~\ref{axion_search} where the yellow band corresponds to the QCD axions in different models. There has been a significant boost in the interest to search for axions or ALPs in the last decade. The reader may consult ~\cite{Sikivie:2020zpn, PhysRevD.98.030001, Graham:2015ouw} for an overview of many novel ideas and experiments that have been proposed. 

\subsubsection{Distinctive features of QCD axions}

In the context of cosmology, QCD axions may have two distinctive features: isocurvature perturbations and axion miniclusters (see \cite{Sikivie:2006ni, Marsh:2015xka} and references therein). Quantum fluctuations of the axion field may cause isocurvature perturbations either on the length scale of the CMB (for the PQ phase transition before inflation), or on the scale of QCD horizons i.e. the horizon size at the time of the QCD phase transition (for the PQ phase transition after inflation). The CMB observations put tight constraints on how much isocurvature perturbations are allowed (see \cite[e.g., ][]{Axenides:1983hj, Seckel:1985tj, Lyth:1989pb}). For the PQ phase transition after inflation, the misalignment angle may be different in different QCD horizons and result into large inhomogeneities in the density of axions. Such inhomogeneities grow in size due to gravitational instabilities and lead to the formation of axion miniclusters with densities many order of magnitude larger than the average density (see, \cite[e.g., ][]{Hogan:1988mp, Kolb:1993zz}). Searches for axions would be more challenging if a substantial fraction of them are present in the form of miniclusters which have typical mass of $10^{-14} \, {\rm M}_\odot$ and typical size of $100 R_\odot$ \cite{Sikivie:2006ni}. They can be probed by gravitational microlensing \cite{Fairbairn:2017sil, 2019arXiv190801773D} and the current constraints on the fraction $f_{mc} ^a$ of axions present in the form of miniclusters is $f_{mc} ^a \lesssim 0.1$.

If the cold axions remain decoupled and their evolution can be described by classical field theory, they behave like CDM on all length scales larger than the correlation length $l(t)$ corresponding to their velocity dispersion \cite{Erken:2011dz}
\begin{equation}
l(t) \sim \frac{1}{m_a \beta_a (t)} \sim 10^{15} \; \text{m} \; \frac{a(t)}{a(t_0)} \left( \frac{6 \times 10^{-6} \; \text{eV}}{m_a} \right)^{\frac{1}{6}} \;\; . \label{correlation_length}
\end{equation}
From the point of view of the large-scale structure, this length is too small to be of observational interest. For ultra-light ALPs of mass $m_a \sim 10^{-22}$ eV, the correlation length would be so large that the ULALPs may be distinguished from ordinary CDM on all length scales. However, numerical simulations suggest that they are indistinguishable in the regime of large-scale structure formation \cite{2014NatPh..10..496S}. Nevertheless, if the QCD axions undergo thermalization and form a Bose-Einstein condensate (BEC), they may have distinctive features like caustic rings and axion stars. 

It has been shown by Sikivie and Yang \cite{2009PhRvL.103k1301S} that, due to their huge phase-space degeneracy and very small velocity dispersion, the axions may thermalize via gravitational self-interactions and form a BEC when the photon temperature reaches about $500 \; \text{eV} \left( {f_a}/10^{12} \text{GeV} \right)^{1/2}$. It has been argued that, as the halo grows in size by accreting more dark matter, the axions continuously rethermalize to track the lowest available energy state \cite{Banik:2013rxa}. This fact supports the necessary initial conditions for the formation of \textit{caustic rings} which are ring-like substructures in the galactic plane \cite{Sikivie:1999jv, Duffy:2008dk}. Caustic rings should have astrophysical signatures in the forms of bumps in the rotation curves, stellar overdensities and characteristic imprints on the density of the interstellar medium \cite{Sikivie:2001fg, Natarajan:2007xh, Chakrabarty:2018gdg}. Several pieces of evidence of the existence of caustic rings have been reported in ~\cite{Sikivie:2001fg}. The existence of vorticities in the axion BEC causes a depletion of dark matter density near the galactic centre \cite{Banik:2013rxa} which may explain why the inner rotation curves can be reconstructed only from baryonic contributions. The rethermalizing axion BEC can also solve \cite{Banik:2013rxa} the \textit{galactic angular momentum problem}  which is the discrepancy between the observed angular momentum distributions in the dwarf galaxies and the predicted ones with ordinary CDM \cite{vandenBosch:2001bp}. 

When the quantum pressure balances the gravitational attraction at small length scales, the axions may form gravitationally bound and stable objects like \textit{axion stars} \cite{Braaten:2015eeu}. Dilute stars from QCD axions are expected to have mass of order $10^{-13} \, {\rm M}_\odot$ and radius of order $10^{-4}R_\odot$. The mass of dense axion stars may range from $10^{-20}\, {\rm M}_\odot$ to ${\rm M}_\odot$. If an axion star collides with a neutron star, the axions may be converted into photons in the strong magnetic field of the neutron star and emit a strong flux of radiation. Interested readers may refer to ~\cite{Braaten:2019knj} and references therein, for a detailed descriptions and observable consequences of such objects. 

\subsubsection{QCD axions and small-scale problems of CDM}

As mentioned above, rethermalizing QCD axions with vortices may solve the \textit{galactic angular momentum problem} and may also explain the depletion of dark matter in the central regions of the halos \cite{Banik:2013rxa}. The caustic ring model, which is an outcome of rethermalizing axion BEC, is found to reproduce the inner and outer rotation curves of the Milky Way \cite{Dumas:2015wba}. Despite these successes, QCD axions ($m_a \sim 10^{-6}$ eV) are hardly explored in the context of the small-scale problems of CDM. In fact, ULALPs of mass $\sim 10^{-22}$ eV (see Section \ref{sec:FuzzyDM}) are attractive candidates to solve the small-scale challenges such as the CCP and the MSP. An attempt to solve the CCP
with QCD axions by simply extrapolating the ULALPs scenario does not work, because, while ULALPs allow a core of size $\sim 1$~kpc, the size of the core from QCD axions would be of $\mathcal{O}(1)$~km \cite{2015MNRAS.451.2479M}. 

Whether QCD axions can address the small-scale problems requires further investigations. From the numerical perspective, resolving small scale structures through N-body simulations of QCD axions with mass $m_a \sim 10^{-6}$ eV is formidable. From the theoretical point of view, QCD axions are usually treated as classical fields (see, for example, \cite{Guth:2014hsa, Chavanis:2016dab}). However, it has been shown that thermalization of axions \cite{2009PhRvL.103k1301S} is a quantum phenomenon and cannot be explained by the classical field theory \cite{Erken:2011dz, Sikivie:2016enz} unless an artificial cut-off wavevector is introduced \cite{Guth:2014hsa}. It has been argued \cite{Chakrabarty:2017fkd} that the wave-mechanics assuming axions as classical fields may not unveil all the physical aspects and the quantum field effects play important roles for self-interacting axions.

\subsection{ Fuzzy Dark Matter}\label{sec:FuzzyDM}

Another excellent alternative to CDM is Fuzzy Dark Matter (FDM) which consists of ultra-light bosons with mass in the range $10^{-23}$--$10^{-20}$ eV \cite{2000PhRvL..85.1158H,2006PhLB..642..192A}. These light bosons are naturally generated from symmetry breaking due to the misalignment mechanism \cite{1983PhLB..120..127P,1983PhLB..120..133A,2006JHEP...06..051S}, and are very common in string theory (for more details see   \cite{2010PhRvD..81l3530A}). FDM is considered to be a real scalar field $\phi$ with mass $m_\phi$ which is minimally coupled to the metric 
\cite{2017PhRvD..95d3541H}.
The field $\phi$ is initially massless until the Universe cools down to some critical temperature \cite{2018arXiv181103771B}. It acquires the mass by rolling down and oscillating about the minimum of a potential generated non-perturbatively \cite{2018arXiv181103771B}. The typical de Broglie wavelength of a FDM particle is a few kpc:
\begin{equation}
\frac{\lambda_{\mathrm {dB}}}{2\pi}=\frac{\hbar}{m_\phi v}=1.92\, {\rm kpc} \left( \frac{10^{-22} {\, \rm eV}}{m_\phi} \right) \left( \frac{10 \, {\rm km/s}}{v} \right) \; .
\label{eq:brog}
\end{equation}
Therefore, the physics of FDM on length scales below $\lambda_{\mathrm {dB}}$ differs from that of the standard CDM. In particular, small density fluctuations  unstable for masses larger than the Jeans mass (see Eq. 42 in \cite{2017PhRvD..95d3541H}) lead to a minimum halo mass of $\sim 10^7 \, {\rm M}_\odot$ for a boson mass of $\sim 10^{-22}$ eV. On the contrary, on a scale above $\lambda_{\mathrm {dB}}$, the large scale structure of FDM is indistinguishable from CDM \cite{2014NatPh..10..496S}.

\subsubsection{Solutions to the observational challenges}

\begin{figure}[htb!]
    \centering
    \includegraphics[width= 0.9\columnwidth]{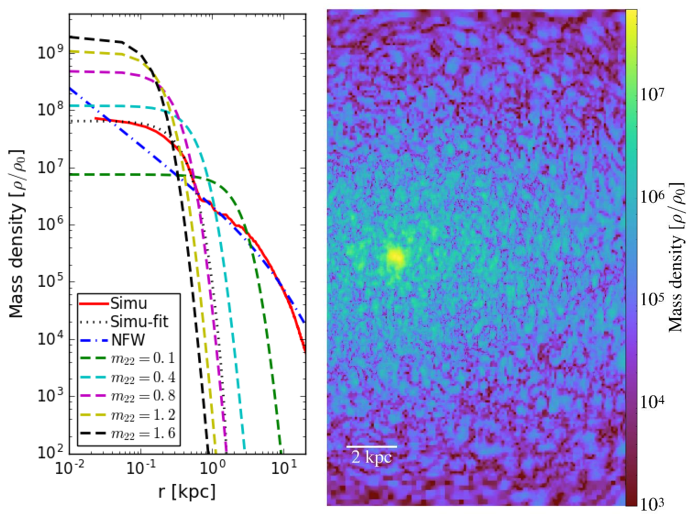}
\caption{Dark matter halos in the FDM scenario. {Left panel:}  dependence of the dark matter density profile on the mass of the boson compared with the NFW profile. Here $m_{22}$ is the boson mass in unit of $10^{-22}$ eV. {Right panel}: an image displaying the soliton in the center of a galaxy with an interference pattern surrounding it.  The figure is reproduced from \cite{2017PhRvL.119v1103D}.}
    \label{fig:wavedm1}
\end{figure}

In 2014, novel N-body simulations with unprecedented  high  resolution showed the rich small scale structures of FDM halos  \cite{2014NatPh..10..496S}. The uniqueness of these N-body simulations was their ability to capture the quantum nature of the dark matter particles by combining the Schr\"odinger's and the Poisson's  equations \cite{1993ApJ...416L..71W}.  Each virialized halo has a core of dark matter in the innermost part, which represents the ground state solution of the Schr\"odinger-Poisson equations. This core is surrounded by an interference pattern represented by fluctuations in the velocity and density fields of the particles. The cores, more often called solitons, exhibit flat density profiles that can naturally explain the wide cores in dwarf galaxies, and match the  NFW density profiles \cite{NFW96} in the outer regions of the halos (see Figure \ref{fig:wavedm1}). As it was shown in \cite{2014NatPh..10..496S}, the solitonic density distribution of the dark matter is well fitted by 
\begin{equation}\label{eq:sol_density}
\rho_c(r) \sim \frac{1.9~(m_\phi/10^{-23}~{\rm eV})^{-2}(r_c/{\rm kpc})^{-4}}{[1+9.1\times10^{-2}(r/r_c)^2]^8}~M_\odot \; {\rm pc}^{-3}\, ,
\end{equation}
where $r_c$ is the core radius of the soliton; $r_c$ also is related to the boson mass and the halo mass $M_{\mathrm {h}}$ by  the scaling relation \cite{2014NatPh..10..496S,2014PhRvL.113z1302S}:
\begin{equation}\label{eq:sol_radius}
r_c=1.6\biggl(\frac{10^{-22}~{\rm eV}}{m_\phi}\biggr)a^{1/2} \left[\frac{\zeta(z)}{\zeta(0)}\right]^{-1/6}\biggl(\frac{M_{\mathrm h}}{10^9 M_\odot}\biggr)^{-1/3}\, ,
\end{equation}
where  $a$ is the cosmic scale factor, $\zeta(z)\equiv(18\pi^2+ 82(\Omega_m(z)-1)-39(\Omega_m(z)-1)^2)/\Omega_m(z)$ is the critical overdensity at redshift $z$. These results have been validated by  N-body simulations \cite{2017MNRAS.471.4559M, 2018PhRvD..98d3509V}.

\begin{figure}[htb!]
    \centering
    \includegraphics[width= 0.48\columnwidth]{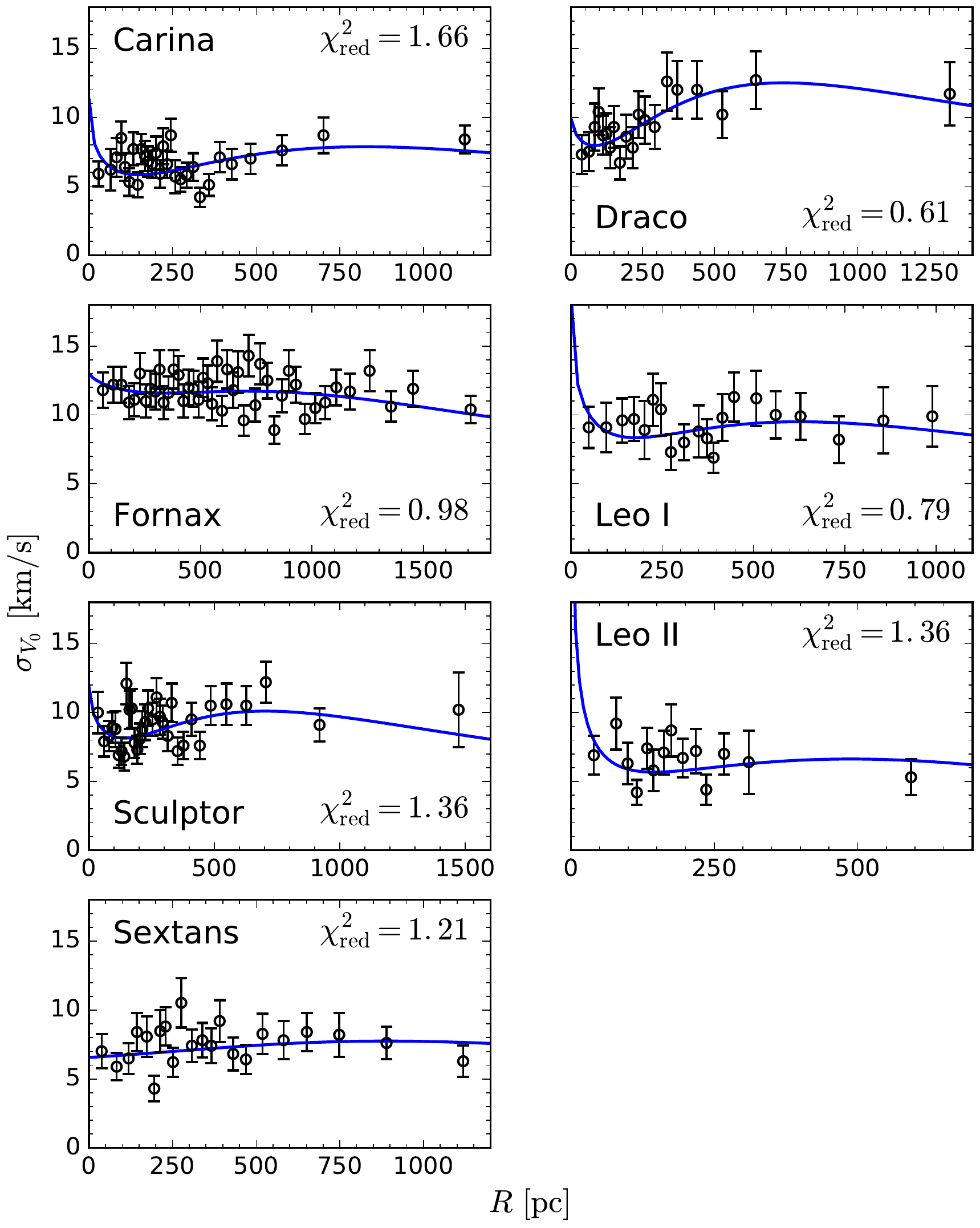}
        \includegraphics[width= 0.49\columnwidth]{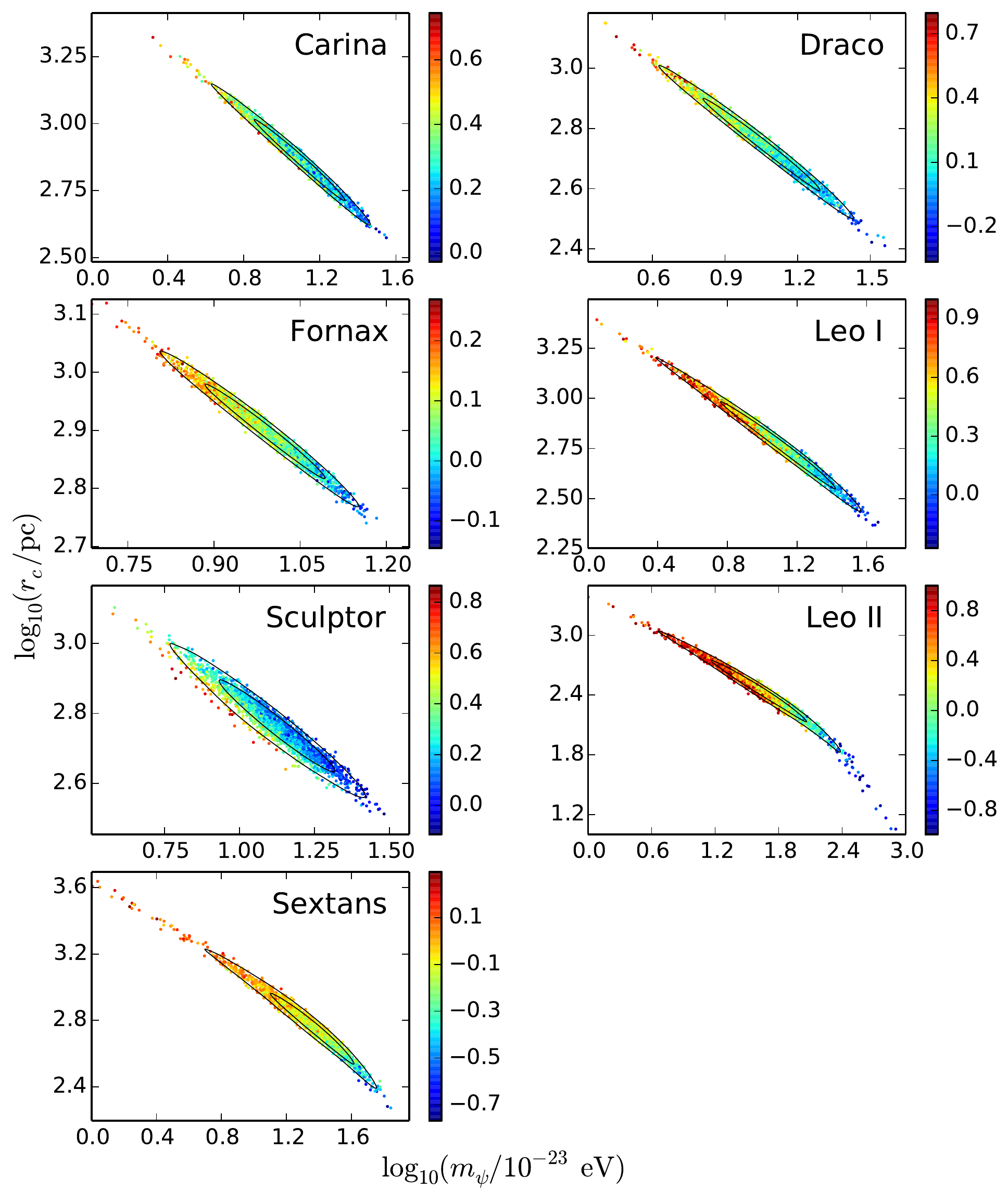}\\
    \caption{{\em Left panels}: best fit models (blue lines) of the velocity dispersion profile in the FDM scenario for a sample of Milky Way dwarf satellites. {\em Right panels}: corresponding core radius-mass relation for each dwarf galaxy sampled from the posterior distribution. The figures are from \cite{2017MNRAS.468.1338C}.}
    \label{fig:wavedm2}
\end{figure}

Since solitons have a constant central density thanks to pressure support, they may potentially solve the CCP. To shed light on this issue, two independent analyses used stellar kinematics data of dwarf spheroidal galaxies by carrying out a Markov Chain Monte Carlo  fitting procedure of the projected velocity dispersion profiles  \cite{2015MNRAS.451.2479M,2016MNRAS.460.4397C,2017MNRAS.472.1346G,2017MNRAS.468.1338C}. Both analyses found that the data prefer  soliton-generated cores (with boson mass $\sim 10^{-22}$ eV) over cuspy NFW profile.  Figure \ref{fig:wavedm2} shows the effectiveness of the FDM halo profile at describing the kinematics of stars in the Milky Way dwarf satellites. 

In addition, a detailed study of the recently discovered ghostly galaxy Antlia II has shown that the solitonic structure of the FDM halo may help to explain the presence of the wide core ($\sim 2.8$ kpc) of this dwarf galaxy \cite{2019arXiv190210488B}, and of other ultra-faint galaxies \cite{2020arXiv200308313P}. However, in contrast to previous results, numerical solutions \cite{2018PhRvD..98b3513D} of the evolution of a scalar field in a spherically symmetric space-time fail to reproduce the scaling relation between the core density and the core radius \cite{Rodrigues_2017}. Nevertheless, it is worth noting that {  this scaling relation} originates from a fitting procedure of the the so-called Burkert profile to the measured rotation curves of disk galaxies. {This approach might not be appropriate, because }, in FDM, the core density and the core radius  depend upon the boson mass and the total halo mass as in the scaling relation of Eq. \eqref{eq:sol_radius}. 
In addition to the cores of ultra-faint galaxies, the solitonic structure can provide a solution to the excessive central velocity dispersion of the stars in the bulge of the Milky Way \cite{DEMARTINO2020100503}. 

\begin{figure}[htb!]
    \centering
    \includegraphics[width= 0.9\columnwidth]{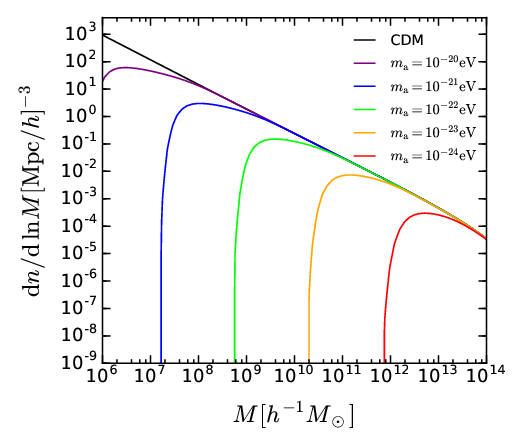}\\
    \caption{Dependence of the cut-off of the halo mass function on the mass of a FDM particle, and comparison with the CDM model. The figure is reproduced from \cite{2017MNRAS.465..941D}.}
    \label{fig:fdm3}
\end{figure}

Finally, N-body simulations show the suppression of the halo number density for mass $\leq 10^{10} \, {\rm M}_\odot $, and how this cut-off depends upon the mass of the FDM particle (Figure \ref{fig:fdm3}). Such suppression may provide a solution to the MSP \cite{2017MNRAS.465..941D}.

The debate on whether or not FDM may be a viable dark matter candidate is still ongoing. The presence of the soliton in every virialized halo can affect the dynamics of the disk by enhancing the circular velocity in the inner part of the rotation curve; this feature can provide a way to probe the model \cite{2019PhRvD..99j3020B}.  One should note that the inner dynamics of a disk galaxy is affected by the baryonic contributions to the gravitational potential, and the breakdown of the spherical symmetry may affect the geometry of the solitons  \cite{2019PhRvD..99j3020B}.

Additionally, the effect of quantum pressure in structure formation, which is suppressed on small scales, is still under investigation due to the poor resolution of the N-body simulations. A box size of at least $\sim 500$~Mpc/h on a side is needed to predict the matter power spectrum, and evaluate how much suppression is introduced \cite{2018FrASS...5...48Z}.
Furthermore, the clustering properties of hot gas at high redshift have been used to constrain the dark matter properties. The analysis of the Lyman-$\alpha$ forest data constrains the boson mass to be larger than $ 7\times10^{-21}$ eV \cite{2016JCAP...08..012B}, which is almost two order of magnitudes larger than the boson mass required to describe dwarf galaxies ($\sim10^{-22}$ eV). Since  the power spectrum of WDM and FDM is the same on large scales \cite{2014MNRAS.437.2652M}, and is suppressed below a certain characteristic scale \cite{Viel_2013,2014MNRAS.437.2652M}, the constraints based on Lyman-$\alpha$ forest data are obtained  by translating the observational bounds on WDM  into the corresponding bounds on FDM by matching the suppression scale \cite{2016JCAP...08..012B}.
However, this analogy between WDM and FDM is still under debate \cite{2017PhRvD..95d3541H,Zhang2018}. Hydrodynamical simulations are required  for FDM to quantify the impact on the matter power spectrum of the small scale  density fluctuations on the de Broglie scale which are absent in WDM \cite{2014NatPh..10..496S,2017MNRAS.471.4559M,2018PhRvD..98d3509V}. These effects may play a fundamental role { in distinguishing between the models} \cite{2017PhRvD..95d3541H}, although the uncertainties on the different thermal histories and underlying reionisation models of the WDM and FDM particles may weaken these constraints \cite{GARZILLI2017258, 2017PhRvD..95d3541H}.

\section{Possible solutions beyond Newtonian dynamics }\label{sec:gravitmodels}

The observational challenges  of the CDM model on galactic scales (see Section \ref{sec:obschallenges}) may also be interpreted as a breakdown of the law of gravity. Modifications of the law of gravity conceived to explain the observed kinematics of visible matter started to be systematically investigated back in the 1980s \cite{1984ApJ...286....7B,1986MNRAS.223..539S,1990A&ARv...2....1S,SandersandMcGaugh02}, although some suggestions were put forward much earlier \cite[e.g., ][]{1963MNRAS.127...21F}. On cosmological scales, observational data from the Planck mission do not seem to provide statistical evidence in favor of any particular theory of  gravity  \cite{2018arXiv180706209P}, whereas at galactic scales, where the physics is complicated by the relevant role of baryons, the issue remains open. 
In the following, we briefly touch upon three modified gravity models, proposed in the literature, that focus on the scales of galaxies.

\subsection{MOND} \label{sec:MOND}

In 1983, Milgrom suggested to explain the mass discrepancy in cosmic structures with a modification of the law of gravity rather than with the presence of dark matter \cite{Milgrom83MOND, 1983ApJ...270..371M, 1983ApJ...270..384M}. His phenomenological proposal rests upon the hypothesis that there is an acceleration scale $a_0\simeq 1.2\times 10^{-10}$~m~s$^{-2}$ above which Newtonian gravity holds and below which Newtonian gravity breaks down. This idea goes beyond the naive idea that gravity should be modified simply beyond a length scale \cite{1963MNRAS.127...21F,McGaughanddeBlok98}.

According to Milgrom's suggestion, the magnitude $a$ of the acceleration experienced by a test particle in a gravitational field is 
\begin{equation}
a = \nu\left(\frac{a_N}{a_0}\right) a_N\; ,
\label{eq:MONDacc}
\end{equation}
where $a_N$ is the magnitude of the gravitational acceleration, estimated in Newtonian gravity, originated by the distribution of the baryonic matter alone, as dark matter is assumed to be nonexistent; $\nu$ is an interpolation function whose asymptotic behaviors are $\nu\to 1$ when $a_N\gg a_0$ and $\nu\to (a_N/a_0)^{-1/2}$ when $a_N\ll a_0$. A number of interpolation functions has been proposed in the literature; one of them is  \cite{2008ApJ...683..137M}
\begin{equation}
\nu\left(\frac{a_N}{a_0}\right) = \left(1-e^{-\sqrt{a_N/a_0}}\right)^{-1}\,. 
\label{eq:MONDinterp}
\end{equation}  

Rather than as a modification of the law of gravity, the introduction of an acceleration scale can be alternatively interpreted as a modification of the law of inertia $F=ma$, where the inertial mass differs from the gravitational mass when $a_N\ll a_0$ \cite{Milgrom_1994,1999PhLA..253..273M,Milgrom_2006}.

In both cases, Milgrom's suggestion yields a MOdified Newtonian Dynamics (MOND) (see \cite{2012LRR....15...10F} for an extensive review).
The introduction of an acceleration scale makes the MOND formulation manifestly purely phenomenological: in General Relativity, the acceleration is linked to the affine connection $\Gamma^{\mu\nu}_{\lambda}$ which is not a tensor; therefore, MOND cannot be easily formulated in a covariant form. 

This drawback has two important consequences: in the MOND framework, we can neither build a cosmological model, which is the most relevant success of the $\Lambda$CDM model, nor quantify the phenomenology of gravitational lensing, which is an important probe of the mass distribution on large scale. 
An additional shortcoming is that MOND is unable to explain the observed mass discrepancy on the scale of galaxy clusters and on larger scales, although the amount of required dark matter is substantially reduced \cite{1983ApJ...270..384M}. 

Attempts to provide MOND with a covariant formulation include, for example, AQUAL \cite{1984ApJ...286....7B}, TeVeS \cite{PhysRevD.70.083509}, and bimetric MOND \cite{PhysRevD.80.123536}.  These theories introduce additional scalar, vector or tensor fields and  reduce to MOND in the non-relativistic limit \cite{2009PhRvD..80f3515C}. Their number shows that a covariant theory that reduces to MOND is not uniquely determined. Therefore, invalidating one of these theories does not necessarily invalidate MOND. 

For example, the detection of gravitational waves originating from the merging of two neutron stars \cite{PhysRevLett.119.161101} combined with the observation of a gamma-ray burst within a few seconds \cite{Goldstein_2017,Savchenko_2017} implies that the speed of light and the speed of gravitational waves coincide within one part in $10^{-15}$. In the original formulation of TeVeS, the speed of gravitational waves in general is different from the speed of light and therefore TeVeS appears to be ruled out \cite{PhysRevD.97.084040}. Nevertheless, there is  a family of tensor-vector-scalar theories, that still reduce to MOND in the non-relativistic limit, where the speed of gravitational waves equals the speed of light \cite{2019PhRvD.100j4013S}.

We can also build a hybrid model that merges the success of MOND on small scales with the properties of the large-scale structure provided by the presence of dark matter. This idea was suggested by Angus in 2009 \cite{2009MNRAS.394..527A}. He assumed MOND as the theory of gravity and added a hot dark matter component made of sterile neutrinos of mass $\sim 11$~eV. The existence of a sterile neutrino still appears to be a solution to the detection of the excess of electron-like events in  short-baseline  neutrino  experiments \cite{Aguilar_Arevalo_2018}. Hot dark matter has the advantage of clumping on scales larger than the scale of galaxies. MOND phenomenology would thus be preserved on small scales, whereas dark matter starts becoming relevant on larger scales, where MOND apparently disagrees with observations (see \cite{2012arXiv1206.6231D} for a review). Unfortunately,  this hybrid model is unable to reproduce the mass function of galaxy clusters \cite{2011MNRAS.417..941A,2013MNRAS.436..202A, 2014JCAP...10..079A} and currently appears unviable. 

\subsubsection{Disk galaxies}

Many of the observational challenges described in the previous sections were predicted by MOND many years before they were actually observed and posed unexpected challenges to the traditional dark matter framework. This feature is specific to MOND and makes it fundamentally different from the other suggested theories of modified gravity: these latter theories  attempt to describe these observations only after they become available and never anticipate them.

Indeed, the predictions of MOND on the scales of galaxies are so distinctive that 
it has become customary to collect them in the so-called MOND phenomenology. These observational facts are clearly independent of the theory of gravity; therefore, other theories, alternative to the standard model, must mimic the MOND phenomenology on these scales. Recent models that explicitly attempt to reproduce the MOND phenomenology include  superfluid dark matter \cite{PhysRevD.92.103510}, dipole dark matter \cite{Blanchet_2007}, refracted gravity \cite{2016arXiv160304943M,2020arXiv200307377C}, and emergent gravity \cite{2017ScPP....2...16V}. 

MOND was motivated by the observations of the flat rotation curves of disk galaxies \cite{Milgrom83MOND}. The normalization of the Tully-Fisher relation \cite{1977A&A....54..661T} sets the magnitude of the acceleration scale $a_0$. As reminded in { Section \ref{sec:RAR}}, the Tully-Fisher relation links the luminosity $L$ of a disk galaxy to the width $W$ of the profile of the global neutral hydrogen line in emission; if $L$ and $W$ are proxies of the galaxy mass and the asymptotic rotation velocity respectively, the Tully-Fisher relation becomes a relation between mass and centripetal acceleration. 

At large $r$, where we can consider that the cumulative baryonic mass $M(<r)=M_d$ is constant, we have, in Newtonian gravity, the acceleration $a_N=GM_d/r^2$, and, from Eq. \eqref{eq:MONDacc}, in the limit $a_N\ll a_0$, we have $a^2=a_Na_0=a_0GM_d/r^2$. At large $r$, we have 
$a=v_{\mathrm{rot}}^2/r$, with $v_{\mathrm{rot}}={\mathrm {const}}$,  and we obtain $v_{\mathrm{rot}}^4=a_0GM_d$. We thus see that the normalization of the Tully-Fisher relation is $a_0G$, and its slope is $4$. In the 1980s, observations provided a considerably uncertain slope, in the range $2.5-5$, as reviewed in \cite{1983ApJ...270..371M}; nowadays, this slope is confirmed to converge to $4$ \cite{McGaugh_2015}. 

In this perspective, the Tully-Fisher relation simply is a law of gravity, similar to the Kepler's laws, and it is not linked to any property of galaxies other than their baryonic mass. Consequently, the Tully-Fisher relation should hold for any self-gravitating system; this prediction is relevant because we should expect the Tully-Fisher relation to hold irrespective of the surface brightness of the galaxy. In other words, MOND predicts the existence of a Baryonic Tully-Fisher relation between the baryonic mass, including stars and gas, and the asymptotic flat rotation velocity. The BTFR is now neatly supported by the data and holds from LSB to HSB galaxies over six orders of magnitude in baryonic mass  \cite{2019arXiv190902011M} with a slope confirmed to be in the range $3.5-4$ \cite{2019MNRAS.484.3267L} (see Figure~\ref{fig:BTFR}).

The BTFR is intimately connected to another MOND prediction: the correlation  between the disk surface brightness $\mu$ and the centripetal acceleration $a=v_{\mathrm {rot}}^2/r$, that directly derives from the  relation $a\sim 2\pi G\mu$.
Accurate spectroscopic measures of the disk galaxies in SPARC \cite{Lellietal16SPARC}  and other data sets \cite[e.g., ][]{2013MNRAS.433L..30L} support this relation.

MOND also implies the Faber-Jackson relation $M\propto \sigma^4$ between the mass $M$, if proportional to luminosity, and the stellar velocity dispersion $\sigma$ of elliptical galaxies \cite{1976ApJ...204..668F}. Unlike the Tully-Fisher relation however, the Faber-Jackson relation would be exact in MOND only if ellipticals were isothermal and their velocity fields were isotropic. In fact, matching the observed Fundamental Plane of ellipticals requires that the velocity anisotropy parameter varies with radius \cite{2000MNRAS.313..767S, 2011MNRAS.412.2617C}.

The fit of the rotation curve in the standard model has two free parameters that describe the dark matter halo density profile when assumed to be spherical, in addition to the disk-to-halo mass ratio. In MOND, there is only one free parameter:
 the mass-to-light ratio $M/L$. For MOND to be viable, the best fit $M/L$ has to be consistent with stellar
population synthesis models. Indeed, MOND fits yield mass-to-light ratios $M/L$ that agree with the expectations and also recover the expected relations between color and $M/L$ \cite{McGaugh_2015}. Claimed failures of MOND fits often derive from inaccurate measures of the galaxy distance and/or disk inclination or inappropriate bulge-disk decompositions \cite{2012A&A...543A..76A}.

By requiring that the velocity of the baryonic matter is directly set by its distribution, MOND makes a clear prediction on the shape of the rotation  curves of disk galaxies: HSB galaxies are expected to have steeply rising rotation curves that flatten at small radii, whereas LSB galaxies are expected to have slowly rising rotation curves that converge to the asymptotic constant velocity at large radii. 
This prediction by Milgrom in 1983 \cite{1983ApJ...270..371M} appeared well before LSB galaxies were known to
be a substantial fraction of the galaxy population and well before their systematic 
observations confirmed the MOND prediction \cite{1998ApJ...499...66M}.

An additional consequence of the absence of dark matter is the expected one-to-one
correspondence between the irregularities in the surface brightness distribution and the
features of the rotation curves. This correspondence is widely observed in disk galaxies \cite{2012MNRAS.425.2299S} and is known as the Renzo's rule \cite{2004IAUS..220..233S}. These features in HSB disk galaxies are also naturally explained in the standard framework, because the stars dominate the gravitational potential at small radii, according to the maximum-disk hypothesis. However, in LSB disk galaxies this hypothesis does not hold and the observed correspondence remains unexplained. MOND naturally explains these observations in both HSB and LSB disk galaxies \cite{2019MNRAS.485..513S}.

The existence of LSB disk galaxies and their properties have, in MOND, a relevant role that needs to be emphasized. In the 1970s, the idea of disk galaxies embedded in a halo of dark matter was introduced to describe the observations of flat rotation curves inferred from the spectroscopy of neutral hydrogen \cite{1973A&A....26..483R}. The community easily accepted this idea because it had recently become clear that bare cold self-gravitating disks, as
spirals appear to be, are violently unstable and rapidly generate a bar \cite{1970ApJ...161..903M, 1971ApJ...168..343H}, unless the disk is embedded in a massive halo \cite{1973ApJ...186..467O}:  
a single massive halo, that was initially  thought to be composed of too faint stars to be observed, made the disk dynamically stable and explained the flat rotation curves.

This consideration has an important consequence on the dynamics and morphology of LSB disk galaxies: the rotation curves of LSB disks indicate that these galaxies are dynamically dominated by dark matter in the standard model even at small radii, where the baryonic matter dominates in HSB disk galaxies \cite{1998ApJ...499...66M}. Consequently, the disk of LSB galaxies is over-stabilized \cite{1997ApJ...477L..79M} and its mass should be smaller than the limit required to generate spiral modes, unlike what happens in HSB disks \cite{1987A&A...179...23A}; nevertheless, spiral arms and bars are observed in LSB disk galaxies \cite{1995AJ....109.2019M, 2006AJ....131..296S, 2011AdAst2011E..12S, 2018MNRAS.476.2938P}. 
As a consequence, explaining the amplitude of the rotation curves and the presence of spiral arms and bars in the standard model requires mass-to-light ratios that are inconsistently larger than the expectations from stellar population synthesis models \cite{2003Ap&SS.284..719F, 2008AN....329..916F, 2013AN....334..785S}. This tension does not appear in MOND, where there is no dark matter and the spiral modes are naturally generated as in the HSB disk galaxies \cite{1989ApJ...338..121M, 1998ApJ...499...66M, 1999ApJ...519..590B}.

At the same time when the introduction of the massive halos was thought to be necessary, Freeman noted that the mean central surface brightness of disk galaxies is $\mu_F= 21.65$~mag~arcsec$^{-2}$ in the $B$-band with a little scatter of $0.30$~mag~arcsec$^{-2}$: this relation became
known as the Freeman law \cite{1970ApJ...160..811F}. However, the Freeman law was the result of an observational bias, as it became clear a few years later \cite{1979ApJ...227...67A}. In fact, 
disks appear in a wide range of central surface brightness \cite{2011ARA&A..49..301V}. In addition, LSB galaxies are actually  more than half of the total galaxy population and the number of galaxies with central surface brightness brighter than $\mu_F=21.65$~mag~arcsec$^{-2}$ drops substantially faster than for a normal distribution \cite{1995AJ....110..573M}: the Freeman law actually is a Freeman limit. 

This observational result obviously makes us wonder what sets the Freeman limit. The Freeman limit is a property of the baryonic matter, whereas dark matter dominates the dynamics of disk galaxies in the standard model. Therefore, the Freeman limit must originate by the interplay between dark matter and baryonic matter. The CDM model has not yet an obvious solution for deriving this limit \cite{1997ApJ...482..659D}.

In MOND, the origin of the Freeman limit simply derives from gravitational dynamics. 
Without dark matter, disks are unstable in Newtonian gravity. However, if MOND is the theory of gravity, the disk becomes stable against the development of bars \cite{1989ApJ...338..121M, 1999ApJ...519..590B, 2008A&A...483..719T, 2014arXiv1406.0537J, 2016arXiv160604942T, 2016MNRAS.462.3918S, 2018arXiv180810545B}. It follows that disks without any dark matter are gravitationally stable only if they are in the MOND regime $a_N<a_0$.  If mass is proportional to luminosity, we have the acceleration $a_N\sim 2\pi G \Sigma$, in Newtonian gravity, where $\Sigma$ is the surface mass density. The limit $a_N<a_0$ becomes
$a_N\sim 2\pi G\Sigma<a_0$; in other words, gravitationally stable disks must have $\Sigma<a_0/(2\pi G)$: this limit $a_0/(2\pi G)\approx 143$~M$_\odot$~pc$^{-2}$ neatly returns 
the Freeman limit $\mu_F= 21.65$~mag~arcsec$^{-2}$ \cite{1983ApJ...270..371M, 1989ApJ...338..121M}. 
This result demonstrates that the acceleration scale $a_0$ enters in another context which is completely different from the Tully-Fisher relation shown above.

In MOND, from the distribution of the baryonic matter, we can also predict the profile of 
the stellar vertical velocity dispersion, namely the velocity 
dispersion perpendicular to plane of the disk, as a function of the radial coordinate in the
disk plane \cite{1983ApJ...270..371M}. Testing this prediction is particularly challenging,
because we have to measure both the vertical velocity dispersion and the rotation curve and, ideally, we would need face-on galaxies for the former and edge-on galaxies for the latter. 
The task was performed by the DiskMass Survey collaboration \cite{2010ApJ...716..198B, 2010ApJ...716..234B, 2011ApJ...739L..47B}, who found that the kinematic properties, when interpreted with Newtonian dynamics, require a stellar population with a mass-to-light ratio $\sim 0.3$~M$_\odot$/L$_\odot$ in the $K$-band, a factor two smaller than expected from stellar population synthesis models \cite{2013A&A...557A.130M, 2013A&A...557A.131M}. In other words, the vertical velocity dispersion is too small compared to the rotation curve and the disks are too cold for their stellar mass.

When interpreted in MOND, the mass-to-light ratio in the $K$-band is $0.55\pm 0.15$~M$_\odot$/L$_\odot$, in agreement with stellar population synthesis models, but the disks are a factor of two thinner than expected from the observations of edge-on galaxies \cite{2015MNRAS.451.3551A}, or, alternatively, the vertical velocity dispersion is overpredicted by $\sim 30$\% \cite{2016JPhCS.718c2001A}.  However, the shape of the vertical velocity dispersion profile in the model is consistent with the observed shape over the entire radial range; this concordance suggests that the measurements might suffer from a systematic bias \cite{2015arXiv151108087M}. Indeed, the estimate of the velocity dispersion could be dominated by the younger, and dynamically colder, stellar population, whereas the estimate of the disk thickness in edge-on galaxies could be dominated by the older, and dynamically hotter, stellar population \cite{2016MNRAS.456.1484A}.

To date, the conundrum remains unsolved, although a number of additional observations might support the MOND framework: superthin disk galaxies, that appear largely self-gravitating \cite{1999AJ....118.2751M, 2000AJ....120.1764M, 2008AJ....135..291M}, would be naturally stabilized by the enhanced MOND acceleration \cite{1998ApJ...499...66M}; similarly, the large vertical velocity dispersion of the gas in the outer region of some disk galaxies, when interpreted in the standard model, would require an embedding disk of dark matter or highly flattened halos that are difficult to reconcile with the conventional framework \cite{2020ApJ...889...10D}. A similar dark matter disk, or alternatively an extended dark matter core with a large core radius of $10$~kpc \cite{2013JCAP...07..016N}, might be required for the Milky Way, according to the kinematics of red clump stars from the RAVE survey \cite{2014A&A...571A..92B}.

Overall, the solidity of the existence of the acceleration scale $a_0$ in the data has been increasing for the last decades { (see however \cite{2020PDU....2800468Z})}. Its existence requires a
natural explanation, irrespective of the correctness of MOND. In addition to the Tully-Fisher relation and the Freeman limit, $a_0$ appears in the mass-discrepancy relation and in the radial acceleration relation.

In the standard model, we estimate the dynamical mass with $M_d\sim rv^2/G$; from the luminosity we can estimate the baryonic mass $M_b$. In MOND, Newtonian gravity holds when the Newtonian
gravitational acceleration generated by the baryonic mass is larger then $a_0$, therefore
we expect $M_d/M_b\sim 1$ in this regime and $M_d/M_b$ increasingly larger than 1 at increasingly smaller accelerations (Figure~\ref{fig:MDAR}). This mass-discrepancy relation \cite{1990A&ARv...2....1S, McG04MDAR}, predicted in 1983 \cite{1983ApJ...270..371M}, implies increasingly large mass-to-light ratios for galaxies with increasingly fainter surface brightness, as it was confirmed years later with the observations of dwarf spheroidals and LSB disks \cite{1991AJ....102..914M, 1998ApJ...499...66M}. 

Similarly, we can see the appearance of $a_0$ in the RAR, between
the observed centripetal acceleration derived from the rotation curve of disk galaxies and the Newtonian gravitational acceleration generated by
the baryonic mass distribution (Figure~\ref{fig:RAR}): the deviation from the one-to-one relation appears at Newtonian
accelerations smaller than $a_0$ and the relation is described by the interpolation function of Eq. \eqref{eq:MONDinterp} \cite{McGetal16RAR}. { Although the agreement shows no scatter  and no dependence of the residuals on the galaxy properties \cite{Lellietal17aRAR, Lietal18RAR}, as expected for a relation driven by gravity alone, additional investigations are required to confirm these results, because some dependence of the residuals on the galaxy properties appears to be present in other galaxy samples different from SPARC \cite{2019ApJ...873..106D, 2020arXiv200307377C}.}

\subsubsection{Dwarf galaxies}

In MOND, the cusp/core problem is absent by definition, because there is no dark matter. MOND can explain the velocity dispersion profiles of  dwarf spheroidals with mass-to-light ratios $M/L$ consistent with stellar population synthesis models for the classical dwarfs, except  Carina \cite{2008MNRAS.387.1481A, 2010A&A...524A..16S}. Carina is the closest dwarf spheroidal to the Milky Way and detailed N-body simulations in MOND show that  tidal forces and the external field effect, an effect that lacks in Newtonian gravity, can only partly alleviate the tension \cite{2014MNRAS.440..746A}: the best-fit $M/L$ in the $V$-band required to match the observed velocity dispersion profile is $M/L\sim 5.3-5.7$~${\rm M}_\odot/{\rm L}_\odot$, a value $\sim 10\%$ greater than the upper limit for the old stellar population of Carina $\sim 5$~${\rm M}_\odot/{\rm L}_\odot$ \cite{2005MNRAS.362..799M}.   However, there might still be the possibility to alleviate this tension both observationally and theoretically: more accurate measurements of the proper motion of the dwarf spheroidals and larger samples of stars with accurate photometry can provide a better understanding of the physical properties of the dwarf spheroidals; similarly, the modelling of these systems can be improved by considering a triaxial three-dimensional stellar  distribution, more sophisticated treatments of stellar binaries and even more accurate stellar population synthesis models. 

{ If MOND is the correct theory of gravity, interpreting the dynamics of dwarfs with Newtonian gravity implies dark matter  distributions that might differ from the standard model expectations. The external field effect in MOND mostly generates (1) an offset of the dark matter distribution from the stellar distribution of the order of the half-mass radius of the dwarf, and (2) possible large concentrations of dark matter along the direction to the Milky Way, especially in Fornax and Sculptor, which have no reason to appear in the standard model \cite{hodson2020}. In principle, accurate measurements of the surface brightness distribution and  measurements of the proper motions of the dwarf stars with future astrometric missions \cite{theia,malbet2019esa} could test these predictions and eventually falsify MOND. }

In the standard model, the problem of the plane of satellites would be most easily solved if these satellites were collapsed tidal debris formed during galaxy interactions \cite{2000A&A...358..819W, Bournaud:2006qz, 2007MNRAS.375..805W}. Unfortunately, the dwarf galaxies appear to be dark matter dominated, whereas these Tidal Dwarf Galaxies (TDGs) are expected to be dark-matter free. In fact, the dark matter halo of the parent galaxy is supported by the velocity dispersion of the dark matter particles and it is dynamically hot, unlike the dynamically cold baryonic galactic disk supported by rotation. Therefore the dark matter halo does not participate in the formation of the TDGs orbiting in a dynamically cold plane: this tidal tail can only originate from the galactic disk. In MOND, this mechanism would work without the complication of the existence of the dark matter, as shown by N-body simulations \cite{2008ASPC..396..259T, Renaud:2016blx}; the mechanism is actually favoured by the enhanced self-gravity of the baryons. 

Observations of recently formed TDGs are still unable to confirm whether these systems have mass discrepancies consistent with MOND or rather no mass discrepancies and are thus consistent with dark-matter free TDGs in Newtonian gravity \cite{2007Sci...316.1166B, 2015A&A...584A.113L, 2016MNRAS.457L..14F}. However, this latter solution would suggest the existence of two dwarf populations in the standard model: dark-matter dominated dwarfs, presumably formed at early time, and more recently formed dark-matter free TDGs. However, the existence of these two populations appears to be inconsistent with the samples currently available \cite{2013MNRAS.429.1858D}.  

In conclusion, MOND predicted, rather than solved, most of the dynamical properties of disk and dwarf galaxies { and might present a viable mechanism for the formation of the plane of satellites. MOND appears mostly consistent with the observed kinematics of dwarf spheroidals and makes additional predictions on their density and velocity fields that can further falsify the theory. Additional theoretical and observational investigations are required to clarify these open issues. In addition,} it still remains to be seen how MOND can be embedded in a more extended theory capable of providing a cosmological model and describing the formation and evolution of the large-scale structure.

\subsection{MOdified Gravity (MOG)}\label{sec:MOG}
Scalar-Vector-Tensor theory of gravity, also renamed  MOdified Gravity (MOG), is built to describe the observational effects related to dark matter. In this model,  scalar, tensor and massive vector fields are added to the standard Hilbert-Einstein action  
\cite{2006JCAP...03..004M}. Thus, the total action reads:
\begin{equation}
\label{action1}
{\cal A}={\cal A}_G+{\cal A}_\phi+{\cal A}_S+{\cal A}_M\,,
\end{equation}
where the first term is
\begin{equation}
{\cal A}_G=\frac{1}{16\pi}\int\frac{1}{G}\left(R+2\Lambda\right)\sqrt{-g}~d^4x,
\end{equation}
with   $\Lambda$ the cosmological constant. The second term  is the action related to the massive vector field:
\begin{eqnarray}
{\cal A}_\phi&=&-\int\Big[\frac{1}{4}B^{\sigma\tau}B_{\sigma\tau}-\frac{1}{2}\mu^2\phi_\beta\phi^\beta\Big]\sqrt{-g}~d^4x\, ,
\end{eqnarray}
where $\phi^\beta$ is the vector field, and $B_{\sigma\tau} = \partial_\sigma\phi_\tau -  \partial_\tau\phi_\sigma$. 
The third term represents the action for the scalar  fields $G$ and $\mu$:
\begin{eqnarray}
{\cal A}_S&=&-\int\frac{1}{G}\Big[\frac{1}{2}g^{\tau\beta}\biggl(\frac{\nabla_\tau G\nabla_\beta G}{G^2}
+\frac{\nabla_\tau\mu\nabla_\beta\mu}{\mu^2}\biggr)+\frac{V_{G}(G)}{G^2}+\frac{V_\mu(\mu)}{\mu^2}\biggr]\sqrt{-g}~d^4x\, ,
\label{scalar}
\end{eqnarray}
where $\nabla_\tau$ is the covariant derivative with respect to the metric 
$g_{\tau\beta}$,
and  $V_G(G)$ and $V_\mu(\mu)$ are self-interaction potentials of the $\mu$ and $G$ fields. 
Finally, the fourth term describes the matter action.

In the weak field limit,  the gravitational potential shows a Yukawa-like correction to the Newtonian term. Such a correction generates a repulsive gravitational force that cancels out
gravity at short range (galactic, sub-galactic, or smaller scales) and stabilizes the system. On the other hand, at larger scales, the repulsive term becomes weaker and one obtains the Newtonian gravity with a larger gravitational constant \cite{2013MNRAS.436.1439M}. The gravitational potential in MOG is given by \cite{2013MNRAS.436.1439M}:
\begin{align}
\Phi_{\rm eff}(\vec x) = - G_N\int\frac{\rho(\vec x')}{|\vec x-\vec x'|}\left[1+\alpha-\alpha e^{-{\mu}|\vec x-\vec x'|}\right]d^3\vec{x}'\, ,
\label{eq:MOG1}
\end{align}
where $\mu$ is the inverse of the characteristic length of the gravitational force 
that acts at a certain scale, and it depends on the total mass of the self-gravitating system \cite{2013MNRAS.436.1439M}, and $\alpha = (G_\infty - G_N)/G_N$ accounts for the modification of
the gravitational constant, where $G_\infty$ is the effective gravitational constant at infinite { distance from the self-gravitating system} { and $G_N$ is the standard gravitational constant} \cite{2009CQGra..26h5002M}. The theory has been tested on a wide range of observational scales. 
At extragalactic and cosmological scales, MOG { describes} a wide variety of phenomena ranging from the gravitational bending of light \cite{2009MNRAS.397.1885M,Rahvar_2018}, to the X-ray and Sunyaev-Zel'dovich emissions of galaxy clusters  \cite{2019MNRAS.483.3754I,2009MNRAS.397.1885M, 2014MNRAS.441.3724M, 2007MNRAS.382...29B,2017PhLB..770..440D}, and the accelerated expansion of space-time   \cite{2006JCAP...05..001M,2015EPJC...75..405R}. 

\subsubsection{Solutions to the observational challenges}

\begin{figure}[htb!]
    \centering
    \includegraphics[width= 0.9\columnwidth]{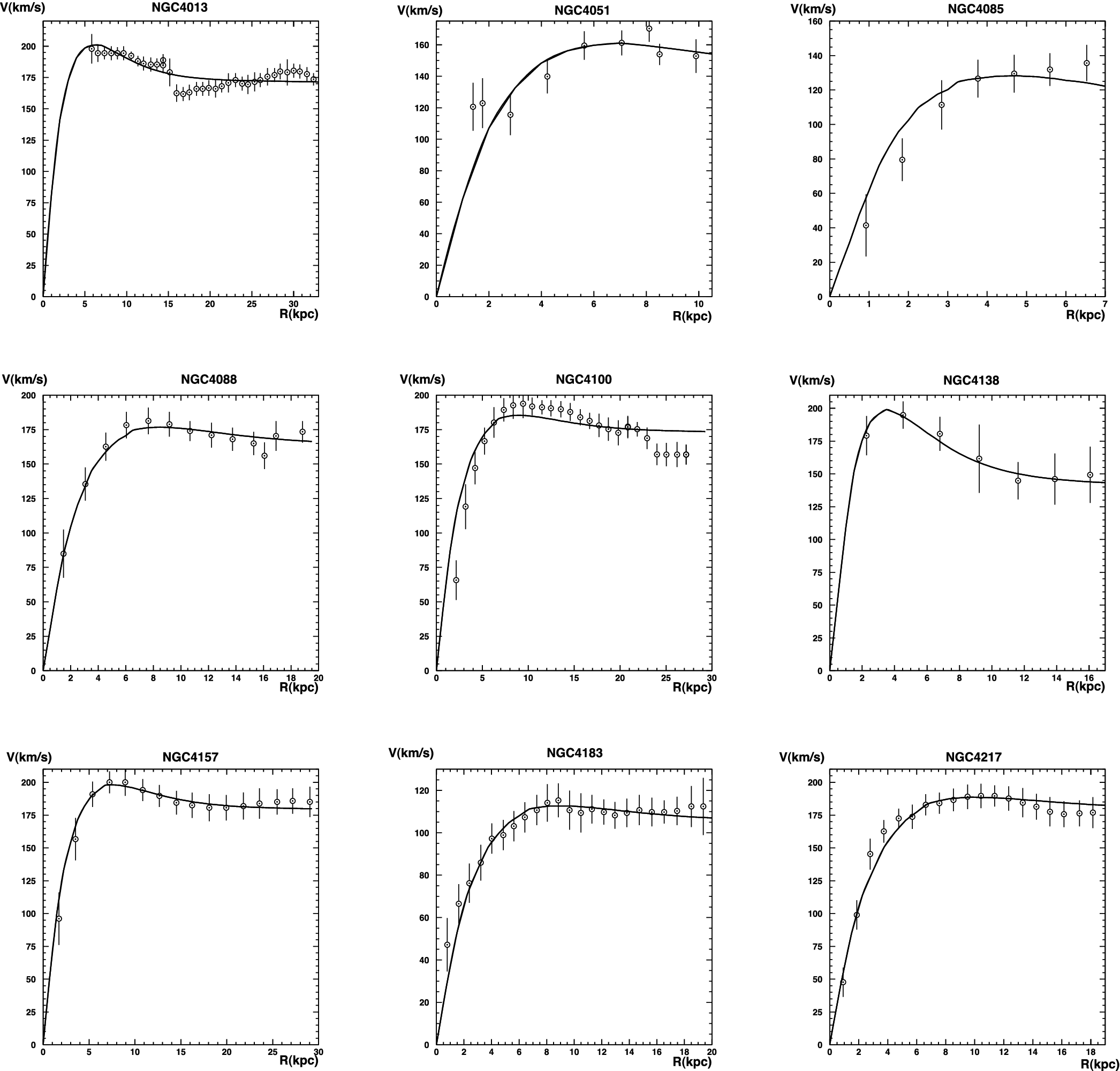}\\
    \caption{ Galaxy rotation curves from the THINGS catalogue fitted with MOG. The best fits are shown as black lines. The figure is reproduced from \cite{2013MNRAS.436.1439M}.}
    \label{fig:mog1}
\end{figure}

In the weak-field limit, the gravitational potential in Eq. \eqref{eq:MOG1} has been used to { fit the observed}  rotation curves 
of both LSB and HSB galaxies from the THINGS catalogue, achieving an excellent agreement between the model and the observations, as shown in Figure \ref{fig:mog1}. This comparison implies a single set of the parameters $\alpha$ and $\mu$ of the gravitational potential: the gravitational strength is $\alpha = 8.89 \pm 0.34$, and the characteristic scale is $\mu = 0.042 \pm 0.004 \, {\rm kpc}^{-1}$ \cite{2013MNRAS.436.1439M}.
MOG also yields a good fit to the rotation curve of the Milky Way with total mass $M\approx 5\times10^{10}\, {\rm M}_\odot$ \cite{2015PhRvD..91d3004M}. Finally, no significant difference with Newtonian dynamics is found by studying the stability of the disk in MOG  \cite{2015ApJ...802....9R}; this result implies that the formation and evolution of self-gravitating systems is not altered. 

Rotation curves of less massive systems also are well accommodated in the MOG theory. The rotation curves of the dwarf galaxies in the LITTLE THINGS catalogue constrain the parameters $(\alpha, \mu)$  with a relative error of 10\% \cite{2017MNRAS.468.4048Z}. Furthermore, MOG is consistent with the RAR observed in the SPARC galaxies \cite{2019PDU....25..323G}. Finally,  the observed velocity dispersion of the ultra-diffuse galaxy NGC1052-DF2  $\sigma =3.2^{+5.5}_{-3.2}$ km/s \cite{2018Natur.555..629V} is consistent with the value $\sigma = 3.9 $ km/s estimated in MOG  \cite{2019MNRAS.482L...1M}. 

\begin{figure}[htb!]
    \centering
    \includegraphics[width= 0.8\columnwidth]{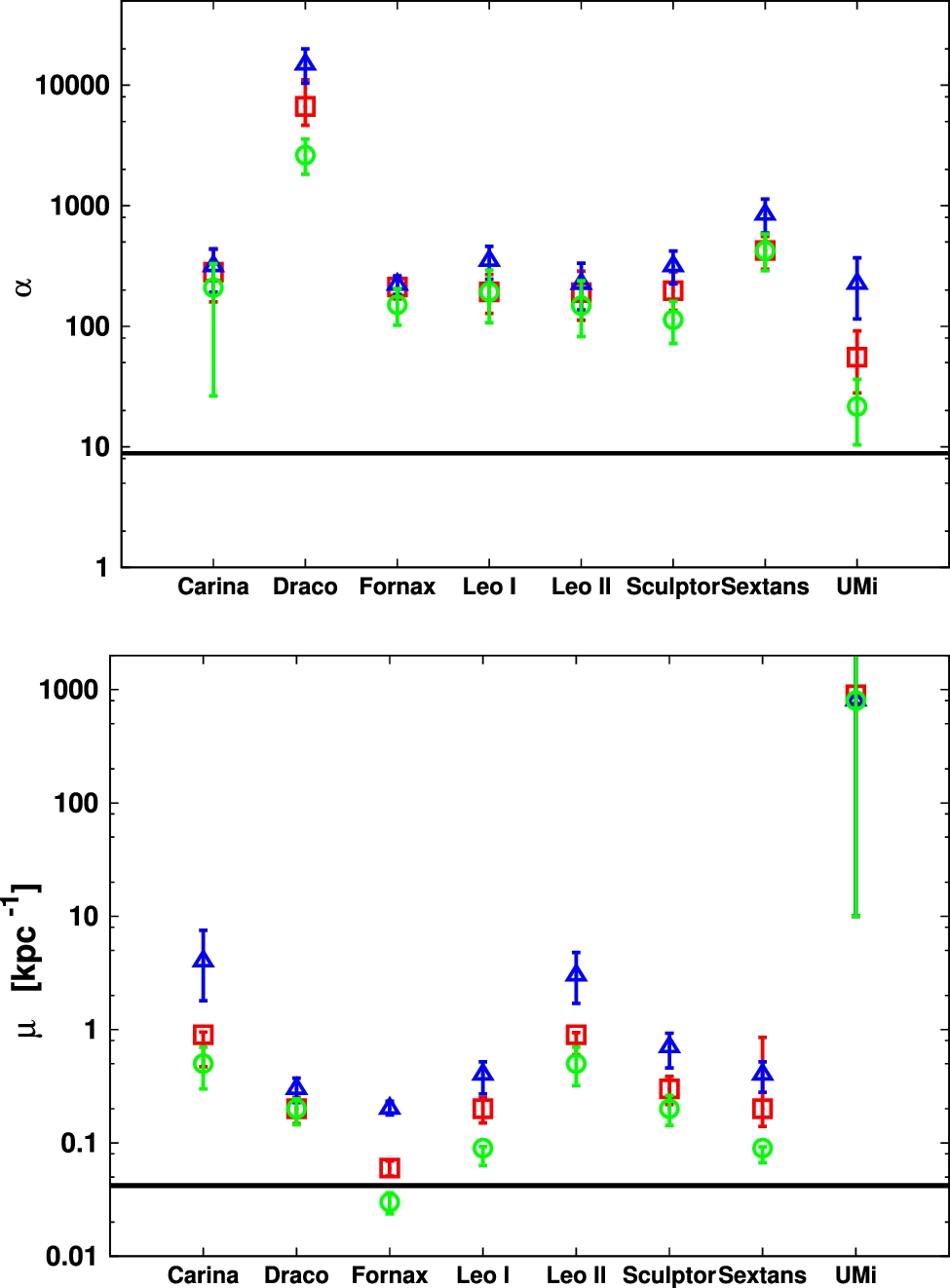}\\
    \caption{ The top and bottom panels show the best-fit $\alpha$ and $\mu$ parameters of the MOG gravitational potential in Eq. \eqref{eq:MOG1} with their uncertainties as obtained from the analysis of dSphs. The solid lines show the values obtained by fitting the rotation curves of more massive galaxies \cite{2013MNRAS.436.1439M}. For each galaxy, three assumptions on the mass-to-light ratio are used, i.e. $M/L_V =[ 0.5, 2, 5]$,  corresponding to blue triangles, red squares, and green circles respectively. The figure is reproduced from \cite{2016MNRAS.463.1944H}.}
    \label{fig:mog2}
\end{figure}

However, other analyses encounter some difficulties. { A recent analysis  shows that the theory fails to reproduce the observed rotation curve of the Milky Way at radii  $<20$ kpc \cite{2018PhRvD..98j4061N}, at odds with previous results showing the capability of MOG to fit the rotation curve of the Milky Way \cite{2015PhRvD..91d3004M}. This inconsistency remains unexplained.} Furthermore, a detailed analysis of the velocity dispersion profile of the Milky Way's dwarf spheroidals leads to mass-to-light ratios in the $V$-band in the range $M_*/L = 5.2^{+1.0}_{-0.9}$~${\rm M}_\odot/{\rm L}_\odot$ for Fornax to $M_*/L=152.3^{+78.8}_{-62.2}$~${\rm M}_\odot/{\rm L}_\odot$ for Draco. This wide interval of values is { inconsistent with the stellar population synthesis models}. Additionally, a single set of the parameters $\alpha$ and $\mu$ 
for all the galaxies in the sample { cannot be} found. These results are summarized in Figure \ref{fig:mog2}; this figure also shows a difference of a few order of magnitudes between the parameters constrained by the dSphs galaxies (colored data points) and  by the more massive galaxies in the THINGS catalogue (black lines) \cite{2016MNRAS.463.1944H}.  

These difficulties lead to argue that MOG may fail at reproducing the dynamics of pressure-supported systems and, thus, at solving the CCP problem. Along the same lines, the results obtained by comparing the predicted velocity dispersion profile of Antlia II with the observed one  \cite{2019MNRAS.488.2743T} show that MOG cannot explain the existence of such a large core ($\sim 2.8$ kpc) without introducing a strongly negative anisotropy parameter $\beta\approx-18$, which means that stars  have a tangential velocity much larger than the radial counterpart \cite{idm2019}. 

In addition, high-resolution N-body simulations show substantial differences between MOG and Newtonian gravity with dark matter. In MOG, the growth rate of the bar { in disk galaxies} is slower, and the final size { of the bar}  is almost an order of magnitude smaller \cite{2018ApJ...854...38R}. Moreover, at cosmological scales, where MOG is supposed to behave like the $\Lambda$CDM model, the growth rate of the perturbations is reduced, and the value of the normalization of the power spectrum $\sigma_8$ is increased to 1.44 \cite{JAMALI2020135238}, which is inconsistent with $\sigma_8=0.802\pm0.018$ obtained by the Planck satellite \cite{2018arXiv180706209P}.
Finally, due to the lack of cosmological N-body simulations in the context of the MOG theory, { the MSP and PSP can not be currently addressed}.

\subsection{ f(R)-gravity}\label{sec:f(R)gravity}

Extensions of General Relativity  are obtained by including  higher-order curvature  invariants in the Lagrangian, such as $R^{2}$, $R_{\mu\nu} R^{\mu\nu}$, 
$R^{\mu\nu\alpha\beta}R_{\mu\nu\alpha\beta}$, $R \,\Box R$, or $R \,\Box^{k}R$, and minimally or non-minimally coupled terms between scalar fields and geometry, such as $\phi^{2}R$ \cite{2010LRR....13....3D,CAPOZZIELLO2011167, 2011PhR...505...59N}. These theories can be roughly classified as \emph{Scalar-Tensor Theories}
if the geometry is non-minimally coupled to some scalar field, and  \emph{Higher-Order Theories}
if the action contains derivatives of the metric components of order higher than two.  Combinations of both types of theories give rise to higher-order/scalar-tensor theories of gravity.

Among them,  $f(R)$-theories extend General Relativity  by replacing the Einstein-Hilbert Lagrangian, which is linear in the Ricci scalar, with a more general function $f(R)$ of the
curvature $R$:
\begin{align}\label{eq:FOGaction}
S\,=\,\int d^{4}x\sqrt{-g}\left[f(R)+\frac{8\pi G}{c^4}\mathcal{L}_m\right]\, ,
\end{align}
where  $\mathcal{L}_m$ is the matter Lagrangian. The field equations thus read
\begin{equation}\label{eq:fieldequationFOG}
{\begin{array}{l}
f'(R)R_{\mu\nu}-\frac{1}{2}f(R)g_{\mu\nu}-f'(R)_{;\mu\nu}+g_{\mu\nu}\Box f'(R)=\frac{8\pi G}{c^4}\,T_{\mu\nu}\,,
\end{array}}
\end{equation}
where $ T_{\mu\nu}$ is  the energy momentum tensor of matter.
Back in the 1980s, A. Starobinsky made the first attempt to describe the acceleration
of the Universe by extending General Relativity { along this line} \cite{Starobinsky1980}. Afterwards, many other $f(R)$ models have been  considered and tested to give an alternative explanation to the cosmological constant \cite{WayneHU2007,Starobinsky2007,2008GReGr..40..357C,Miranda2009,Li2007,Amendola2008,2009JCAP...02..034G, 2015Univ....1..123D,2019EPJC...79...93C,2019PDU....25..305L}. However, there is no statistical evidence favouring any of these models over  $\Lambda$CDM \cite{Lazkoz_2018,2019arXiv191105983H}.

{ Usually, $f(R)$ gravity is invoked to replace either the Einstein-Hilbert Lagrangian with a cosmological constant or dark energy models. However, $f(R)$ gravity can also partly remove the need for  dark matter on small scales.  
In 1977, Stelle already noted that, in the weak-field limit, $R^2$-gravity gives rise to Yukawa-like corrections to the Newtonian gravitational potential} \cite{PhysRevD.16.953}. Under the hypothesis that the $f(R)$ Lagrangian is expandable in Taylor's series \cite{2012AnP...524..545C}:
\begin{equation}
\label{eq:tay}
    f(R)= \sum_n \dfrac{f^n(R_0)}{n!}(R-R_0)^n  \simeq f_0+f_0'R+f_0''R^2+...\,,
\end{equation}
the gravitational potential can be recast as follows: 
\begin{equation}\label{eq:gravpot1}
\Phi(r) = -\frac{G M(r)}{
(1+\delta) r}\left(1+\delta e^{-\frac{r}{L}}\right)\,,
\end{equation}
where the parameter $L$ represents the effective scale length  above which the corrections to the gravitational potential are relevant, and the parameter $\delta$ is related to the strength of the gravitational force  \cite{2012AnP...524..545C}. These parameters are related to the coefficients of Taylor's series as: $\delta=f_0'-1$, and $L= [-6f''_0/f'_0]^{1/2}$.
These corrections may affect the astrophysical scales of galaxies and galaxy clusters while being negligible at the Solar systems scale \cite{2004PhLA..326..292C,2007MNRAS.375.1423C,2007MNRAS.381.1103F,2009A&A...505...21C, 2011MNRAS.414.1301C,2012AnP...524..545C,2012PhRvD..85d4022C,2013MNRAS.431..741D,2014MNRAS.442..921D,2015IJGMM..1250040D,2016PhRvD..93l4043D,2018PhRvD..97j4068D,2018PhRvD..97j4067D,2018EPJC...78..916D}. Therefore, geometric modifications of this type may serve also to fully or partially replace dark matter in the energy-density content of the Universe.

\subsubsection{Solutions to the observational challenges}

Back in the 1980s, R.H. Sanders found the Yukawa-correction term to be able to { describe} the rotation curves of several spiral galaxies \cite{1986MNRAS.223..539S}. More recently, it was shown that $R^n$-gravity models are able to describe the kinematics of the stars in spiral galaxies without resorting to dark matter  \cite{2004PhLA..326..292C,2007MNRAS.375.1423C,2009A&A...505...21C,2007MNRAS.381.1103F,2014IJMPD..2342005S}.  Hybrid models combining $f(R)$-gravity and dark matter have also been proposed: studies of galactic rotation curves find an $8\sigma$ Bayesian evidence favouring $f(R)$-gravity plus a NFW dark matter halo over the CDM model \cite{2018JCAP...08..012D}.
In addition, the rotation curves expected in $f(R)$-gravity combined with cuspy NFW density profiles are favored over the rotation curves expected in CDM  profiles with a core, offering a potential solution to the CCP  \cite{2011MNRAS.414.1301C}. 

\begin{figure}
    \centering
    \includegraphics[width= 0.49\columnwidth]{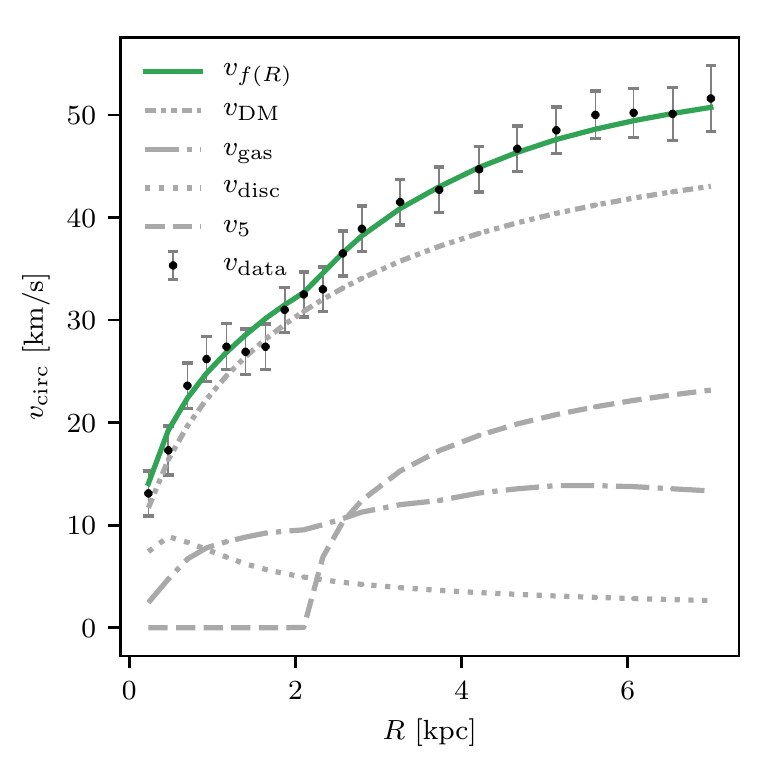}
        \includegraphics[width= 0.49\columnwidth]{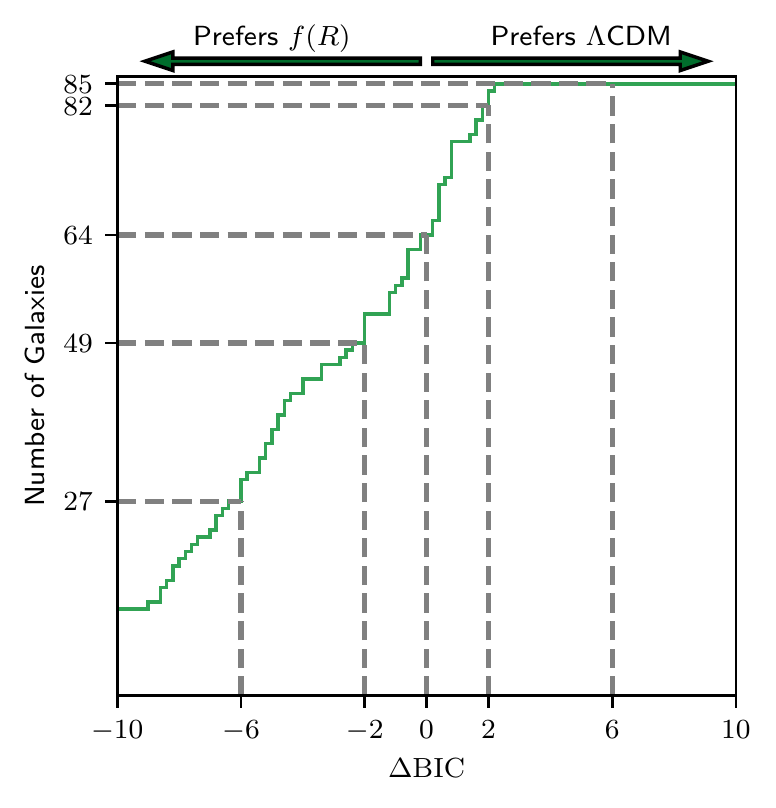}\\
    \caption{{\em Left panel}: rotation curve of NGC 3741 and the corresponding best fit with { chameleon-$f(R)$ gravity combined with a cuspy NFW dark matter halo} (green curve). The grey lines show the contributions of the different galactic components to the $f(R)$-model as labelled.  {\em Right panel}: Cumulative distribution function of   $\Delta BIC$, where BIC is the Bayesian  Information  Criterion,     across SPARC galaxies. Most galaxies in the sample prefer the $f(R)$ model, but not necessarily with the same value of the model parameters. The figure is reproduced from \cite{2019MNRAS.489..771N}.}
    \label{fig:fR}
\end{figure}

A recent study of the chameleon-$f(R)$ gravity \cite[e.g.][]{Brax_2008}, where higher order curvature terms can be recast as a scalar field strongly coupled to matter, tightly constrains the parameters of the model with HI/H$\alpha$  rotation curves  of the disk galaxies in the SPARC sample. 
{Specifically,} the chameleon-$f(R)$ gravity with a cuspy NFW dark matter halo fits well the galaxy rotation curves; the left panel of   Figure \ref{fig:fR} shows { the example} of NGC 3741. Nevertheless, { in contrast with previous results \cite{2018JCAP...08..012D}, there is no statistical evidence favouring this model over Newtonian gravity with a NFW dark matter halo with a core
\cite{2019MNRAS.489..771N}, as illustrated by the Bayesian Information Criteria estimator shown in the right panel of Figure \ref{fig:fR}.} 

Whereas rotation curves are well accommodated in $f(R)$-gravity, there is a lack of studies of pressure supported systems such as dwarf galaxies. Nevertheless, due to the similarities between the gravitational potential in $f(R)$-gravity and in MOG (see Eq. \eqref{eq:MOG1} in Section \ref{sec:MOG}), one may expect to encounter difficulties similar to MOG to explain the dynamics of dwarf galaxies, and to provide a solution to the CCP.  It is worth  mentioning that the possibility that these systems might not be in dynamical equilibrium in the context of these theories may invalidate the constraints \cite{2000ApJ...541..556B,2010ApJ...722..248M,2016ApJ...832L...8M}.

Finally, N-body simulations of $f(R)$-gravity alone (see, for example, \cite{2012JCAP...01..051L,2013MNRAS.436..348P,2014A&A...562A..78L}), and of $f(R)$-gravity with dark matter, such as  massive neutrinos or  WDM particles, indicate the existence of a degeneracy between these models and the standard $\Lambda$CDM paradigm \cite{2014MNRAS.440...75B,10.1093/mnras/stx2594, Arnold_2019}. 
For instance, N-body simulations of $f(R)$ gravity  with $f_{R,0} = -1 \times 10^{-4}$ plus massive neutrinos with a total neutrino mass of $\Sigma_i m_{\nu_i} = 0.4$ eV show a  matter power spectrum and a halo mass function consistent with the predictions of $\Lambda$CDM at 10\% and 20\% accuracy levels, respectively \cite{2014MNRAS.440...75B}.

\section{Summary and discussion}\label{sec:summary}

The cosmological parameters of the $\Lambda$CDM model  are measured at an accuracy of $\sim 1\%$ or smaller \cite{2013ApJS..208...19H,2018arXiv180706209P}, effectively confirming the capability of the model to describe the cosmological evolution of the Universe. In this model, dark matter is $\sim 85\%$ of the total matter density of the Universe \cite{2018arXiv180706209P}, but its fundamental nature  is still unknown \cite{2018RvMP...90d5002B}. WIMPs are the most promising candidate for CDM \cite{Jungman_1996,1998pesu.book....1M}.  WIMPs are  massive and weakly interacting particles, expected in supersymmetric theories, which decoupled from the primordial plasma when they were non-relativistic.  Collider, direct, and indirect searches for such  particles are ongoing \cite{2017PhRvL.118b1303A,2011PhRvL.107m1302A,PhysRevLett.114.141301,2019ICRC...36..506B,2015APh....62...12A,2013PhRvL.111q1101B}. Nevertheless, LHC has not provided any evidence  of supersymmetry so far \cite{PhysRevD.98.030001}, and the claimed DAMA/LIBRA/CoGeNT annual modulation is still under debate \cite{Bernabei_2013, 2014EPJWC..7000043B,PhysRevLett.114.151301} (for further details see Section \ref{sec:overviewCDM}). 

The lack of a direct detection of the dark matter particles is not the only issue encountered by the CDM model: the issues on galactic scales  discussed in Section \ref{sec:obschallenges} are difficult to accommodate in the context of the CDM model \cite{1994Natur.370..629M,10.1046/j.1365-8711.1999.03039.x,Boylan_Kolchin_2011,Bullock_2017,2017Galax...5...17D}. Supernovae feedback and dynamical friction from baryonic clumps may help to explain the transition from the cuspy dark matter profiles, predicted by CDM N-body simulations, to the cores suggested by the observed kinematic properties of LSB, dwarf and ultra-faint galaxies \cite{10.1093/mnras/stw713}. In principle, these baryonic mechanisms may  also help to reduce the overabundance of subhalos and describe the observed radial acceleration relation of disk galaxies. However, the effectiveness of baryonic feedback is still under debate, as discussed in Section \ref{sec:obschallenges}.

At this point, a question arises: are the lack of detection of a CDM particle and the observational issues at galactic scales indicating a breakdown of the model? The current state of the debate is inconclusive and puzzling. A change of the dark matter paradigm or modified gravity models can help to solve some of the issues but they do not offer a definitive answer to the question. 
{ A visual overview of whether the models discussed in Sections \ref{sec:particlemodels} and \ref{sec:gravitmodels}  either solve or do not display the challenges of the CDM model is given in Table \ref{tab:table_summary}; for the sake of completeness, Table \ref{tab:table_summary} also displays the ability of each model to explain the large-scale structure of the Universe, successfully described by the $\Lambda$CDM model.} 

\begin{table}[htb!]
\centering
\caption{Summary of whether alternative dark matter (DM) and gravity models  { either solve or do not display} the challenges of the CDM model discussed in this work.
Here, we refer to a problem as ``solved'' when the model does not currently display major tensions with the observations.
The ability of the models to explain the large-scale structure of the Universe is also included.}
  \begin{tabular}{lcccccc}
    \toprule
    { } & {Rotation curves} & {Cusp/core} & Missing    & Too-big-     & Planes of   & Large scale \\
    { } & {and scaling}     & {problem}   & satellites & to-fail     & satellites  & structure and \\
    { } & {relations}       &  ~~~~       & problem    &  problem    & problem     & cosmological scales\\
    \midrule
      Warm DM &{\textcolor{green}{\ding{52}}} & {\textcolor{red}{\ding{56}}}  & {\textcolor{green}{\ding{52}}}  & {\textcolor{green}{\ding{52}}} & {{\scalebox{0.6}{\bcloupe}}} &{{\scalebox{0.6}{\bcloupe}}} \\ [0.1cm]
     Self-interacting DM & {\textcolor{green}{\ding{52}}} & {\textcolor{green}{\ding{52}}} & {{\scalebox{0.6}{\bcloupe}}} & {\textcolor{green}{\ding{52}}} & {{\scalebox{0.6}{\bcloupe}}} & {\textcolor{green}{\ding{52}}}  \\ [0.1cm]     
     QCD axions & {\textcolor{green}{\ding{52}}} & {{\scalebox{0.6}{\bcloupe}}} & {{\scalebox{0.6}{\bcloupe}}} & {{\scalebox{0.6}{\bcloupe}}} & {{\scalebox{0.6}{\bcloupe}}} & {\textcolor{green}{\ding{52}}} \\ [0.1cm]
     Fuzzy DM & {{\scalebox{0.6}{\bcloupe}}} & {\textcolor{green}{\ding{52}}}  &  {\textcolor{green}{\ding{52}}} & {{\scalebox{0.6}{\bcloupe}}} &  {{\scalebox{0.6}{\bcloupe}}}  & {{\scalebox{0.6}{\bcloupe}}} \\ [0.1cm]
     MOND & {\textcolor{green}{\ding{52}}} & {\textcolor{green}{\ding{52}}} & {\textcolor{green}{\ding{52}}} & {\textcolor{green}{\ding{52}}}  & {{\scalebox{0.6}{\bcloupe}}} & {\textcolor{red}{\ding{56}}} \\ [0.1cm]
    MOG & {\textcolor{green}{\ding{52}}} & {\textcolor{red}{\ding{56}}} & {{\scalebox{0.6}{\bcloupe}}} & {{\scalebox{0.6}{\bcloupe}}} &  {{\scalebox{0.6}{\bcloupe}}}  &  {\textcolor{green}{\ding{52}}} \\ [0.1cm]
     $f(R)$-gravity & {\textcolor{green}{\ding{52}}} & {{\scalebox{0.6}{\bcloupe}}} &  {\textcolor{red}{\ding{56}}}  & {{\scalebox{0.6}{\bcloupe}}}  &  {{\scalebox{0.6}{\bcloupe}}}  & {\textcolor{green}{\ding{52}}}\\ [0.1cm]
    \bottomrule
  \end{tabular}\\
  \begin{tabular}{ccccc}
  {\textcolor{green}{\ding{52}}} = Solved or { Not present }& ~~~ & {\textcolor{red}{\ding{56}}} = Not solved & ~~~ &   {\textcolor{red}{\scalebox{0.6}{\bcloupe}}} = Under investigation\\ 
  \end{tabular}
  \label{tab:table_summary} 
\end{table}

WDM, SIDM, ALPs, and FDM,  discussed in Section \ref{sec:particlemodels}, allow the existence of a dark matter core in dwarf galaxies, 
hence,  providing, in principle, a solution to the CCP. Moreover, { WDM and FDM} show a cut-off in the matter power spectrum that suppresses the formation of halos below a given mass threshold \cite{PhysRevLett.42.407,2014NatPh..10..496S}. However, WDM does not alleviate any of the CDM issues at galactic scale when the constraints from the Lyman-$\alpha$ forest or the gravitational lensed quasars are taken into account. { The cores in the WDM halos are indeed too small to solve the CCP, although they might contribute to fix the TBTF problem}. It thus remains unclear whether WDM models may represent a viable solution \cite{Schneider_2014}. SIDM  solves the CCP, but its ability to solve the MSP and PSP requires further investigations that take into account the baryonic feedback. QCD axions are highly motivated dark matter candidates from the perspectives of both particle physics and cosmology. However, whether they can solve the small-scale issues of CDM has not been properly investigated yet, because the role of the quantum nature of thermalizing QCD axions on cosmic small scales demands a more rigorous theoretical framework. Finally, FDM also encounters its own difficulties. The boson mass, $\sim 10^{-22}$ eV, required to explain the dynamics of the dwarf galaxies and to solve the CCP and the MSP  is almost two orders of magnitude lower than the boson mass, $\sim 7\times 10^{-21}$ eV, needed to account for the Lyman-$\alpha$ data  \cite{2016JCAP...08..012B}.  Nevertheless, the debate is far from being settled \cite{2017PhRvD..95d3541H}. Uncertainties on the thermal histories and the underlying reionisation model may invalidate these constraints \cite{GARZILLI2017258, 2017PhRvD..96l3514K}

While the CCP and MSP may usually be solved in paradigms beyond the standard model of particle physics, both are hardly addressed in modified theories of gravity. Both MOG and $f(R)$-gravity are able to reproduce the rotation curve of the giant spiral galaxies \cite{2013MNRAS.436.1439M,2015PhRvD..91d3004M,2004PhLA..326..292C,2007MNRAS.375.1423C,2009A&A...505...21C,2007MNRAS.381.1103F,2014IJMPD..2342005S}, { whereas} a deeper investigation of dwarf and ultra-faint galaxies, which appear to be dark-matter dominated and  show the presence of wide cores, is still missing. Recent results obtained in the context of MOG have shown that, although this modified gravity model has been built to specifically replace dark matter, it is unable to self-consistently reproduce the dynamics of the dwarf galaxies orbiting the Milky Way \cite{2017MNRAS.468.4048Z,idm2019} (Section \ref{sec:MOG}). While similar studies for $f(R)$-gravity are missing, one may expect similar results on the basis of the theoretical analogies of the weak field limits of MOG and $f(R)$. Nevertheless, one must be careful when comparing predictions from modified gravity models to the observational data. In fact, the assumption that all real systems are in dynamical equilibrium  may not hold and, { if so, their amount of dark matter would be substantially overestimated } \cite{2000ApJ...541..556B,2010ApJ...722..248M,2016ApJ...832L...8M}. 

MOND is fundamentally different from the other suggested solutions that attempt to remove the requirement of dark matter (Section \ref{sec:MOND}). MOND predicted many of the challenges of the conventional dark matter paradigm related to disk and dwarf galaxies many years before these challenges were actually observed. MOND  { might also have  a natural solution for the formation of the plane of satellites, although it  is yet to be demonstrated that this is indeed the case. In addition,}  MOND remains a phenomenological model  whose extension  providing a cosmological model and describing the formation and evolution of the large-scale structure is not yet available.

To sum up, no unambiguous signature of dark matter has been found yet. The standard CDM paradigm encounters several issues at galactic scales, and its capability to solve them is still unclear, as discussed in Section \ref{sec:obschallenges}.    Ideally, the next generation facilities, such as the Maunakea Spectroscopic Explorer (MSE, \cite{2019arXiv190303155L}) or the Large Synoptic Survey Telescope (LSST, \cite{2019arXiv190201055D}), will achieve the sensitivity needed to discover other ultra-faint and ultra-diffuse galaxies, whose observed properties can strongly constrain the { theoretical models}. In fact, understanding the inner structure of these galaxies has a fundamental role in our comprehension of the nature of dark matter.  Additional observational constraints may come from the next generation astrometric missions, such as the proposed Theia satellite \cite{theia,malbet2019esa}. Theia is expected to measure the proper motion of stars in the Milky Way dwarf satellites that, together with high precision measurements of the position of their stars, would shed light on the dynamics of these galaxies. We can thus either constrain the theory of gravity or the nature of the dark matter, if  the  presence of either cores or cusps in their dark matter density profile is definitely confirmed. { This evidence will complement the constraints from other astrophysical and cosmological probes and from indirect or direct searches of dark matter.}


\vspace{6pt}

\noindent{\bf Author Contributions:} All authors contributed equally to this review article.

\vspace{6pt}
\noindent{\bf Funding:} This research received no external funding.

\vspace{6pt}
\noindent{\bf Acknowledgments:} We sincerely thank  three referees whose suggestions and corrections improved our review. Possible residual inaccuracies remain our full responsibility. IDM and SSC are supported by the grant ``The Milky Way and Dwarf Weights with Space Scales" funded by University of Torino and Compagnia di S. Paolo (UniTO-CSP). We also thank Compagnia
di San Paolo (CSP) for funding the fellowship of VC and the Italian Ministry of Education, University and Research (MIUR) under the Departments of Excellence grant L.232/2016 for funding the fellowship of AG. We also acknowledge partial support from the INFN grant InDark. This research has made use of NASA's Astrophysics Data System Bibliographic Services.

\vspace{6pt}
\noindent{\bf Conflicts of Interest:} The authors declare no conflict of  interest.

\vspace{6pt}
\noindent{\bf Abbreviations}\\
The following abbreviations are used in this manuscript:\\

\noindent 
\begin{tabular}{@{}ll}
ALPs & Axion-Like Particles \\ 
AM & Abundance Matching\\
BEC & Bose-Einstein Condensate \\ 
BTFR & Baryonic Tully-Fisher Relation\\
CASP & Centaurus A satellite plane\\
CCP & Cusp/Core Problem\\
CDM & Cold Dark Matter\\
CenA & Centaurus A \\
CMBR & Cosmic Microwave Background Radiation\\
dSph & Dwarf Spheroidal\\
FDM & Fuzzy Dark Matter\\
GPoA & Giant Plane of Andromeda\\
HSB & High Surface Brightness\\
IMF & Initial Mass Functions\\
$\Lambda$CDM & $\Lambda$-Cold Dark Matter\\
MACHOs &  Massive Astrophysical Compact Halo Object\\
MDAR & Mass Discrepancy Acceleration Relation\\
MOND & Modified Newtonian Dynamics\\
NFW & Navarro-Frenk-White\\
PQ & Peccei-Quinn \\
QCD & Quantum Chromodynamics \\ 
RAR &  Radial Acceleration Relation\\
SPS & Stellar-Population-Synthesis\\
VPOS & Vast Polar Structure\\
TBTF & Too-Big-To-Fail\\
TDGs & Tidal Dwarf Galaxies\\
ULALPs & Ultra-Light ALPs\\
WMAP &  Wilkinson Microwave Anisotropy Probe\\
WDM & Warm Dark Matter\\
WIMPs & Weakly Interacting Massive Particles\\
\end{tabular}


\end{document}